\documentclass[12pt,preprint]{aastex}

\newcommand{\AAA}{${\rm\AA \;}$ }

\newcommand{\HdA}{H\delta_A }
\newcommand{\Dbreak}{D_{4000} }

\slugcomment{}

\author{
Timothy A. Reichard\altaffilmark{1}, Timothy
M. Heckman\altaffilmark{1}, Gregory Rudnick\altaffilmark{2,5}, Jarle
Brinchmann\altaffilmark{3}, Guinevere Kauffmann\altaffilmark{4}, Vivienne Wild\altaffilmark{4}}

\altaffiltext{1}{Department of Physics and Astronomy, The Johns Hopkins University, 3400 North Charles Street, Baltimore, MD 21218-2686.}
\altaffiltext{2}{National Optical Astronomy Observatory (NOAO), 950 N. Cherry Ave., Tucson, AZ 85719, USA}
\altaffiltext{3}{Leiden Observatory, Leiden University, P.O. Box 9513, 2300 RA, Leiden, The Netherlands.}
\altaffiltext{4}{Max-Planck-Institut f\"{u}r Astrophysik, D-85748 Garching, Germany.}
\altaffiltext{5}{Goldberg Fellow}

\begin{document}

\title{The Lopsidedness of Present-Day Galaxies: Connections to the Formation of Stars, the Chemical Evolution of Galaxies, and the Growth of Black Holes}

\begin{abstract}
A global lopsidedness in the distribution of the stars and gas is
common in galaxies. It is believed to trace a non-equilibrium
dynamical state caused by mergers, tidal interactions, asymmetric
accretion of gas, or asymmetries related to the dark matter halo. We
have used the Sloan Digital Sky Survey (SDSS) to undertake an
investigation of lopsidedness in a sample of $\sim$25,000 nearby
galaxies ($z <$ 0.06). We use the $m=1$ azimuthal Fourier mode between
the 50\% and 90\% light radii as our measure of lopsidedness. The
SDSS spectra are used to measure the properties of the stars, gas, and
black hole in the central-most few-kpc-scale region. We show that
there is a strong link between lopsidedness in the outer parts of the
galactic disk and the youth of the stellar population in the central
region. This link is independent of the other structural properties of
the galaxy. These results provide a robust statistical
characterization of the connections between
accretion/interactions/mergers and the resulting star formation. We
also show that residuals in the galaxy mass-metallicity relation
correlate with lopsidedness (at fixed mass, the more metal-poor
galaxies are more lopsided). This suggests that the events causing
lopsidedness and enhanced star formation deliver lower metallicity gas
into the galaxy's central region. Finally, we find that there is a
trend for the more powerful active galactic nuclei (the more rapidly
growing black holes) to be hosted by more lopsided galaxies (at fixed
galaxy mass, density, or concentration). However if we compare samples
matched to have both the same structures {\it and central stellar
populations}, we then find no difference in lopsidedness between
active and non-active galaxies. Indeed the correlation between the
youth of the stellar population and the rate of black hole growth is
stronger than the correlation between lopsidedness and either of these
other two properties. This leads to the following picture. The
presence of cold gas in the central region of a galaxy (irrespective
of its origin) is essential for both star-formation and black hole
growth. The delivery of cold gas is aided by the processes that
produce lopsidedness. Other processes on scales smaller than we can
probe with our data are required to transport the gas to the black
hole. 
\end{abstract}

\section{Introduction}

In the standard $\Lambda CDM$ universe, galaxies continue to grow
significantly at the current epoch, accreting an average of roughly
5\% of their mass per Giga-year (e.g., \citealt{k+05,k+06}). Minor
mergers, tidal interactions with close companions, and asymmetric
accretion of cold gas can all perturb the underlying structure of the
dark matter halo, the stars, and the gas in galaxies for time-scales
of-order a Giga-year (e.g., \citealt{zr97,lev+98,jog99,kr+02,bou+05,dim07,map+08,cox08}).

A characteristic observational signature of these types of
non-equilibrium situations is a global asymmetry (lopsidedness) in the
distribution of the stars and/or gas. Indeed, galaxies commonly
exhibit asymmetric structures. Asymmetry in the global \ion{H}{1} 21cm
emission-line profile is present in half or more of disk galaxies
\citep{ric+94,mat+98}. Lopsided
distributions of the \ion{H}{1} gas are revealed by spatial maps of
galaxies in the field (e.g., \citealt{swa+99}) and in groups
\citep{ang+06,ang+07}.  The stellar mass distributions of disk
galaxies are often lopsided as well. \citet{zr97} showed
that $\sim$30\% of a sample of 60 field spiral galaxies had a
significant $m=1$ azimuthal Fourier component in the stellar
mass. \citet{rud+98} showed that lopsidedness in the stellar mass distribution was common in early-type disk galaxies. Simulations of spiral galaxies have shown that the typical measured amplitudes of lopsidedness cannot be explained by internal dynamical mechanisms alone (e.g. \citealt{bou+05}). The external processes described above are required.

It has long been recognized that tidal interactions and mergers can
act as triggers for star-formation \citep{too+72,lar+78}.  The resulting non-axisymmetric time-dependent
gravitational potential can transport angular momentum out of the gas
and also lead to shocks in which the gas is compressed and kinetic
energy is ultimately converted into radiation and lost to the system
(e.g., \citealt{mih+96,dim07,cox08}). These processes lead to the inflow
of gas, an increase in gas density (either globally or locally), and
hence an increase in the star formation rate \citep{ken98}.

\citet{li+08a} have investigated the link between interactions
and star formation in the largest sample of galaxies studied to
date. They found a strong correlation between the proximity of a near
neighbor (closer than 100 kpc) and the average specific star formation
rate. This result pertains to the relatively early stages of an
eventual merger. Using lopsidedness as a probe would allow us in
principal to measure the history of the enhancement in star formation
(in an time-averaged sense) all the way through to the post-merger
phase. \citet{zr97} exploited this idea, and found a
correlation between lopsidedness and excess $B$-band luminosity
emitted by a young stellar population.  \citet{rud+00} showed
that both recent (the past 1 Gyr) and ongoing star formation are
correlated with lopsidedness, and estimated that as much as 10\% of
stellar mass in typical galaxies can be formed in such
events. However, these papers investigated very small samples compared
to the \citet{li+08a} study.

An additional link between lopsidedness and star formation is
indicated by the results in \citet{k+05} and \citet{bou+05}. The
former authors use SPH simulations to conclude that the star formation
history of galaxies is primarily regulated by the accretion rate of
cold gas, rather than by mergers. The accretion of this cold gas will
not occur in a spherically symmetric fashion, and \citet{bou+05} argue
that this will lead to long-lived lopsidedness in disk galaxies.

Large samples are crucial to the investigation of the connection
between lopsidedness and the recent star formation history. In our
recent paper (\citealt{paper1}, hereafter Paper I) we reported on our
analysis of multi-color Sloan Digital Sky Survey (SDSS) images of
$\sim$25000 low-redshift ($z < 0.06$) galaxies. We confirmed that
Lopsidedness in the stellar mass distribution is indeed very common, and quantified its distribution as a function of the principal structural properties of the galaxies. We
found that galaxies of lower mass, lower density, and lower
concentration are systematically more lopsided. Similarly,
\citet{kau+03a,kau+03b} showed that galaxies with lower mass, density,
and concentration have younger stellar populations on-average. Taking
these results together, the causal connections between star formation,
lopsidedness, and other galaxy structural properties are
ambiguous. Unraveling this web of mutual correlations requires the
careful analysis of a large and homogeneous sample.

The strong correlation between the stellar mass of a galaxy and the
metallicity of its interstellar medium \citep{tre+04} provides a
powerful constraint on the chemical evolution of galaxies and the
intergalactic medium (e.g., \citealt{dal07}). If lopsidedness in a
galaxy has been recently induced by either a minor merger with a
gas-rich low mass companion galaxy or the accretion of cold
intergalactic gas, then one signature would be a decrease in the
metallicity of the interstellar medium in lopsided galaxies compared
to other galaxies of the same mass. This can be tested with a large
sample of galaxies.

Interactions and mergers are also believed to be an important
mechanism for fueling the formation and growth of a central
supermassive black hole. This idea dates back at least to Toomre \&
Toomre (1972), and is one of the cornerstones of contemporary models
that attempt to explain the co-evolution of galaxies and black holes
within the context of the hierarchical build-up of structure (e.g.,
\citealt{hop+07,dim+07}). However, the observational verification of
this idea remains insecure. At high-redshift the data are still too
sparse to be conclusive, while at low-redshift there have been many
past studies that have led to contradictory results. \citet{li+08b}
have undertaken an analysis of the largest low-redshift sample to
date, based on $10^5$ galaxies in the SDSS. They find a strong link between close companions and star
formation, but none between close companions and AGN. One possibility
is that the black hole growth occurs only late in a merger, after the
companion galaxy has been captured. This idea can be tested using lopsidedness and a similarly large sample since lopsidedness can be used to trace later stages of interaction than that seen in pair studies.

In this study, we use the large data-set described in Paper I to try
to understand how lopsidedness may be linked to star formation,
metallicity, and the AGN phenomenon. We describe our specific galaxy
sample in \S\ref{sec:data} and summarize our measurements of the
structural parameters, star formation indicators, metallicity, and AGN
indicators. Next, in \S\ref{sec:sfh}, we compare the correlations
between star formation and structure with an emphasis on separating
out the dependence of star formation on lopsidedness versus other
galaxy structural properties. Next, in \S\ref{sec:met} we explore
whether the residuals in the mass-metallicity relation correlate with
lopsidedness, and test whether this is a fundamental correlation, or
only a secondary one induced through a mutual dependence of
metallicity and lopsidedness on galaxy structure. Finally, in
\S\ref{sec:agn}, we compare lopsidedness and the structural and
stellar population properties for the host galaxies of AGN, to test
for a possible link between interactions/mergers/accretion and the
growth of black holes.  We discuss the implications of our conclusions
in \S\ref{sec:conclusions}.

\section{Data \label{sec:data}}

For the convenience of the reader, we briefly summarize the data, our
sample selection, and our methodology here. The details are discussed
in Paper I.  The sample of galaxies was taken from the Sloan Digital
Sky Survey \citep{y+00,s+02}, a large survey of photometric and
spectroscopic data across $\pi$ sr. of the northern sky. The sample is
derived from SDSS Data Release 4 \citep{dr4}. The survey's dedicated
2.5 m telescope \citep{g+06} at Apache Point Observatory uses a unique
CCD camera \citep{g+98} and drift-scanning to obtain $u$-, $g$-, $r$-,
$i$-, and $z$-band photometry \citep{f+06,h+01,i+04,sm+02,t+06}. The
pixel scale is 0\farcs396/px.  Fiber spectroscopy is obtained using
3\arcsec~fibers and results in wavelength coverage between
3800$-$9200~\AA~at a resolution $R=\lambda/\delta\lambda = 1850-2200$.

Lopsidedness is defined here as the radially averaged amplitude of the
$m=1$ azimuthal Fourier mode between the radii in the SDSS images
enclosing 50\% and 90\% of the galaxy light ($R_{50}$ and $R_{90}$
respectively). In Paper I we showed that galaxies on-average exhibit
virtually identical distributions of $r$- and $i$-band
lopsidedness (and only slightly larger values in the $g$-band). We
showed that weakness of the color dependence of lopsidedness implied
that lopsidedness primarily traces the asymmetry in the underlying
stellar mass distribution. Since any spatial variations in
mass-to-light ratio will be smaller at longer wavelengths, we use the
$i$-band lopsidedness $A_1^i$ throughout this paper.

The calculation of lopsidedness is subject to a handful of
observational systematic errors.  Insufficient spatial resolution, low
signal-to-noise, and highly inclined galaxies in images lead to
systematically incorrect measurements of $A_1$.  Galaxies subjected to
these systematic effects were eliminated by requiring the following
observational properties: $z < 0.06$, $2R_{50}/($PSF FWHM$) > 4$,
total $S/N > 30$ in the region between $R_{50}$ and $R_{90}$, and $b/a
> 0.4$. These restrictions yield a sample of 25155 galaxies that we
use in our analysis below.  These systematic effects prevent a
reliable determination of lopsidedness in the central region of the galaxies (where the star formation properties are calculated). However, it has been shown that the lopsidedness in mergers is very high in the central as well as the outer regions \citep{jog+06}.

We make extensive use of other galaxy structural parameters derived
from the SDSS data (see \citealt{kau+03a}). The galaxy stellar mass
($M_*$) is derived from the SDSS $z-$band luminosity and a
mass-to-light ratio derived from the spectra. The stellar surface mass
density $\mu_*$ is defined as the mean value within the galaxy
half-light radius ($\mu_* = M_*/2\pi R_{50}^2$). We also use the
galaxy stellar velocity dispersion $\sigma_*$, and the concentration
parameter $C = R_{90}/R_{50}$ (the latter serving as a rough proxy for
Hubble Type, increasing from $\sim2$ for late type galaxies to $\sim$3
for early types).

Most of the quantities we use to characterize the properties of the
stars, ionized gas, and the AGN are based on the SDSS spectra. The
SDSS fiber diameter is 3\arcsec, which for our redshift-limited sample
corresponds to a typical projected diameter of about 3 kpc. Our
measures of lopsidedness are made using the annulus between the radii
enclosing 50\% and 90\% of the light, a region lying entirely outside
the SDSS fiber. { \it Thus, it is important to emphasize that we are
essentially relating the lopsidedness of the outer part of the galaxy to the
central properties of the galaxy.}

We use a variety of measurements to provide information about the star
formation history in this central region \citep{kau+03a}. The
4000~\AA~break, $D_{4000}$, indicates the current luminosity-weighted
mean stellar age.  Young stellar populations produce almost no metal
absorption just shortward of 4000 \AA, giving $D_{4000} \sim 1$, while
older populations have strong metal line absorption, and $D_{4000}$
increases to $\sim$2 for an age of $\sim10^{10}$ years. We use the
narrow definition of $D_{4000}$ as defined by \citet{b83}: the ratio
of the flux density $F_{\nu}$ in the ranges 4050$-$4250 and
3750$-$3950 \AA. Reddening affects this narrow definition less than
other, wider definitions.

The H$\delta$ absorption-line index $H\delta_A$ reaches maximum
strength in a simple stellar population with an age of 0.1-1 Gyr when
A-stars dominate the continuum. We use the $H\delta_A$ index as
defined by \citet{wo97} where an equivalent width is calculated
between two pseudocontinuum bandpasses. The combination of $D_{4000}$
and $H\delta_A$ allows us to recognize objects that have undergone
bursts of star formation within the past Gyr (galaxies having
abnormally strong $H\delta_A$ for their value of $D_{4000}$).

Recently, \citet{wil+07} have applied the methodology of Principal
Component Analysis to the near-UV spectral region using synthetic
spectra derived from a large library of star formation histories. The
amplitudes of the first two principal components correspond closely to
higher signal-to-noise measurements of $D_{4000}$ ($PC_1$: mean age)
and the excess in $H\delta_A$ for a given $D_{4000}$ ($PC_2$: a
post-burst parameter). We use these parameters in our paper.
\footnote{This definition of "post-starburst" is made without regards to the emission-line properties (unlike some other common definitions). This definition allows the inclusion of both AGN and post-starbursts with some residual star-formation.} 
   
We also use the specific SFR ($SSFR$ or $SFR/M_*$: the star formation
rate per solar mass of stars in the galaxy) as computed by
\citet{bri+04} based on the nebular emission-lines. Note that this is
essentially an instantaneous measure of the central star formation
rate (averaged over $\sim10^7$ years), and is hence complementary to
the indicators above which have much longer time-constants.

As we do not have direct measurements of the {\it global} stellar
ages, we use the age-sensitive Petrosian $g-i$ color from the SDSS
images K-corrected to a redshift $z =$ 0.1. This approach has the
disadvantage that the color can be affected by dust extinction as well
as by the mean stellar age.

Type 2 (obscured) AGN are recognized and characterized in the SDSS
sample using the methodology in \citet{kau+03c} and
\citet{hec+04}. This is based on the detection of the signature of the
narrow emission-lines that are excited by the AGN. The
extinction-corrected luminosity of the [\ion{O}{3}]$\lambda$5007
emission-line is taken as an indicator of the AGN bolometric
luminosity and black hole accretion rate (see \citealt{wil+07} for a
discussion of the extinction correction). The AGN hosts are early-type
galaxies with a significant bulge component \citep{kau+03c}, and so
the black hole mass ($M_{BH}$) is estimated from $\sigma_*$ assuming
the relation determined for bulges in \cite{tre+02}. As discussed by
\citet{kau+03c}, for Type 2 AGN the SDSS images and spectra are
dominated by starlight and so the properties of the host galaxies can
be measured in the same way as for non-AGN. The exception is that
because the AGN contaminate the nebular emission-lines, the adopted
SSFR in these cases is taken from \citet{bri+04} based on the mean
relation between $D_{4000}$ and SSFR measured for non-AGN.

The gas-phase metallicities for the star forming galaxies (non-AGN)
are determined from fitting an extensive library of model galaxy
spectra to the strong nebular emission-lines (see \citealt{tre+04} for
details).

\section{Star Formation History and Lopsidedness \label{sec:sfh}}

Our plan for this section is to start with the simplest measures of the relationship between the star formation history and lopsidedness. However, since both lopsidedness (Paper I) and star formation history \citep{kau+03b,bri+04} correlate separately with other galaxy structural properties, we follow this with a more sophisticated multivariate approach.

\subsection{Simple Correlations with Star Formation History}

We begin with the Petrosian $g-i$ color of the entire galaxy. In Fig. 1 there is a clear trend of increasing lopsidedness with bluer $g-i$ color, with the median lopsidedness rising from 0.05 to 0.25 as $g-i$ drops from 1.4 to 0.1. This correlation establishes that there is a strong connection between the lopsidedness in the outer region of the galaxy, and the star formation history of the galaxy on global scales (as shown by \citealt{rud+00}). In a future paper we will use GALEX plus SDSS photometry to relate lopsidedness to the global star-formation rate.

In the present paper our emphasis is on the central properties of our galaxies, so we now turn to stellar age indicators measured within the fiber. Each of our measures has a somewhat different time dependence. We consider them first individually and then together.

\begin{figure}[ht]
\epsscale{1.1}
\plottwo{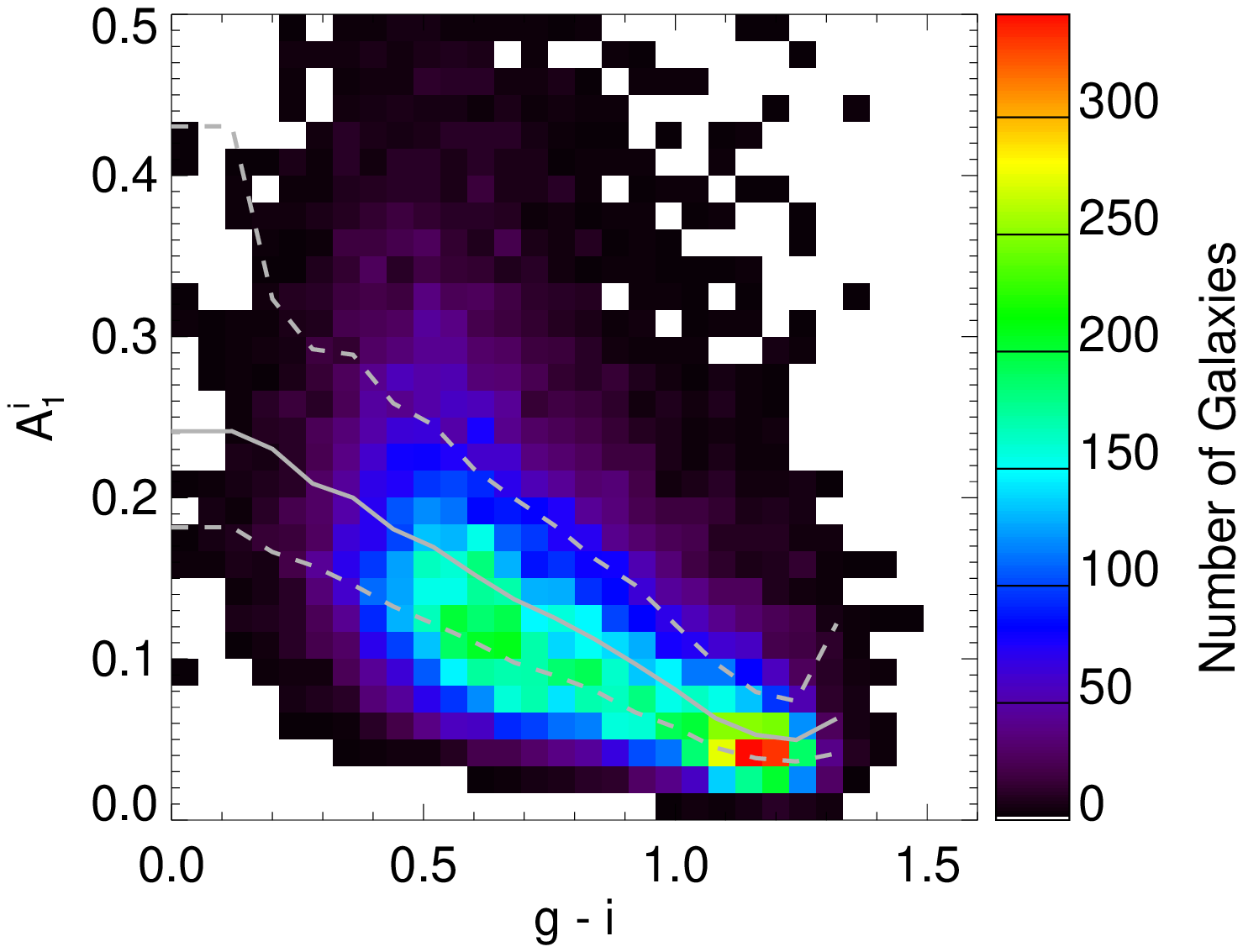}{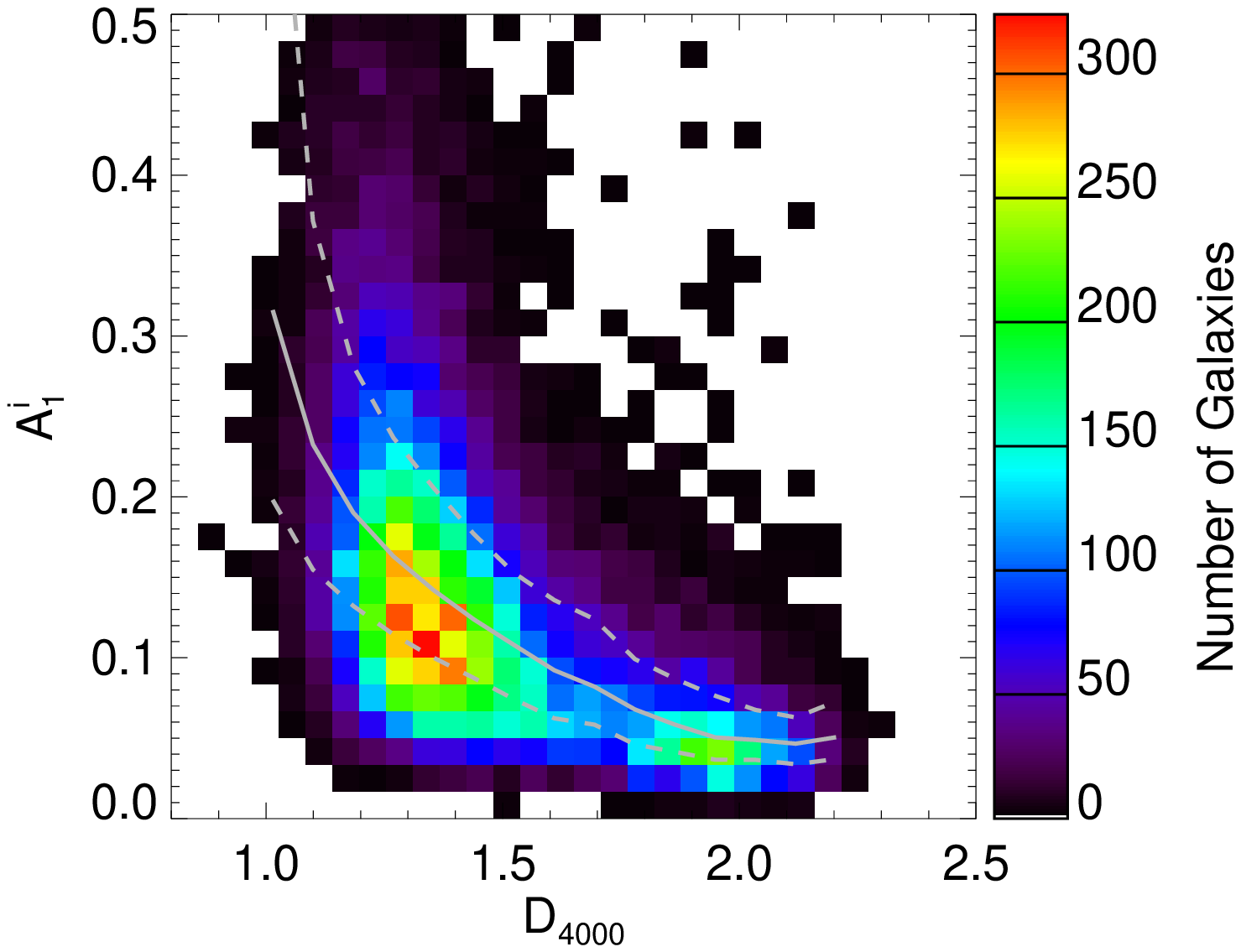}
\plottwo{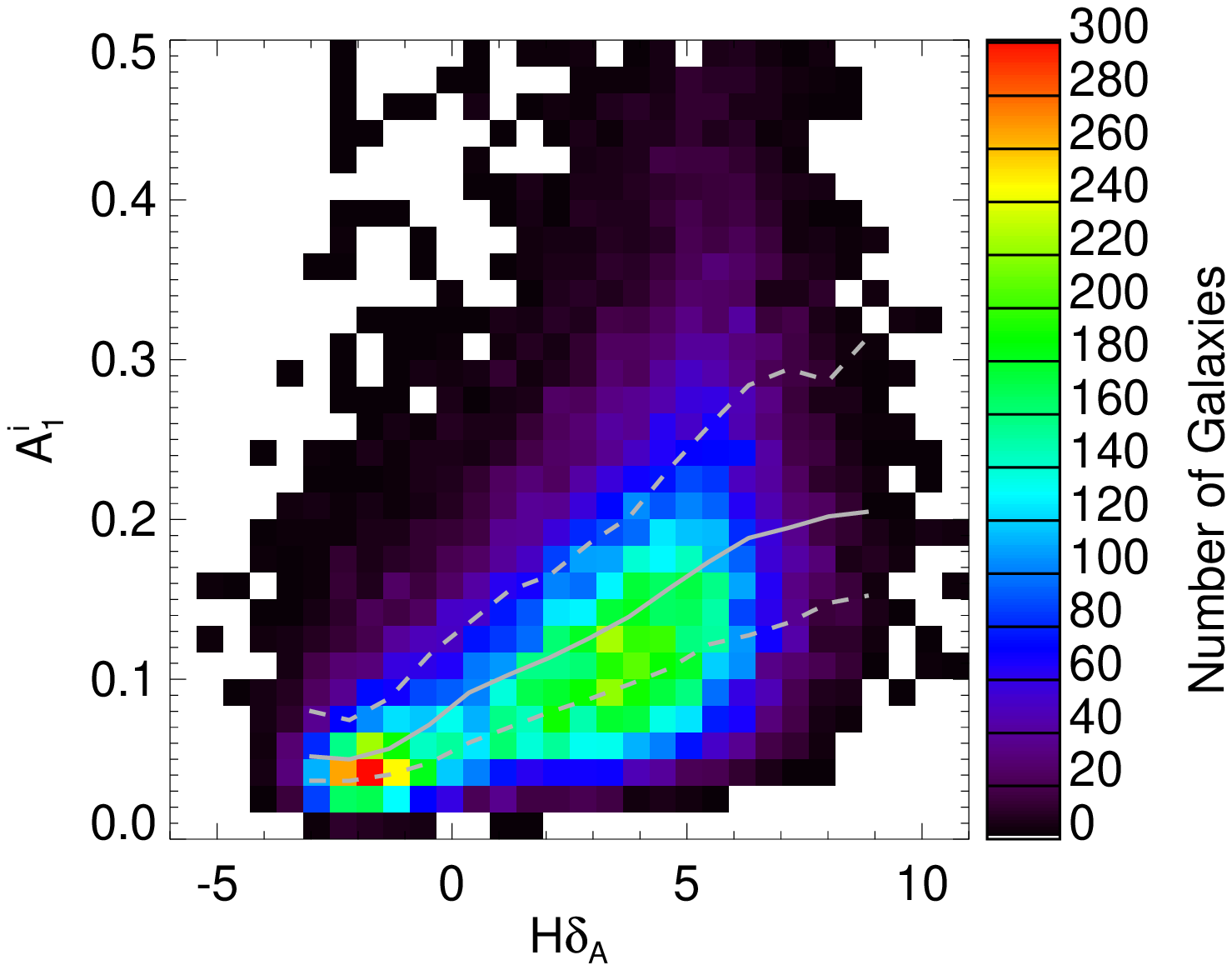}{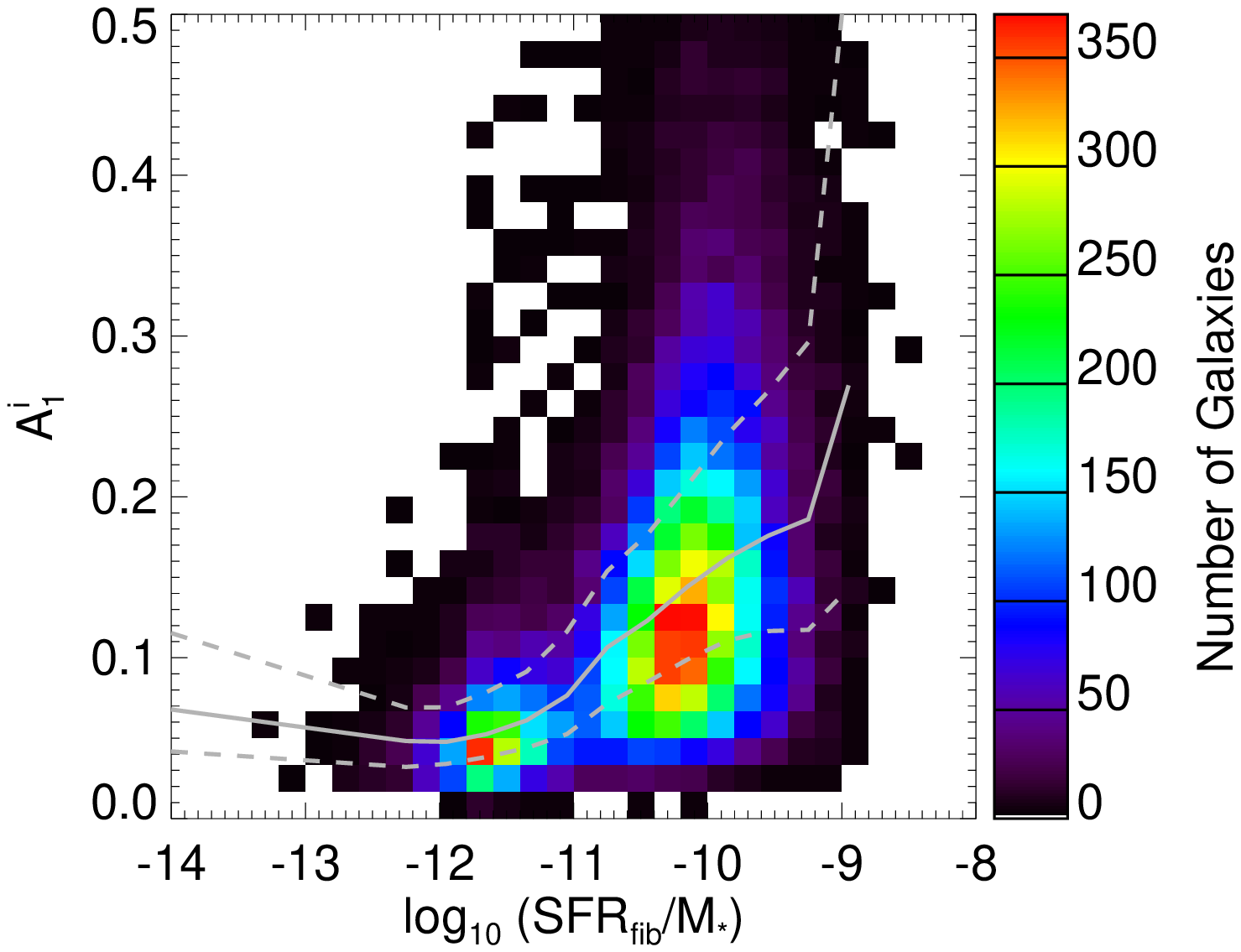}
\caption{Two-dimensional distributions of $A_1^i$ and star formation properties: $g-i$ color, 4000 \AAA break, $H\delta$ absorption index, and specific star formation rate. The distributions of lopsidedness as functions of these properties are overlaid in gray: median {\em (solid line)} and quartiles {\em (dashed lines)}. Lopsidedness tends to increase with bluer color, younger stellar population, and higher SSFR.}
\label{fig:lop1dsfh}
\end{figure}

We plot the distribution of $A_1^i$ against $D_{4000}$ in the upper right panel
of Fig.~\ref{fig:lop1dsfh}. \citet{kau+03b} showed that the local galaxy population is bimodal in the distribution of $D_{4000}$, dividing into young, late-type galaxies ($D_{4000} < 1.5$) and old early-type galaxies ($D_{4000} > 1.7$). For the young galaxies, the lopsidedness increases strongly with decreasing age: the median rises from 0.1 (mildly lopsided) at $D_{4000}$ = 1.5 to over 0.3 (strongly lopsided) at $D_{4000} \sim$ 1. Values this low for $D_{4000}$ correspond to extremely young stellar populations (strong starbursts, younger than a few tens of Myr). Most older galaxies ($D_{4000} > 1.7$) uniformly have rather small values of $A_1^i < 0.1$. 

In the lower left panel of Fig.~\ref{fig:lop1dsfh}, we plot the lopsidedness distribution as a function of $H\delta_A$.  There is a correlation between median $A_1^i$ and $H\delta_A$ for nearly the entire range of $H\delta_A$, with the median value for $A_1^i$ rising from near 0.05 at $H\delta_A = -1$ to 0.20 at $H\delta_A = 9$.  Because values of $H\delta_A$ higher than about 7 can only originate about 0.1 to 1 Gyr after a strong burst of star-formation, we can connect such post-starbursts to events that have produced significant perturbations to the stellar mass distribution.
 
In the lower right panel of Fig.~\ref{fig:lop1dsfh}, we show the distribution of lopsidedness as a function the specific star formation rate (SSFR). Above SSFR $\sim 10^{-11}$ year$^{-1}$ (or below a mass-doubling time of $M_*/SFR = 10^2$ Gyr), there is a strong correlation between star formation and lopsidedness. Typical star-forming galaxies with SSFR $\sim10^{-11}$ to $\sim10^{-10}$ year$^{-1}$ are mildly lopsided (median value of about 0.1) while strong starbursts (SSFR $\sim 10^{-9}$ year$^{-1}$) are very lopsided ($A_1^i >$ 0.2). The smallest SSFR that can be reliably measured from the spectra is about $10^{-12}$ year$^{-1}$. Such galaxies are almost uniformly symmetrical (median $A_1^i \sim$ 0.05).

\clearpage

\begin{figure}[ht]
\plotone{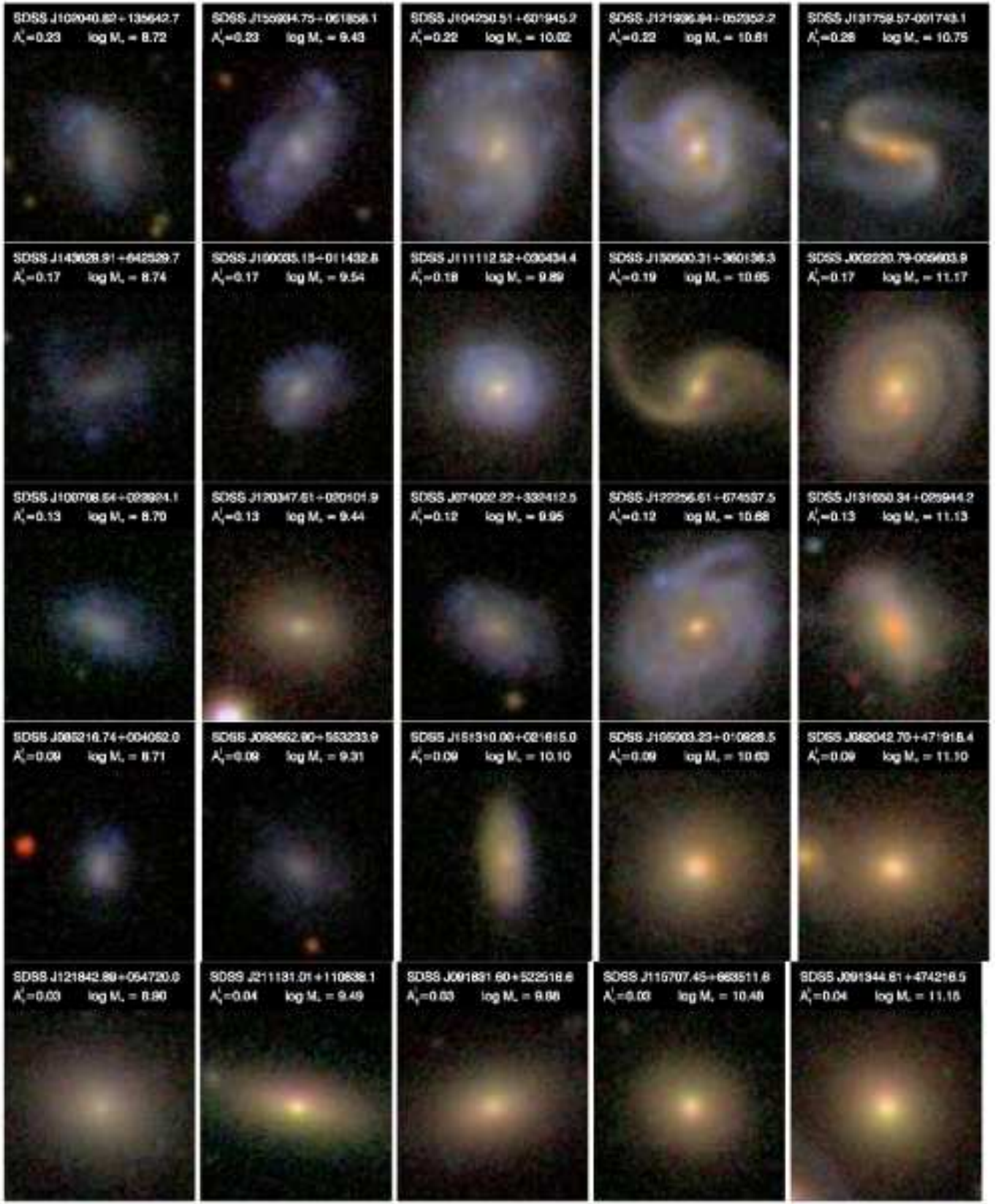}
\caption{Twenty-five SDSS galaxies with increasing lopsidedness from bottom to top and increasing stellar mass from left to right. Each image is 30\arcsec $\times$ 30\arcsec, or about 23 kpc $\times$ 23 kpc at $z = 0.04$, a typical redshift of the sample.}
\label{fig:m-a1-mosaic}
\end{figure}

To orient the reader, in Fig.~\ref{fig:m-a1-mosaic} we show a montage
of galaxies selected to cover the range in mass and lopsidedness
spanned by our sample of star-forming galaxies.  The lopsided galaxies
shown ($A_1^i \le 0.23$) are not extremely distorted like classical major mergers. 

\clearpage

\begin{figure}[ht]
\epsscale{1.1}
\plottwo{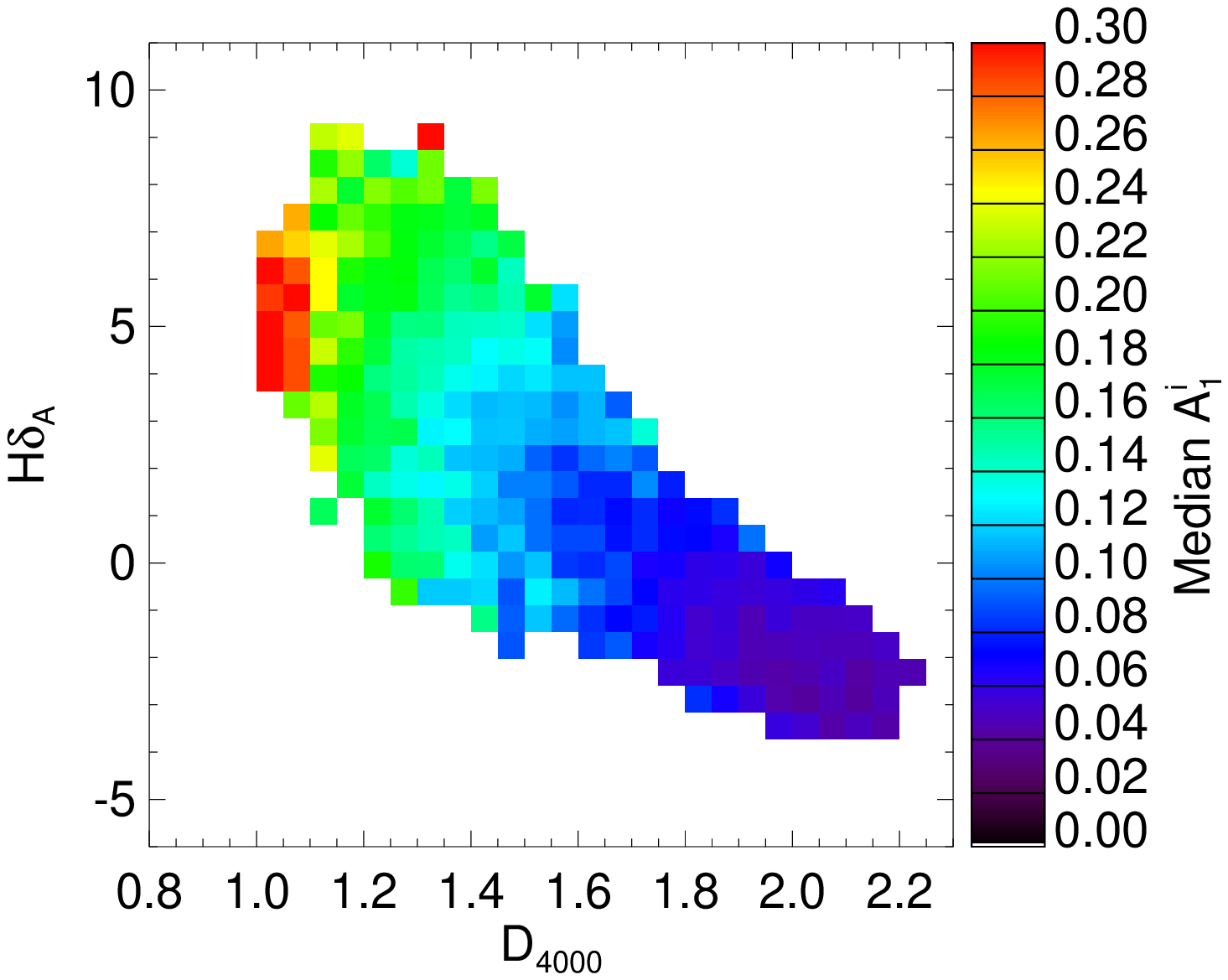}{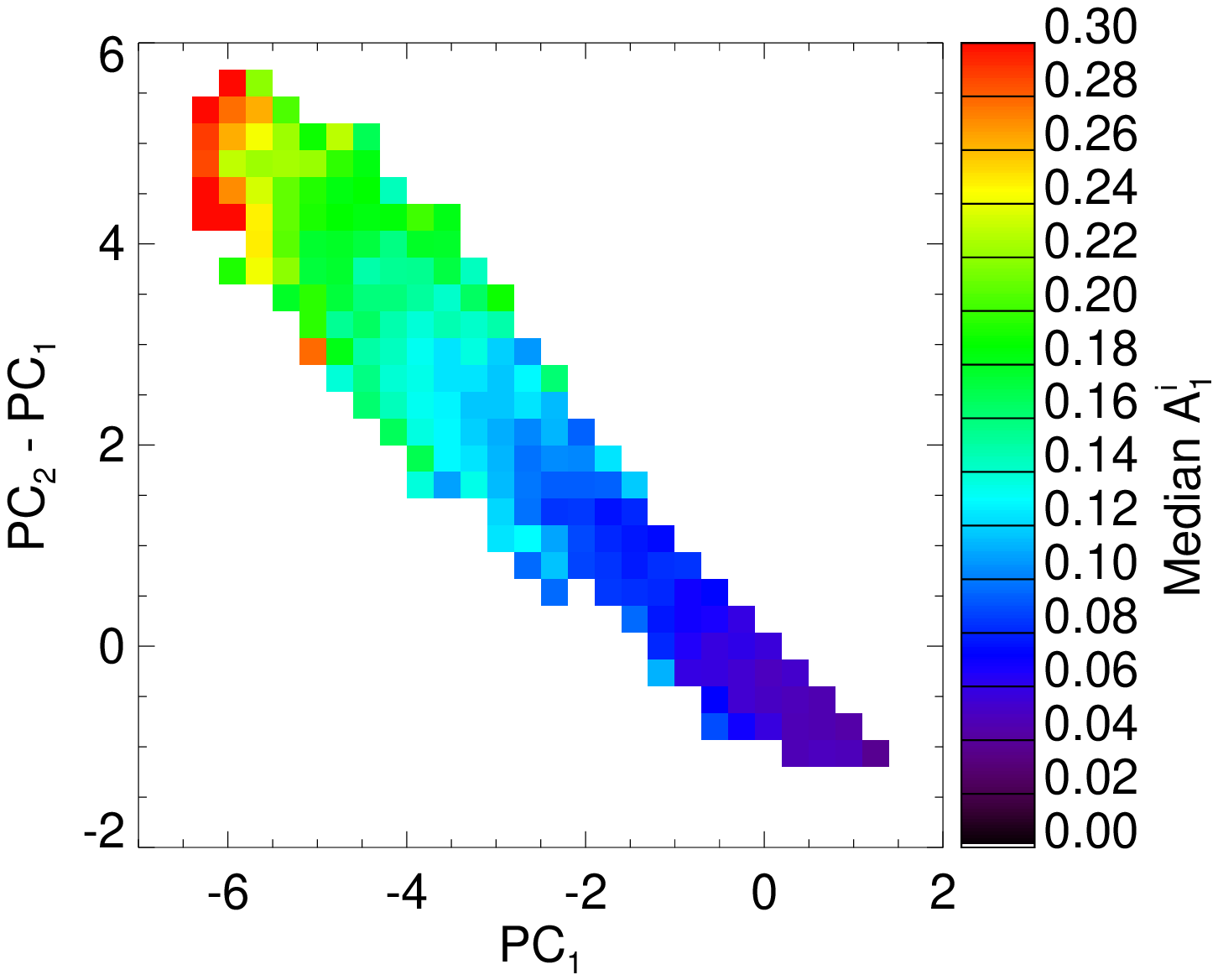}
\caption{Median lopsidedness as a function of the star formation indicators and principal spectral components. Starbust and post-starburst galaxies are typically the most lopsided, while galaxies with little star formation are typically the most symmetric.}
\label{fig:d4000-hda}
\end{figure}

As noted above, the two age indicators $\Dbreak$ and $\HdA$ used
together can identify those galaxies that have undergone a recent ($<$
Gyr) starburst, and separate them from the more common galaxies
experiencing more-or-less continuous star formation
\citep{kau+03a}. The parameters $PC_1$ and $PC_2 - PC_1$ from
\citet{wil+07} provide similar information at higher
signal-to-noise. In Fig.~\ref{fig:d4000-hda}, we examine the joint
dependence of lopsidedness on both $\Dbreak$ and $\HdA$ in the left
panel and on $PC_1$ and $PC_2 - PC_1$ in the right. These panels both
show that while the main correlation with lopsidedness is between the
overall luminosity-weighted mean age ($D_{4000}$ or $PC_1$), there is
also a weaker trend for bursty galaxies at fixed mean age to be more
lopsided. This trend can be seen in both plots, as the regions
color-coded by median lopsidedness slant from lower left to upper
right (implying increasing lopsidedness with the increasing strength
of the post-burst signature for a fixed mean age). This result is
confirmed in Table~\ref{tab:parcor-sfr} where we list the correlation
coefficient of lopsidedness with $PC_2$, and also the partial
correlation coefficient for the residual dependence of lopsidedness on
$\HdA$ after removing the mutual dependence on $\Dbreak$.  The sample
size is sufficient to give these correlations a large statistical
significance.

The highest mean values of lopsidedness ($A_1^i > 0.2$) are found
along the far left edge of these panels. This region lies along the
locus of a starburst evolving to a post-starburst descendant over a
period of about 100 Myr \citep{wil+07}. Galaxies in the region
occupied by older post-starbursts (100 Myr to 1 Gyr) are significantly
less lopsided ($A_1^i \sim 0.1 $ to $ 0.2$). This is reasonable, given
that the characteristic dynamical time over which the galaxy should
relax after a merger or tidal perturbation is a few hundred
Myr. Indeed, the minor merger calculations of \citet{bou+05} show that
values for lopsidedness as high as $\sim$0.2 last for only a few
hundred Myr.
	
These results are qualitatively consistent with those in
\citet{wil+07} who investigated a sample of relatively massive,
bulge-dominated SDSS galaxies (required to have $\mu_* > 10^{8.5}
M_{\odot}/$kpc$^2$ and $\sigma_* > 70 $ km s$^{-1}$). They also showed
that the largest average values of lopsidedness are found in galaxies
lying along the locus of starbursts/post-starbursts younger than about 100
Myr. However, the average value of lopsidedness they found for objects
lying along this locus was only about 60\% as large as what we find in
the same region of Figure~\ref{fig:d4000-hda} for our sample. Since we
make no cut on $\mu_*$ or $M_*$, this difference is presumably related
to the strong inverse correlations we found in Paper I between $A_1^i$
and both $\mu_*$ and $M_*$.

\clearpage


\subsection{Star-Formation History: Lopsidedness vs. Other Structural Properties}
We have shown above that there is a clear correlation between
lopsidedness and all our indicators of the star formation history of
galaxies. However, in order to establish a true causal connection it
is necessary to demonstrate that this correlation is not simply
induced by the known strong correlations of the basic galaxy
structural parameters (mass, density, and concentration) with both
star formation history \citep{kau+03b} and with lopsidedness (Paper
I).

\begin{figure}[ht]
\epsscale{1.1}
\plottwo{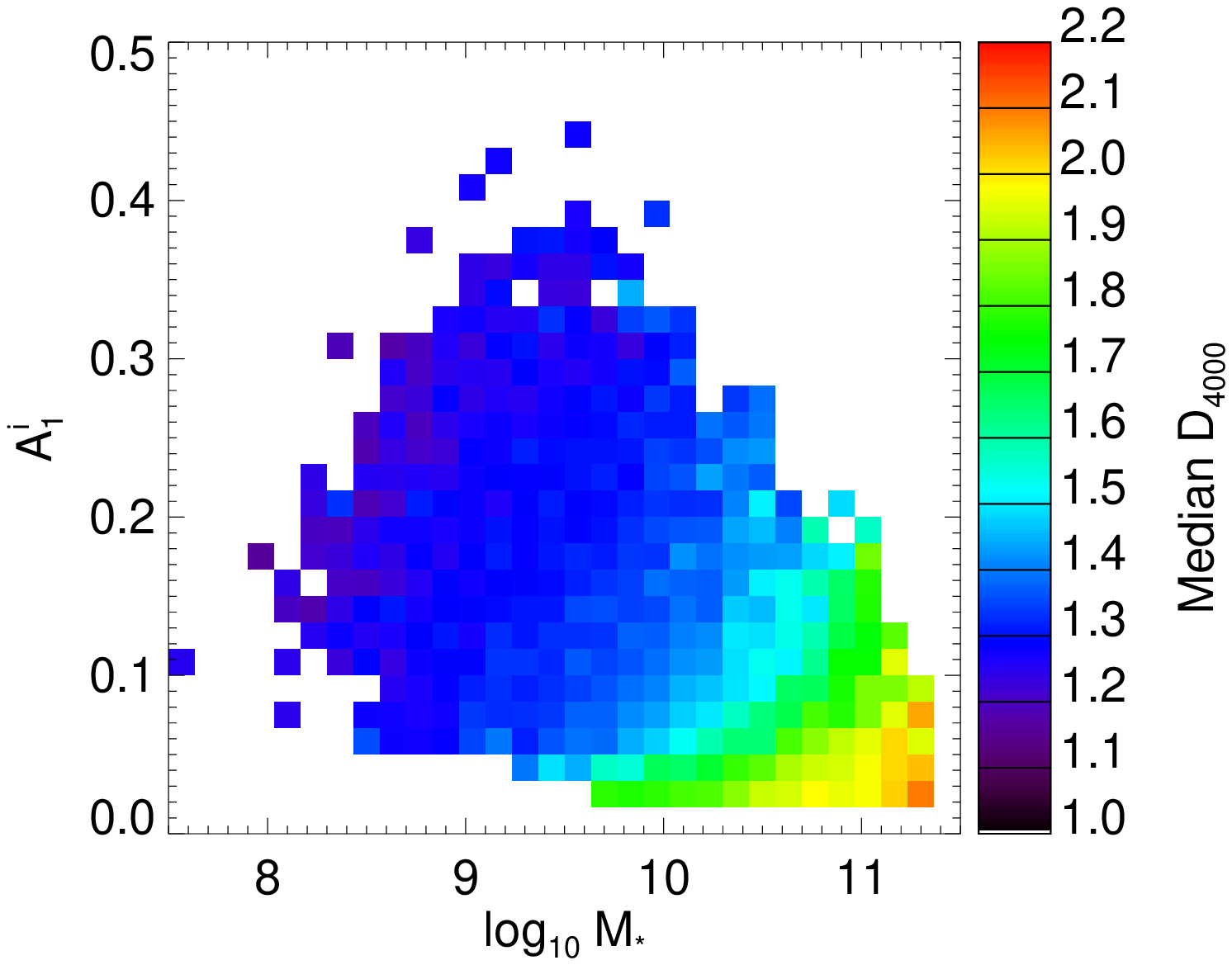}{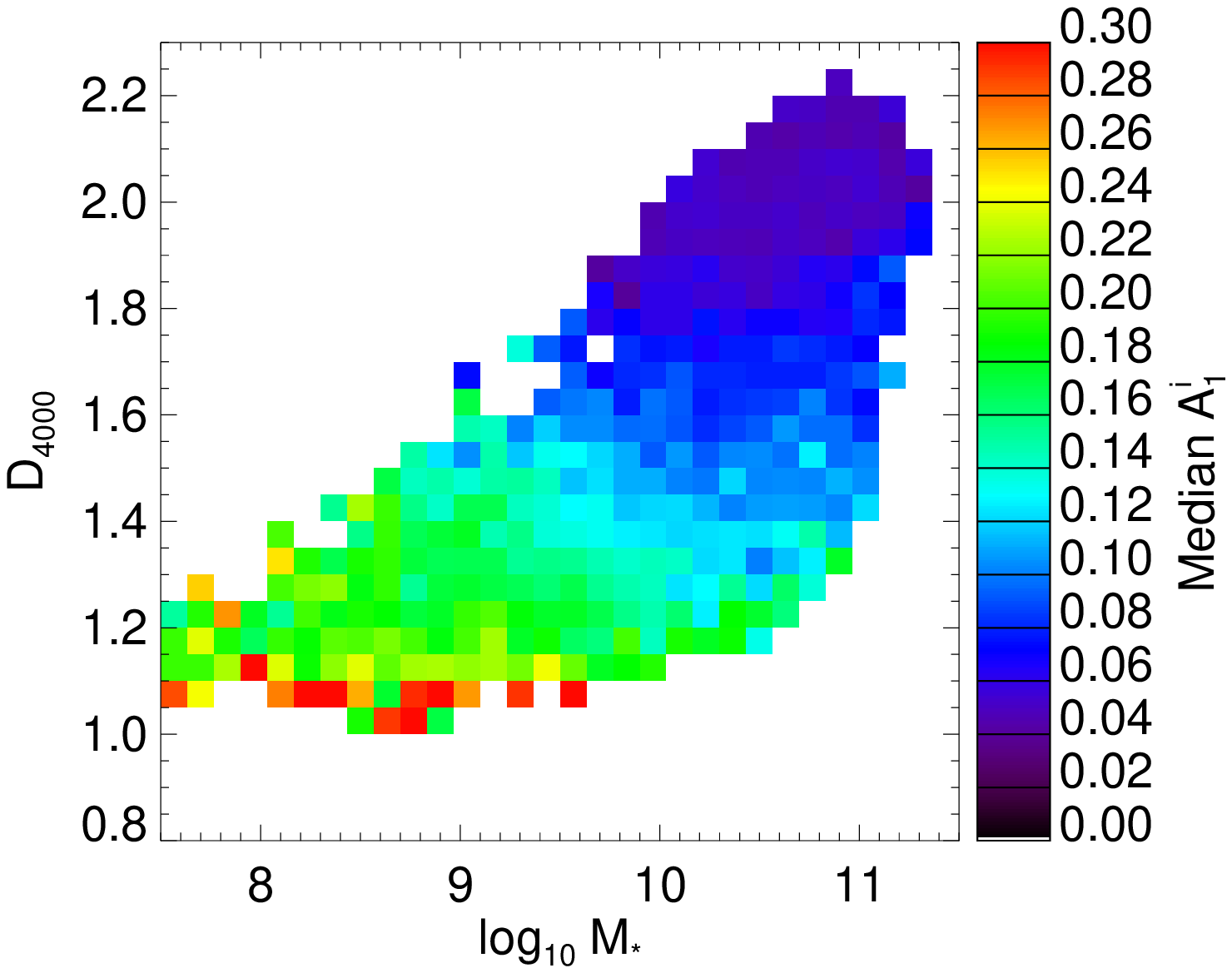}
\plottwo{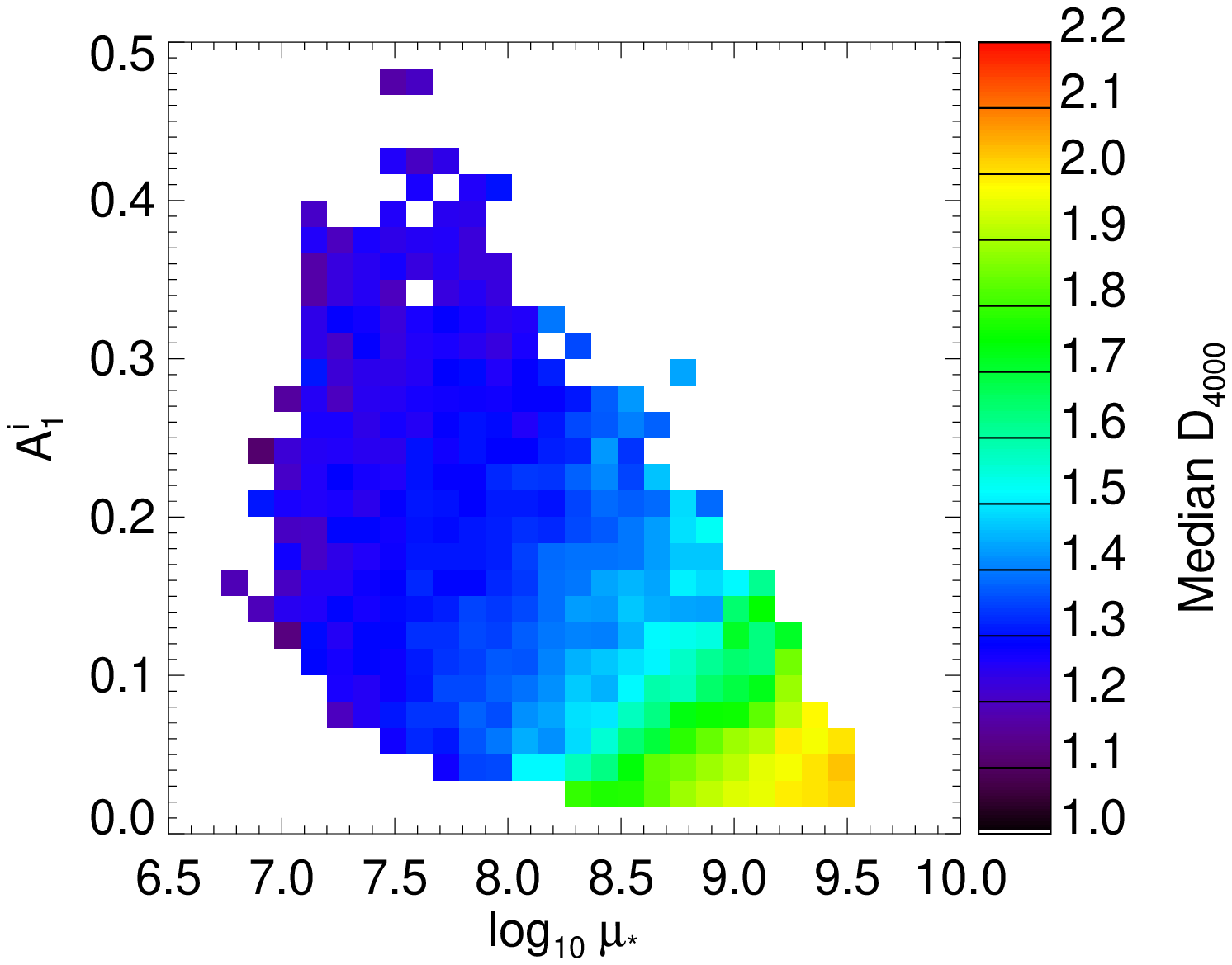}{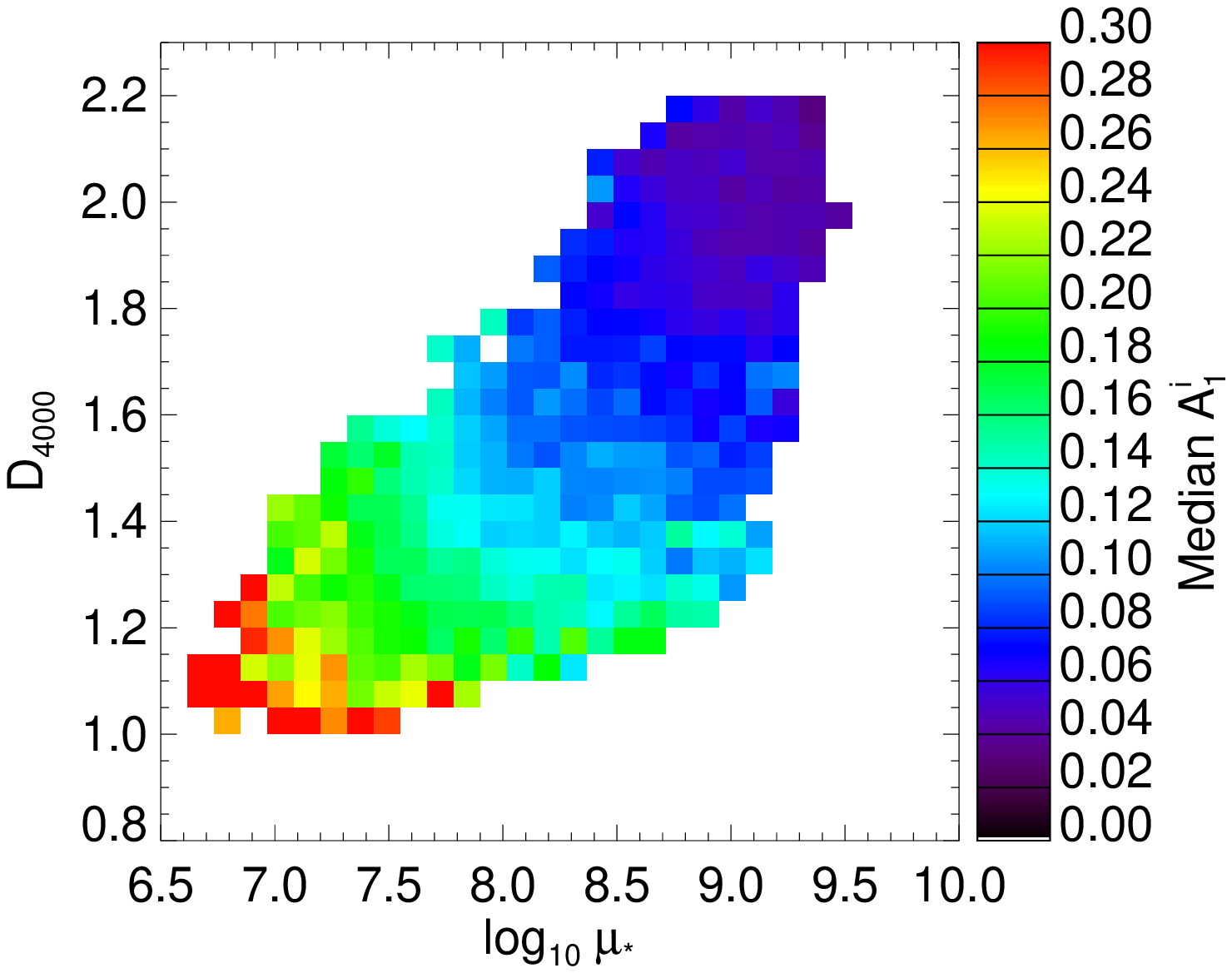}
\plottwo{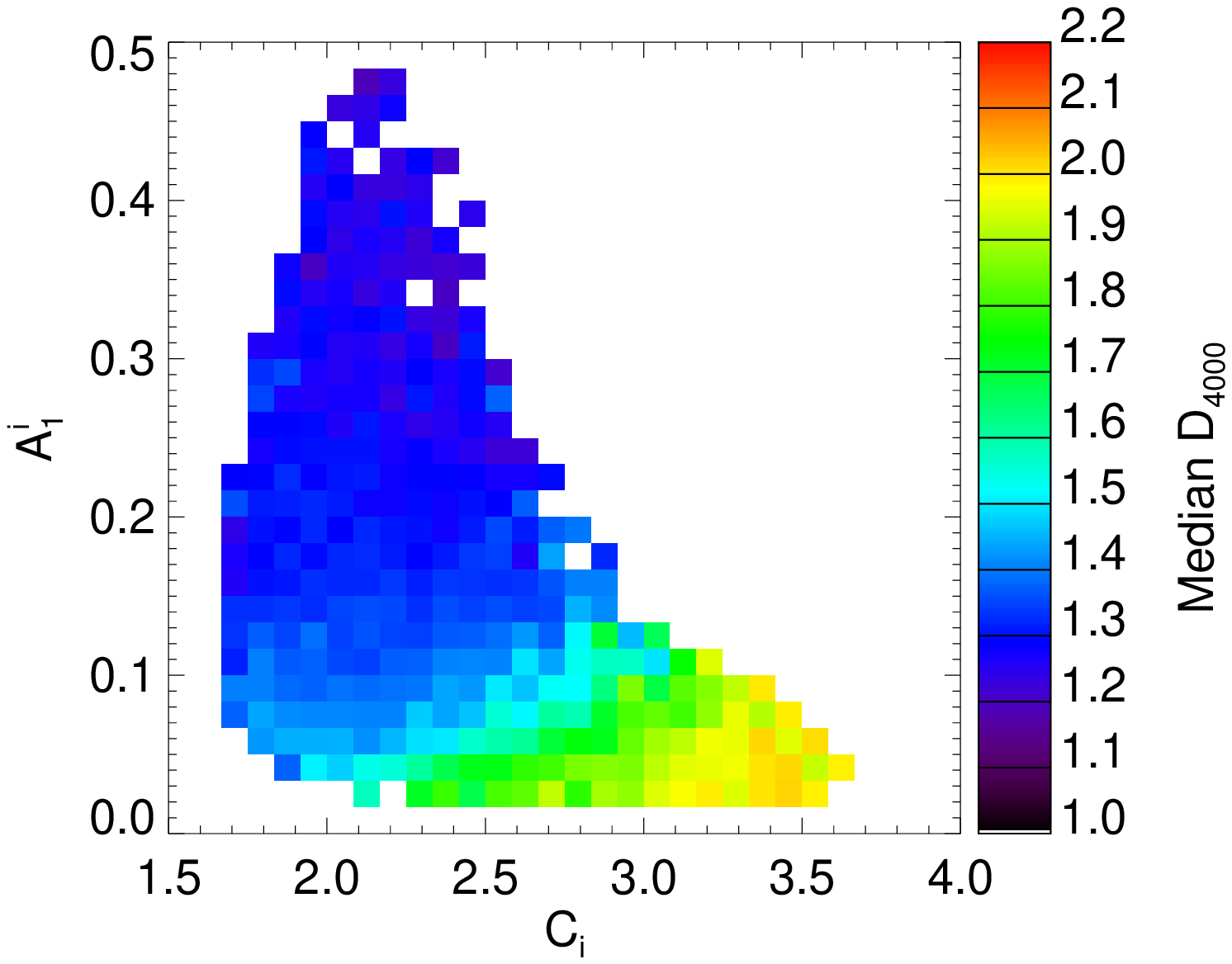}{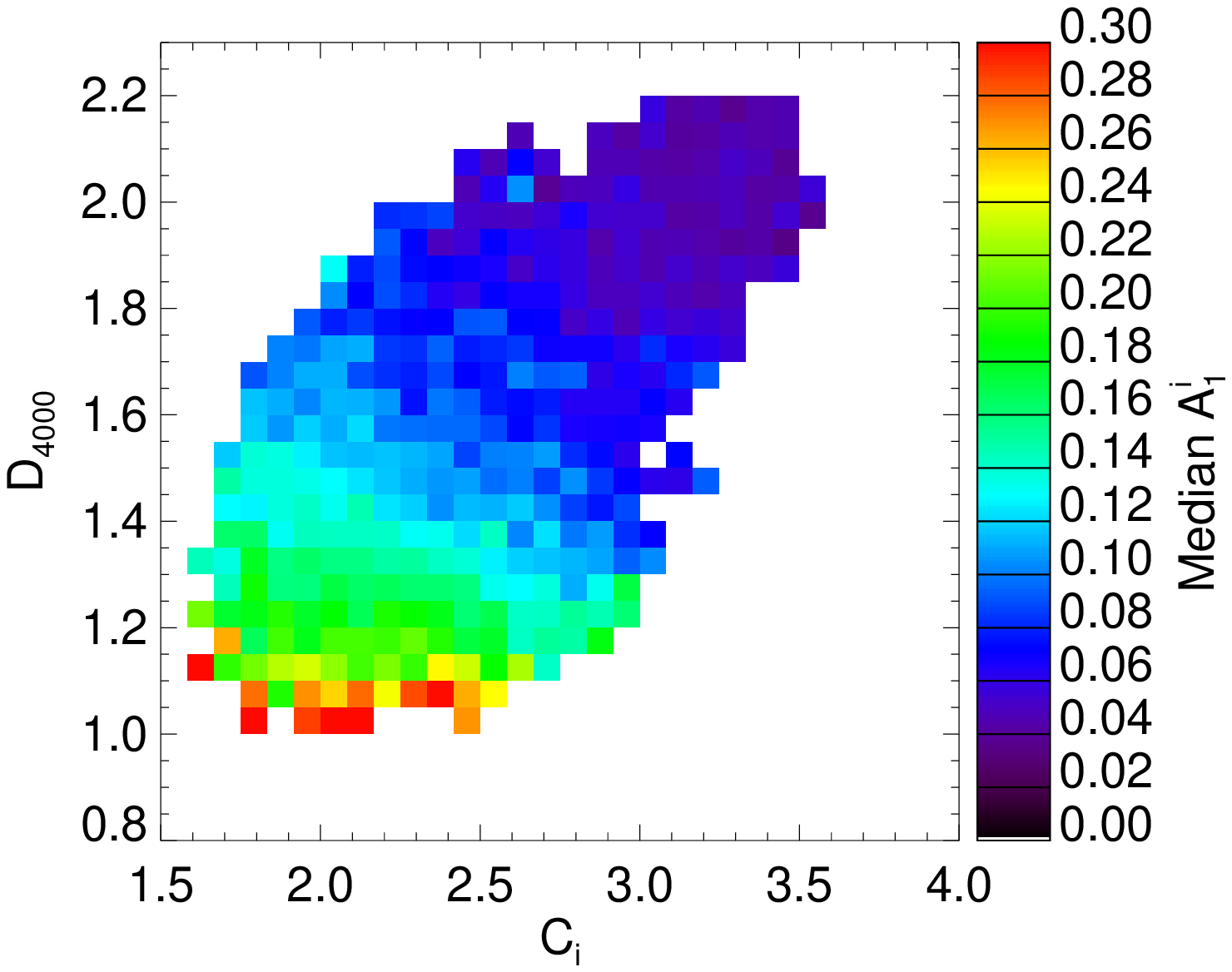}
\caption{Relationships between stellar age and structure shown in several slices of this parameter space.  The color-coding of these two-dimensional histograms indicates the median $\Dbreak$ or $A_1^i$. Galaxies with higher lopsidedness and smaller values of the other structural properties tend to have younger stellar populations.}
\label{fig:s-d4000}
\end{figure}

\begin{figure}[ht]
\epsscale{1.1}
\plottwo{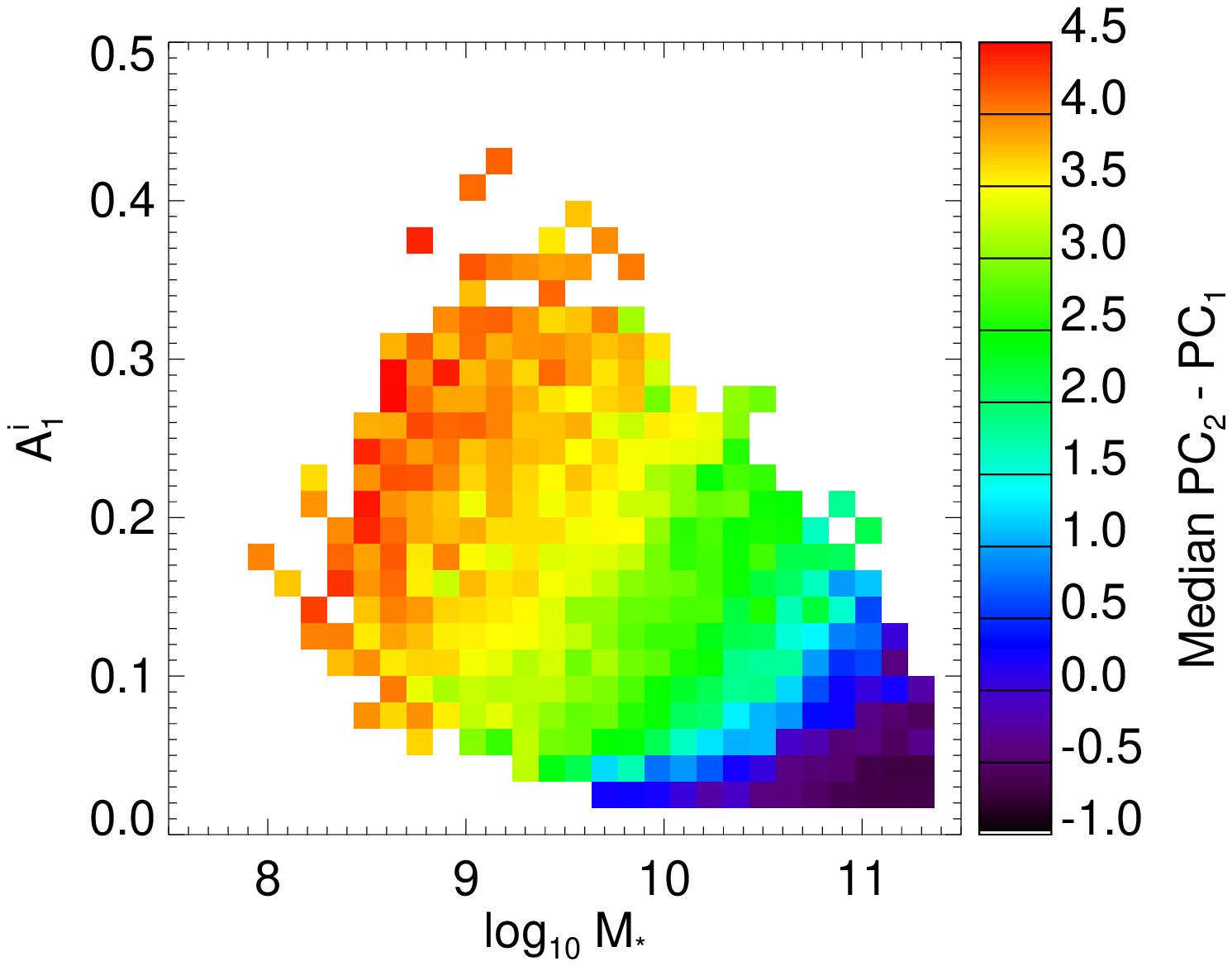}{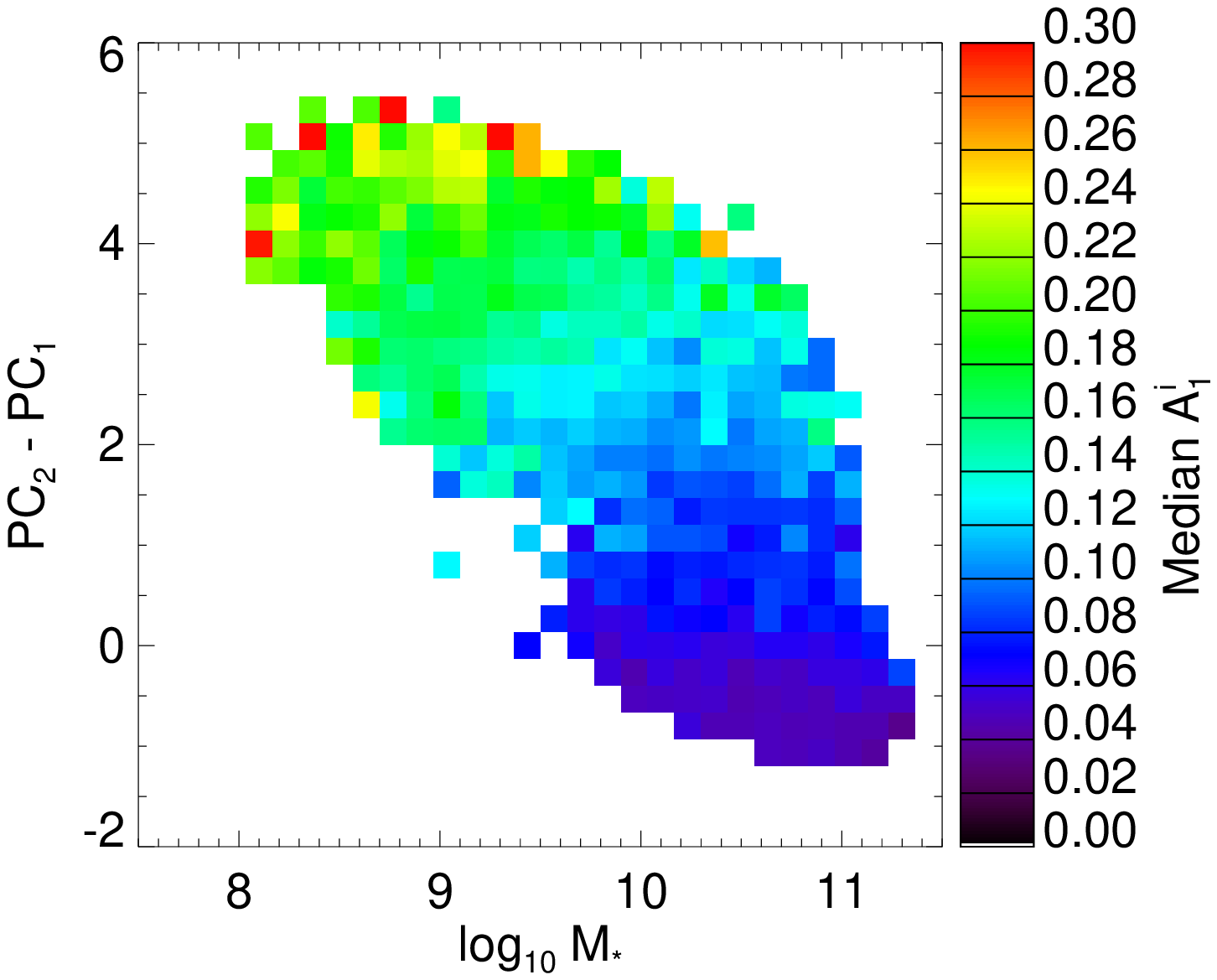}
\plottwo{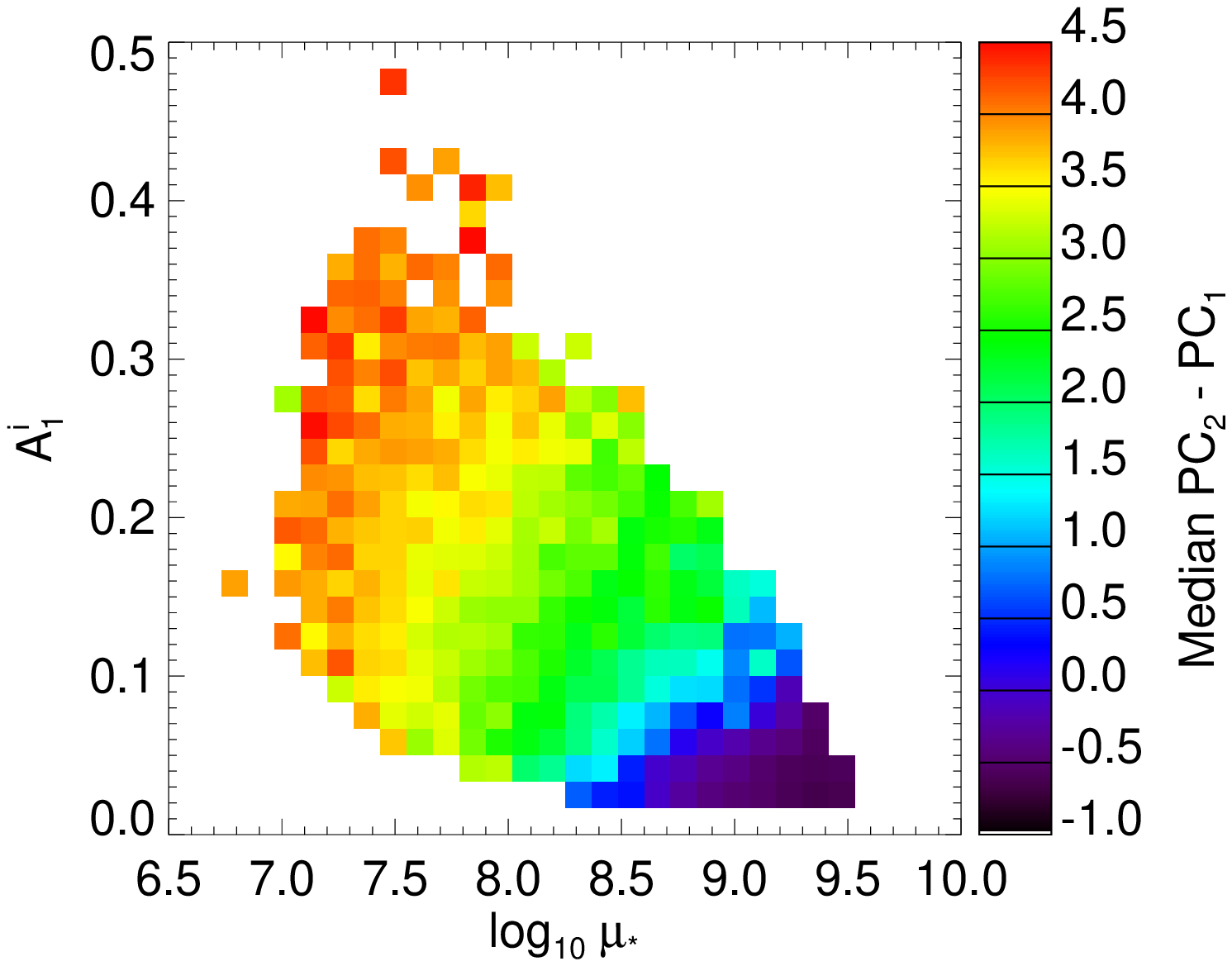}{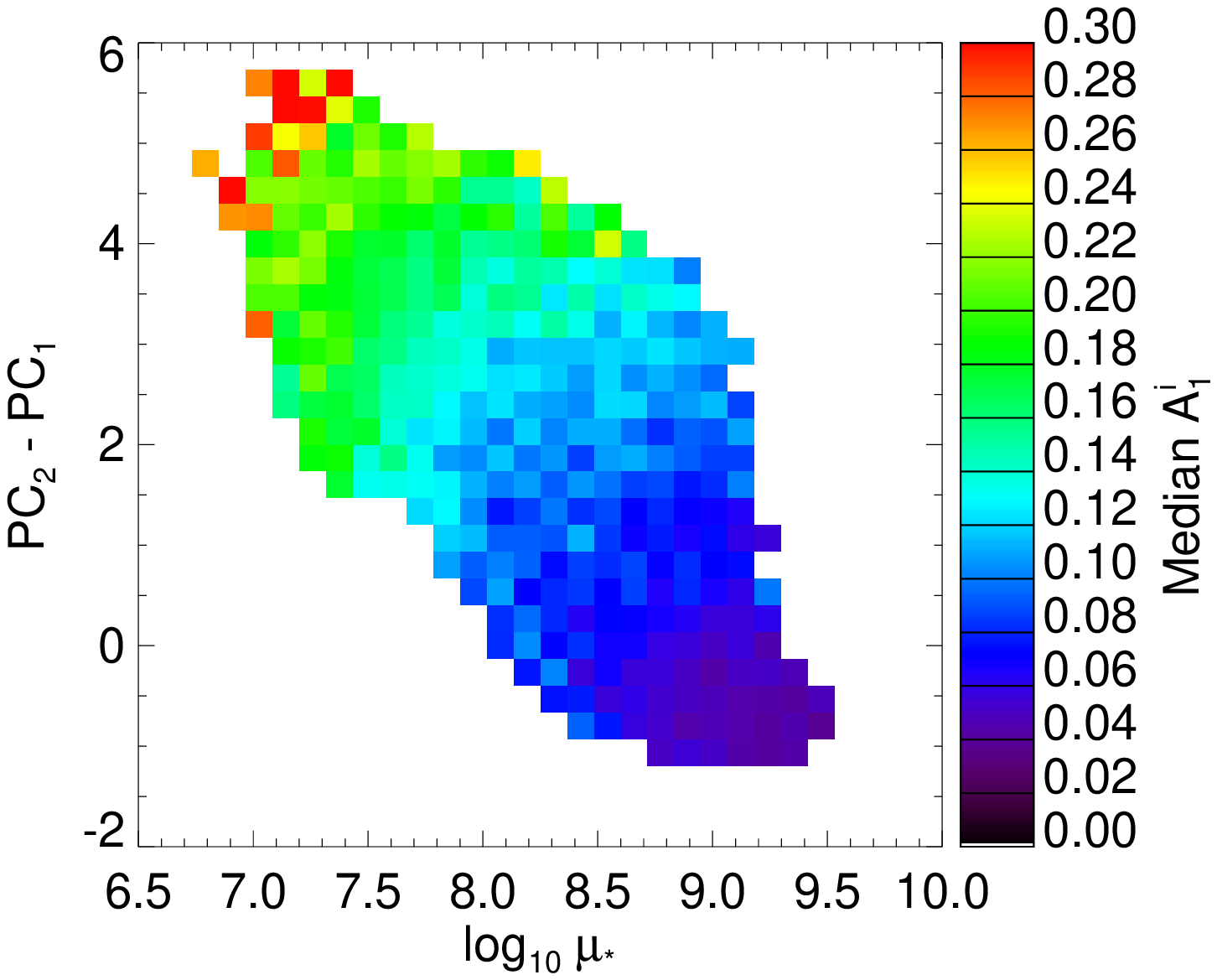}
\plottwo{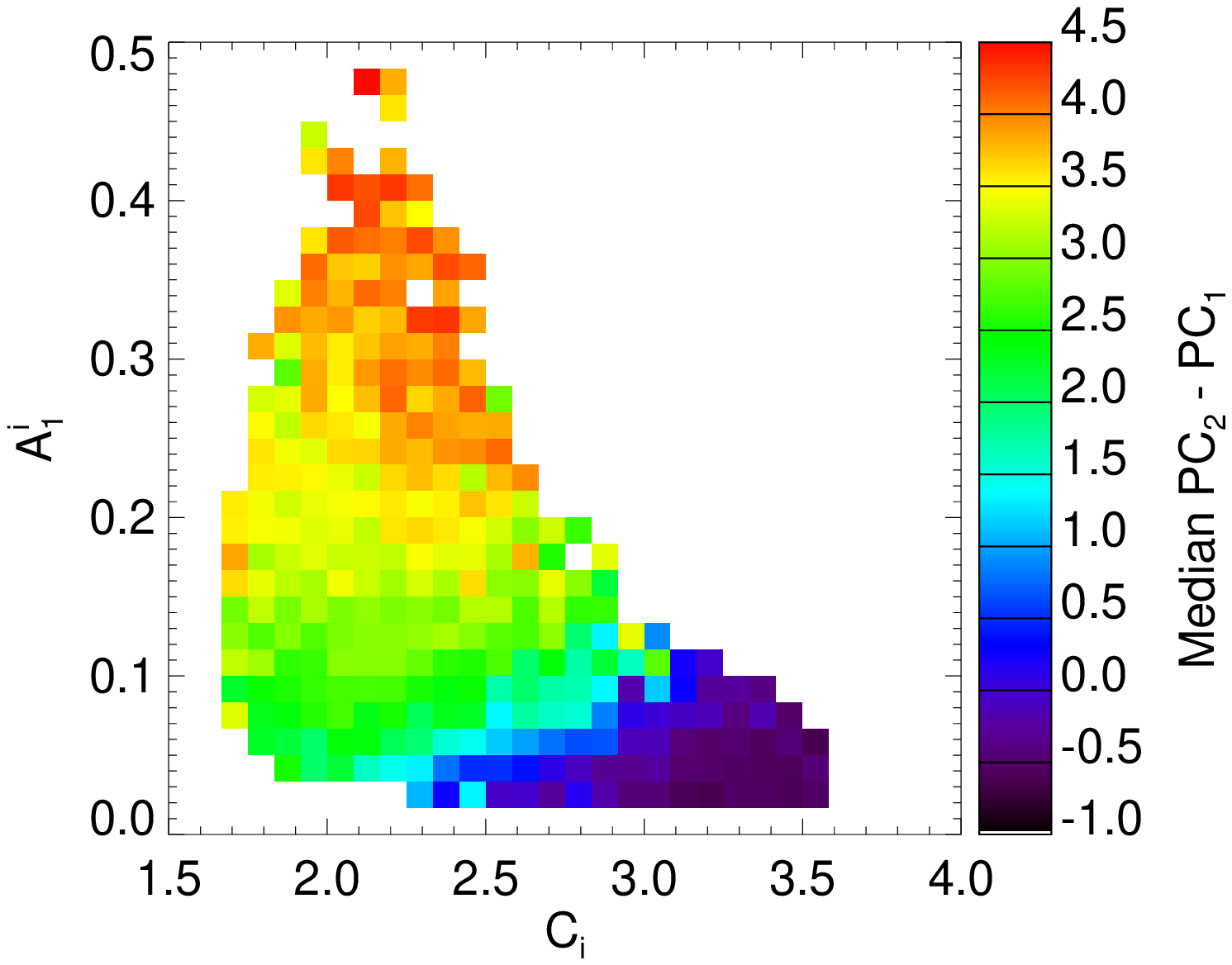}{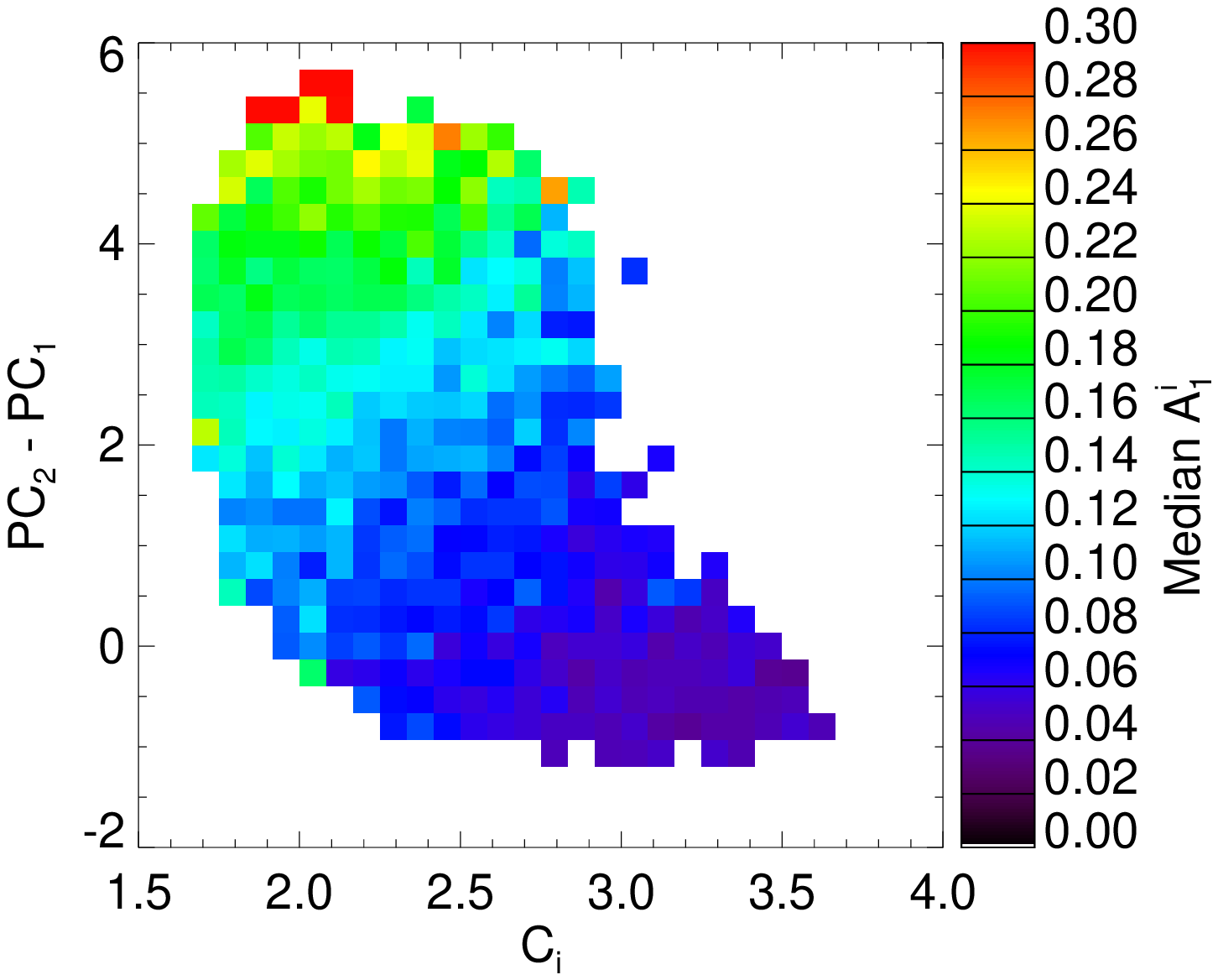}
\caption{Relationships between $PC_2-PC_1$  and structure shown in several slices of this parameter space.  The color-coding of these two-dimensional histograms indicates the median $PC_2-PC_1$ or $A_1^i$. Galaxies with higher lopsidedness and smaller values of the other structural properties tend to have greater values of $PC_2-PC_1$ and thus indicate recent star formation.}
\label{fig:s-pcs}
\end{figure}

\begin{figure}[ht]
\epsscale{1.1}
\plottwo{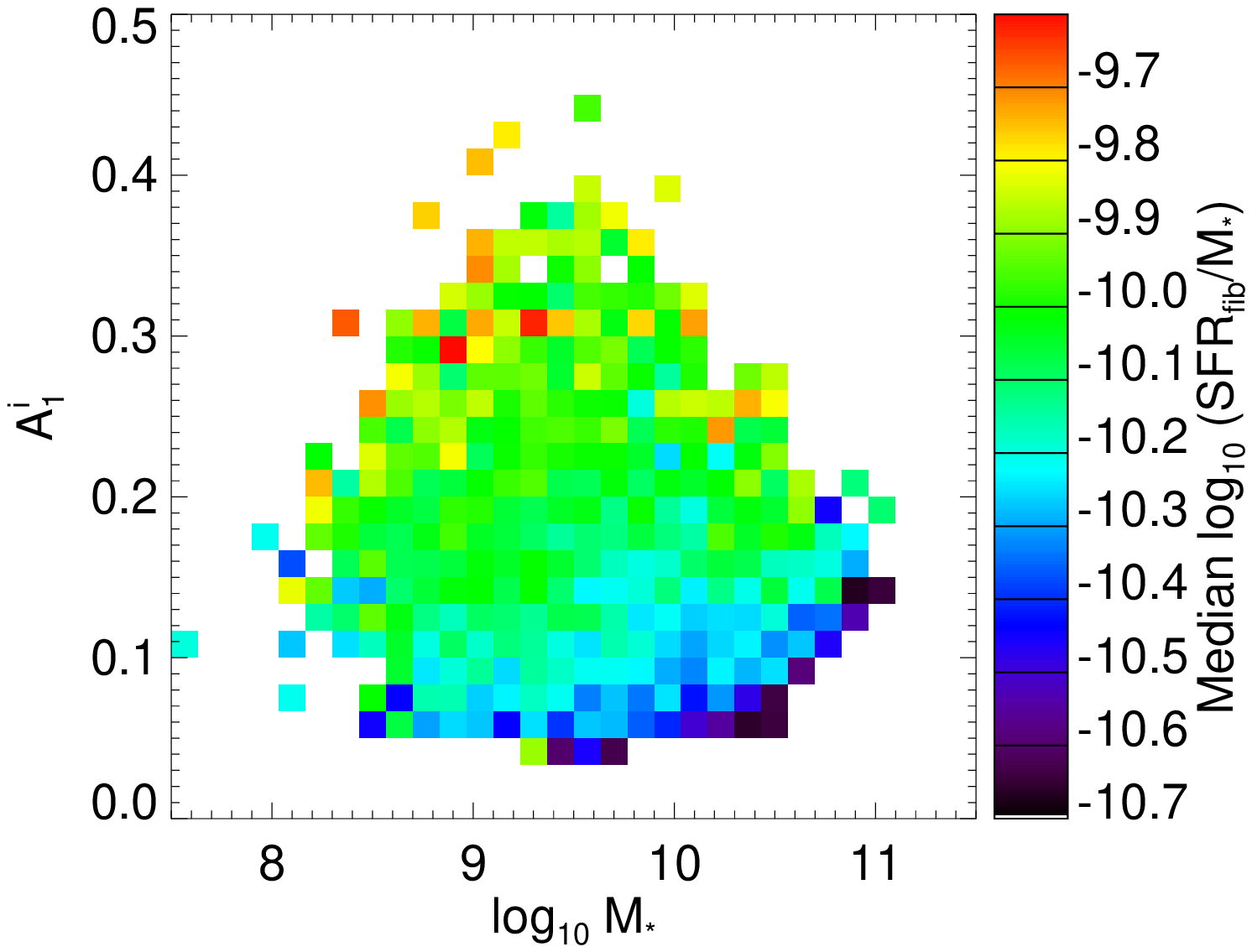}{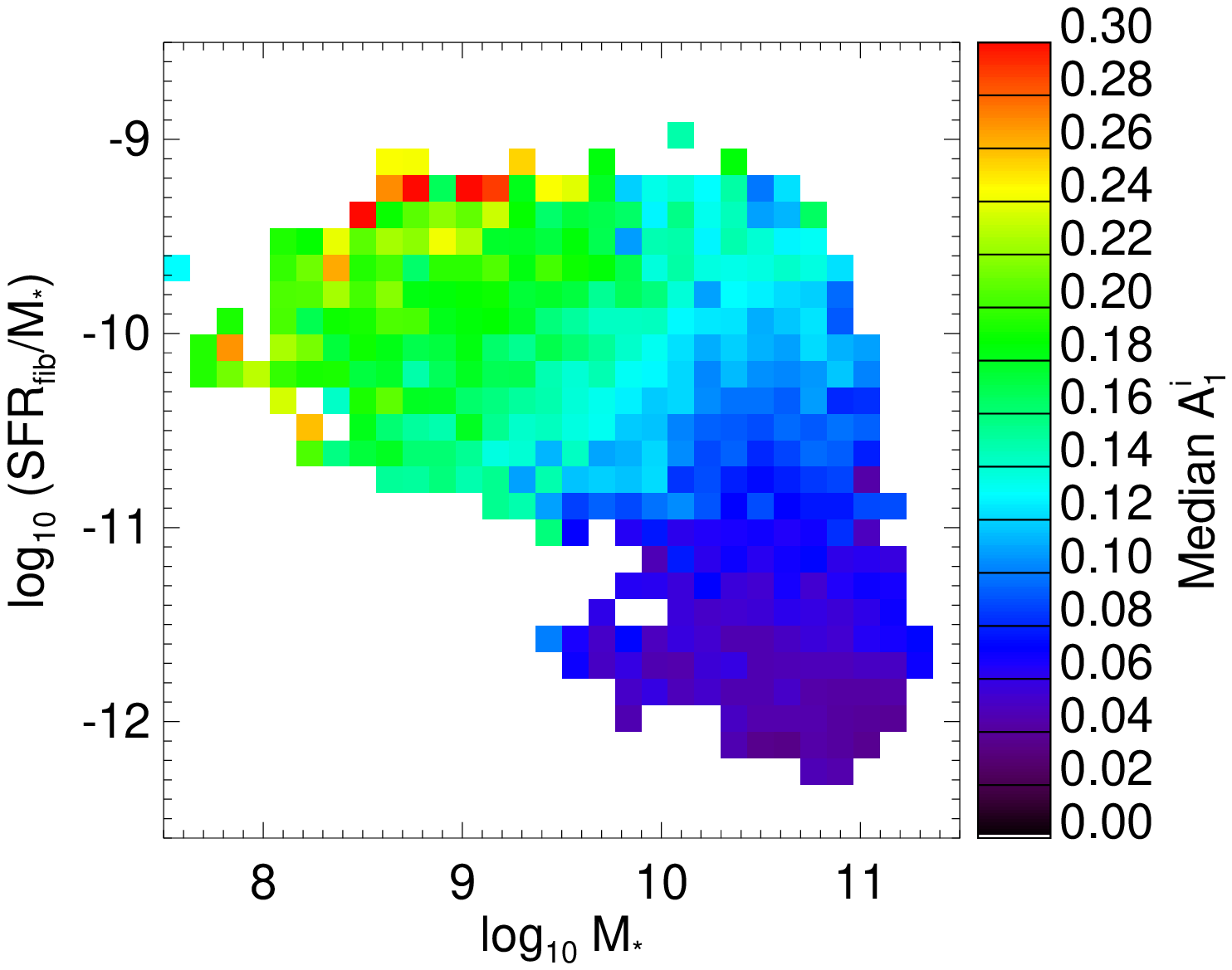}
\plottwo{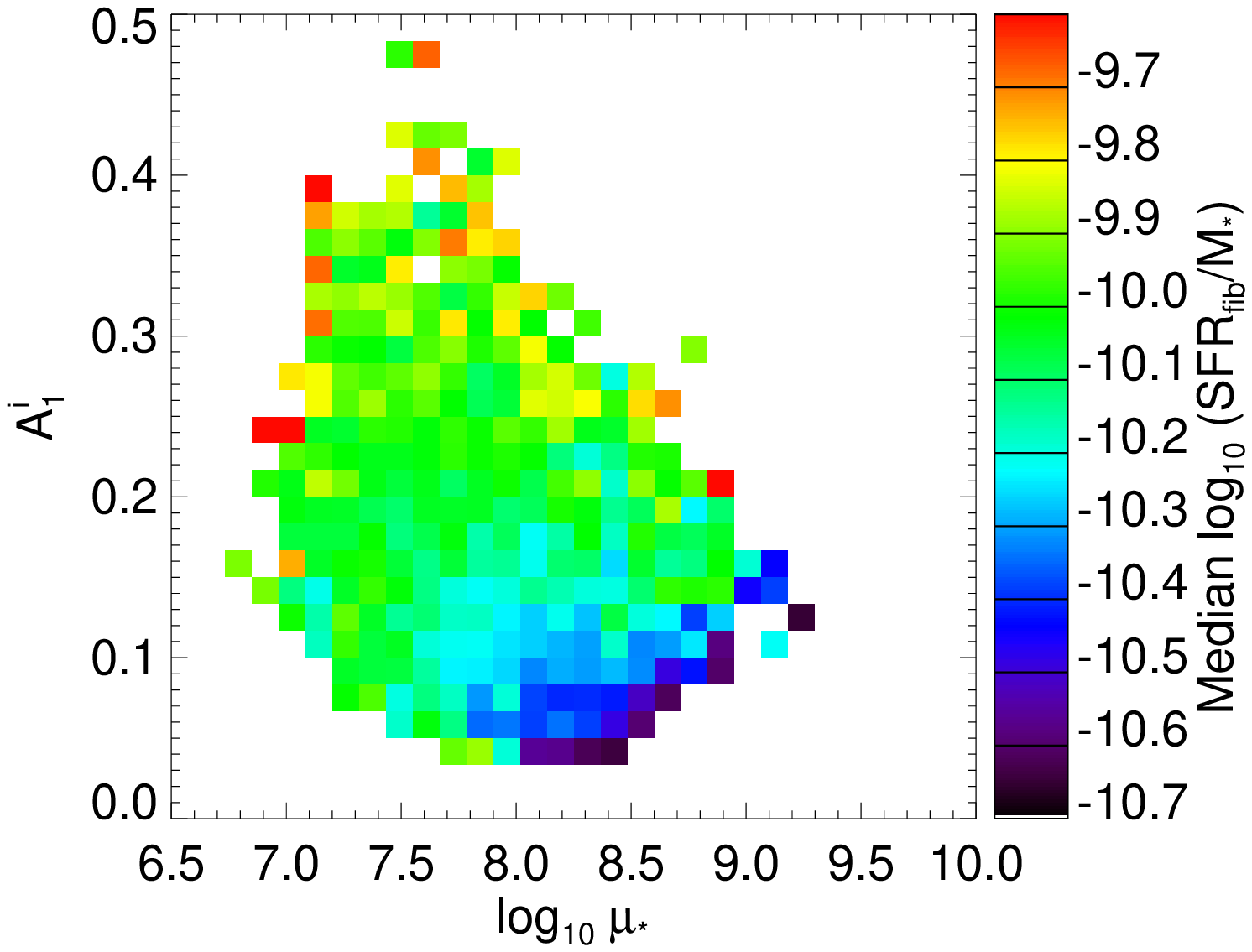}{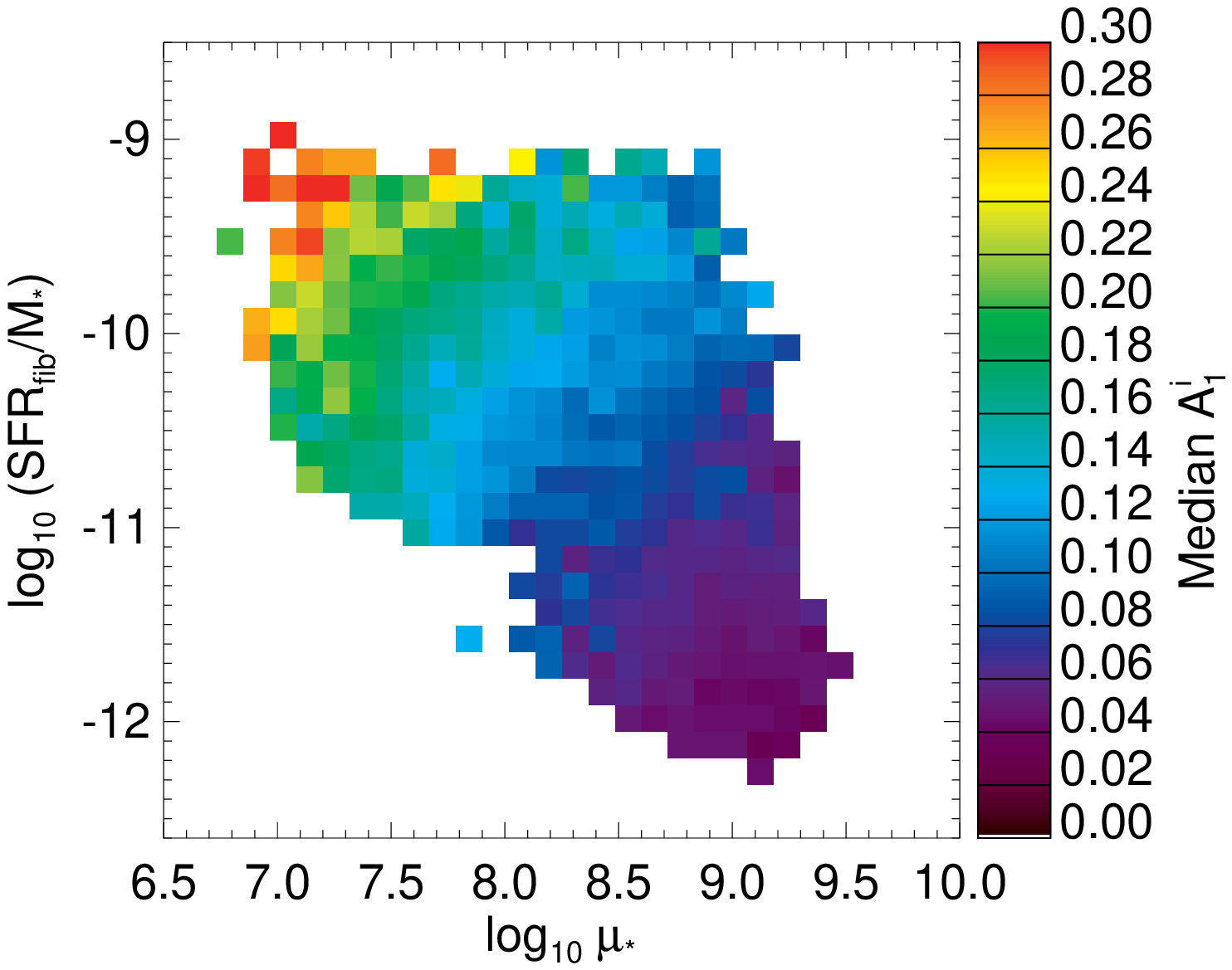}
\plottwo{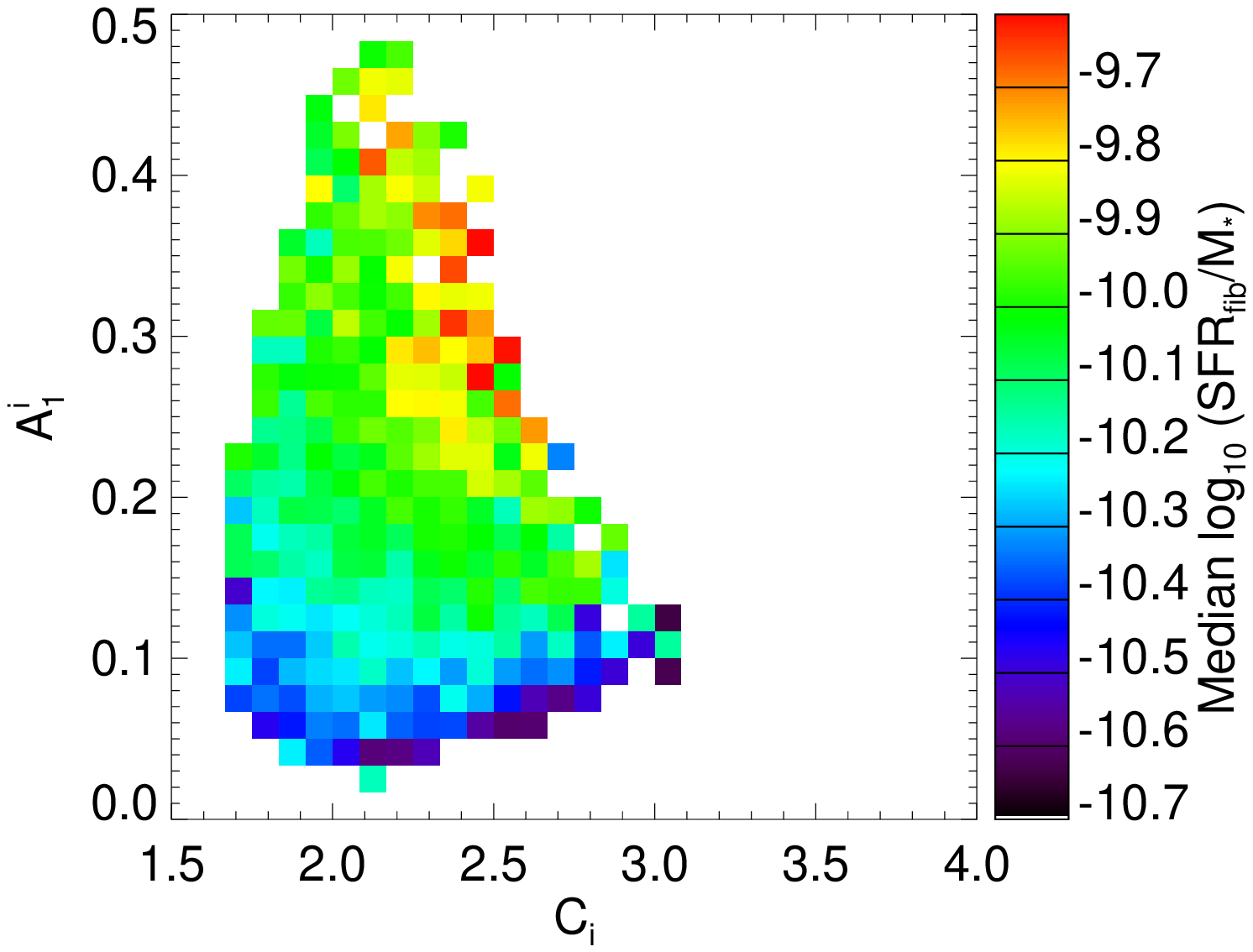}{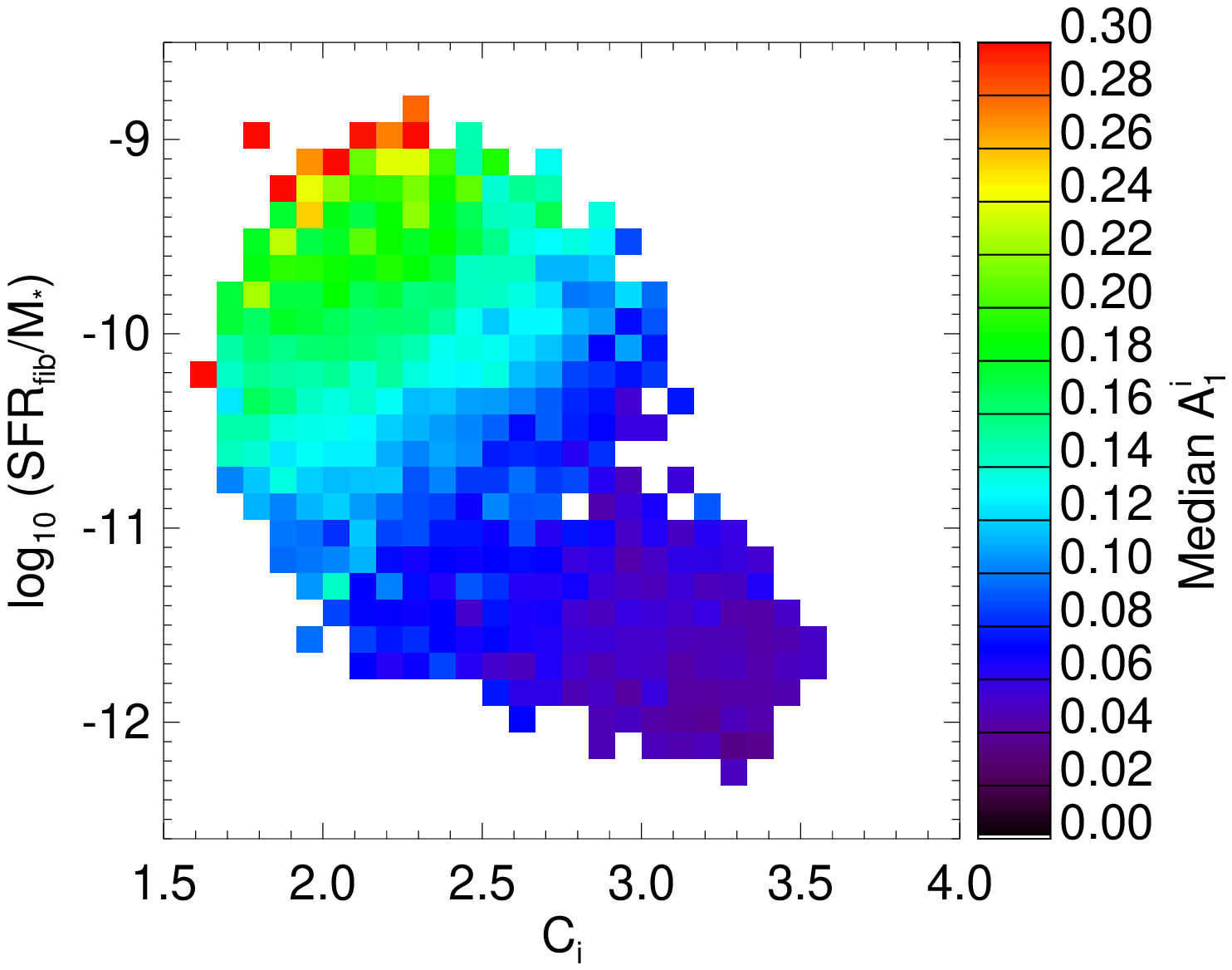}
\caption{Relationships between specific SFR and structure shown in several slices of this parameter space.  The color-coding of these two-dimensional histograms indicates the median SSFR or $A_1^i$. Galaxies with higher lopsidedness and smaller values of the other structural properties tend to have higher SSFR, though the relationship with concentration is more complex.}
\label{fig:a1-ssfr}
\end{figure}

\begin{figure}[ht]
\epsscale{1.1}
\plottwo{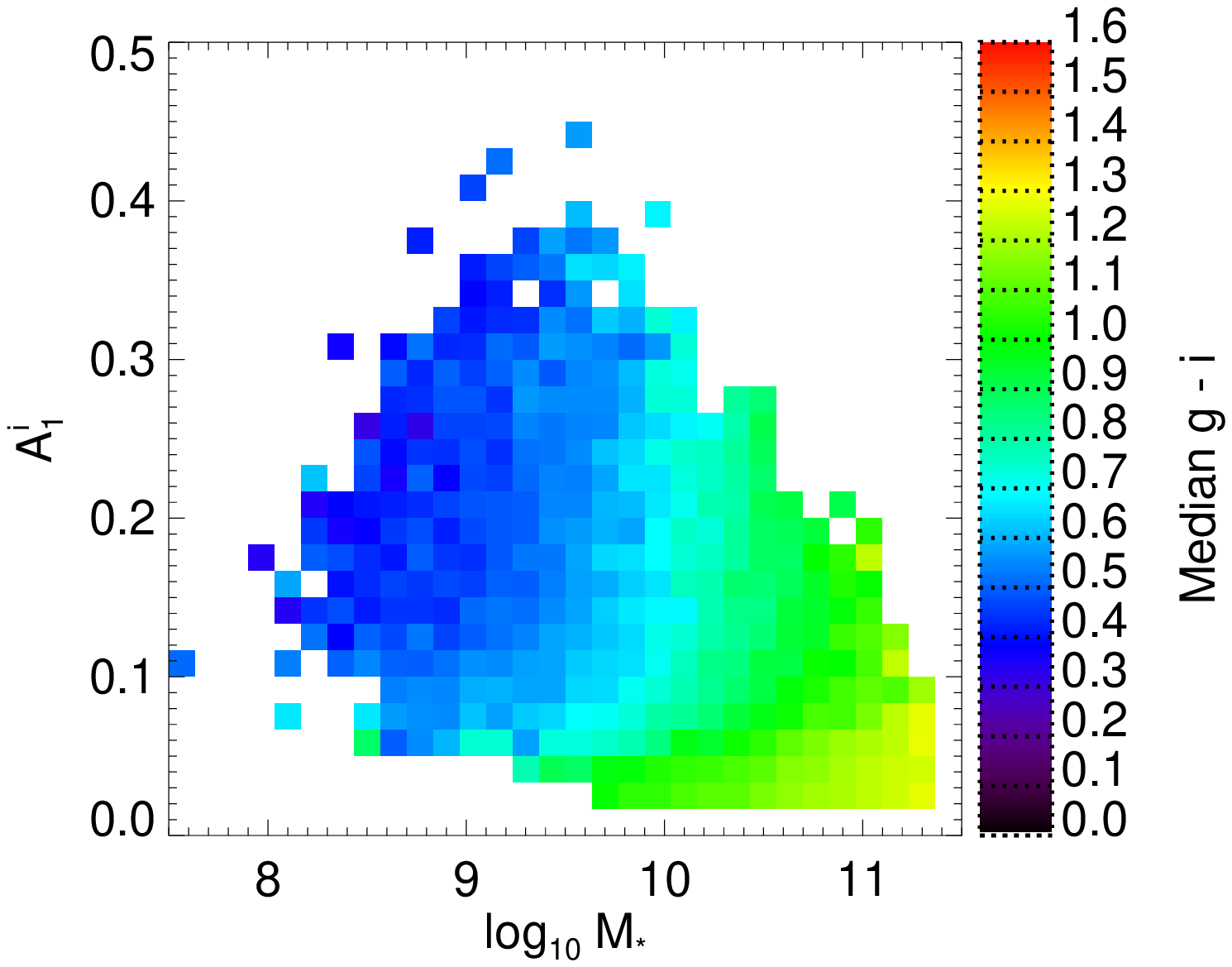}{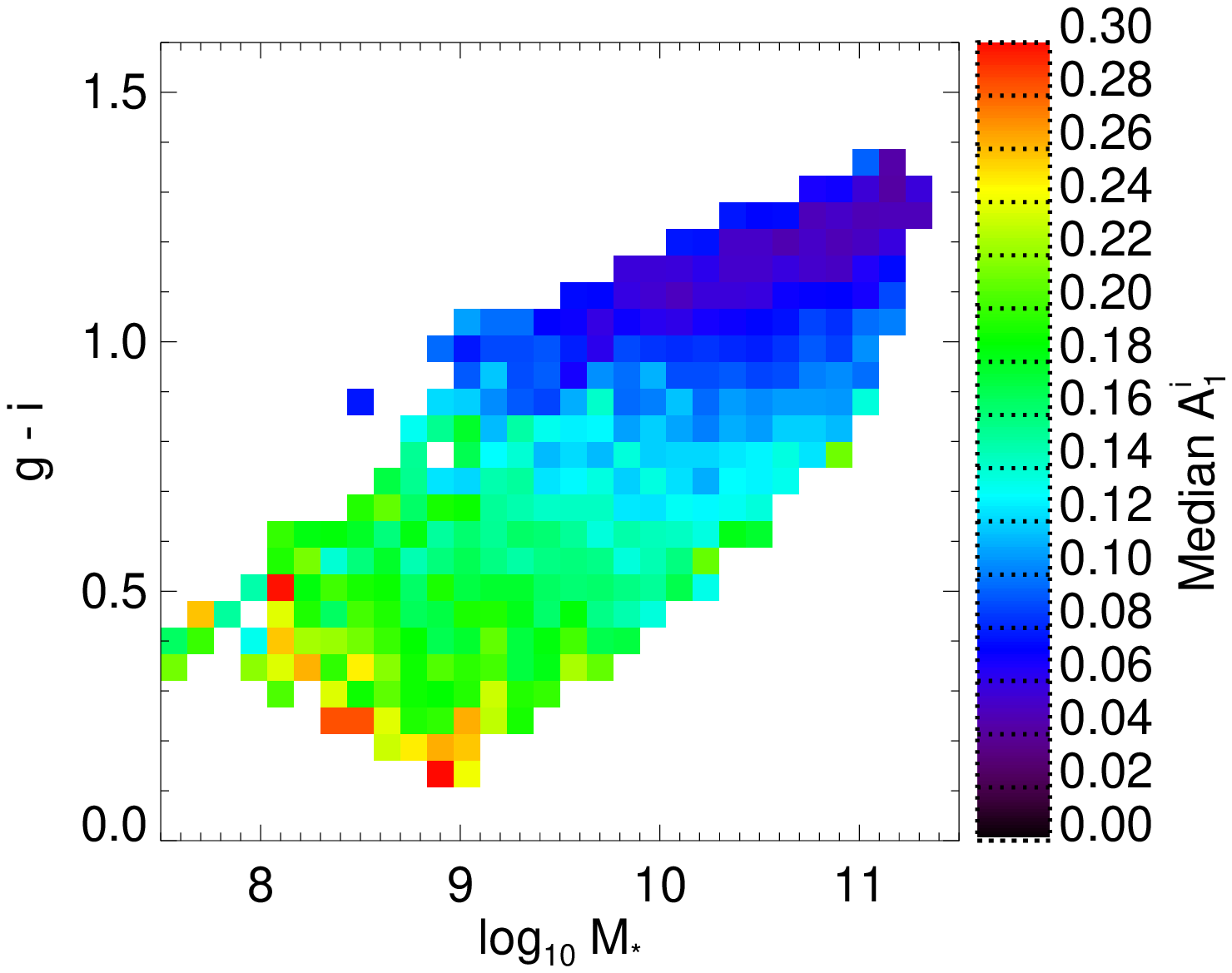}
\plottwo{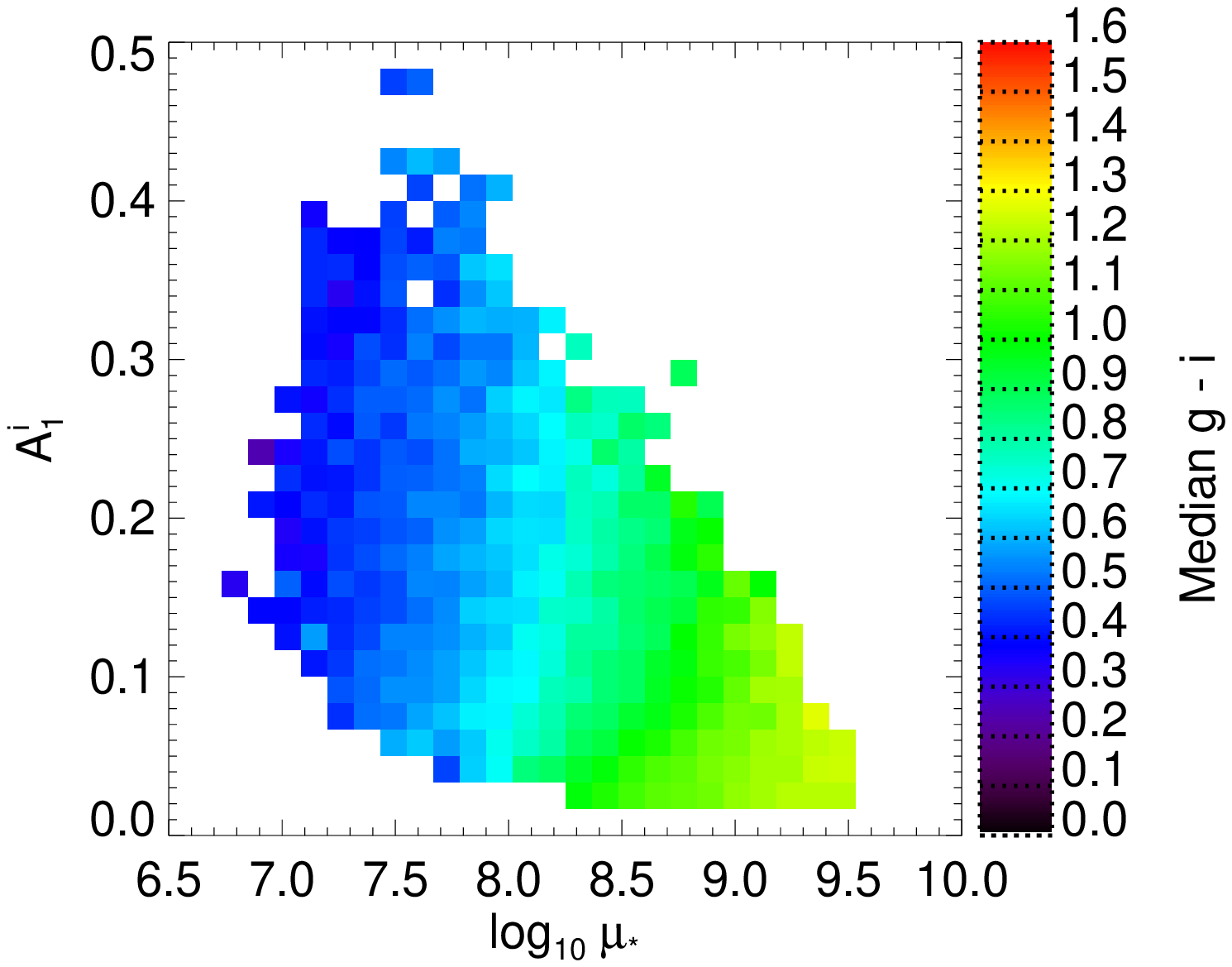}{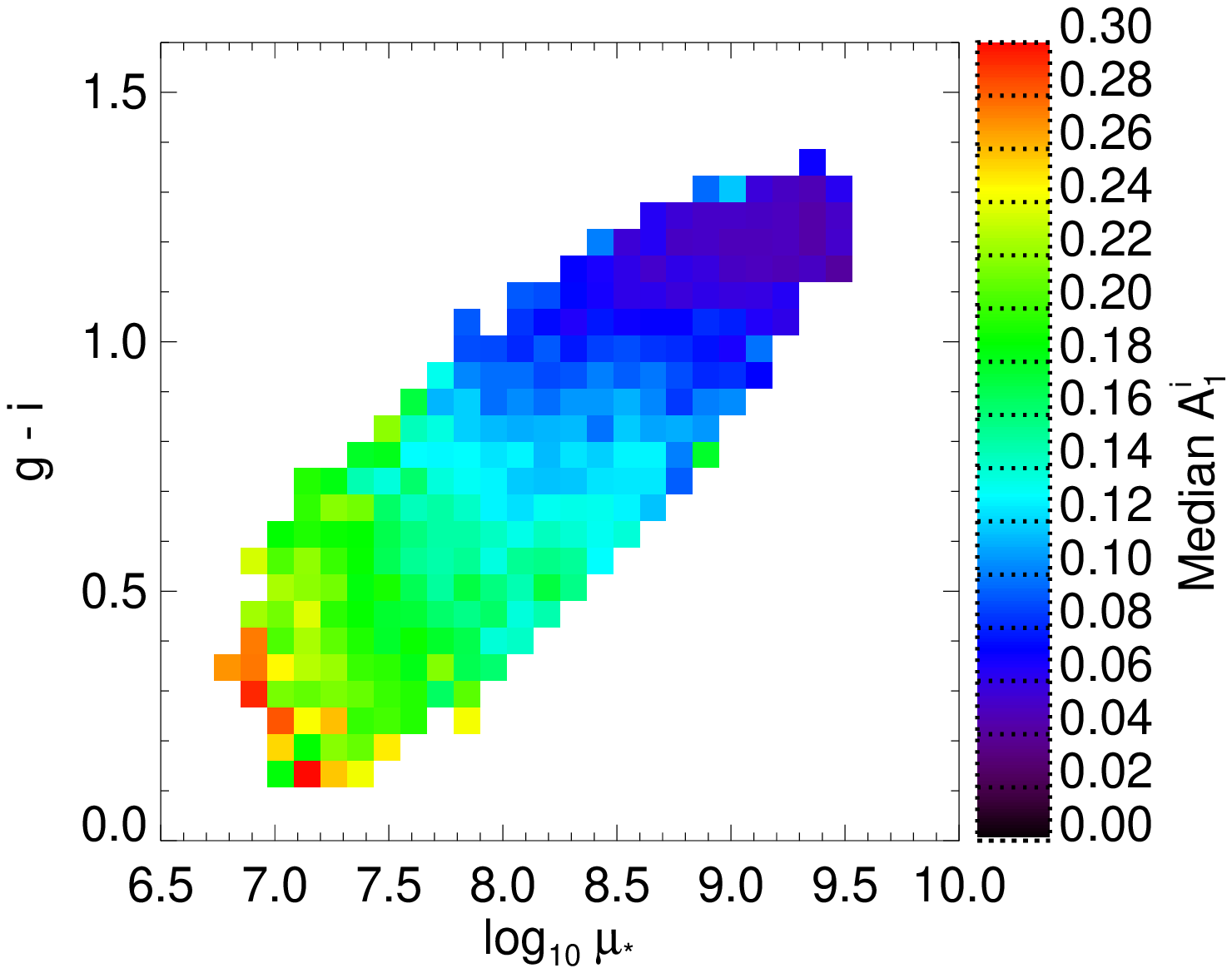}
\plottwo{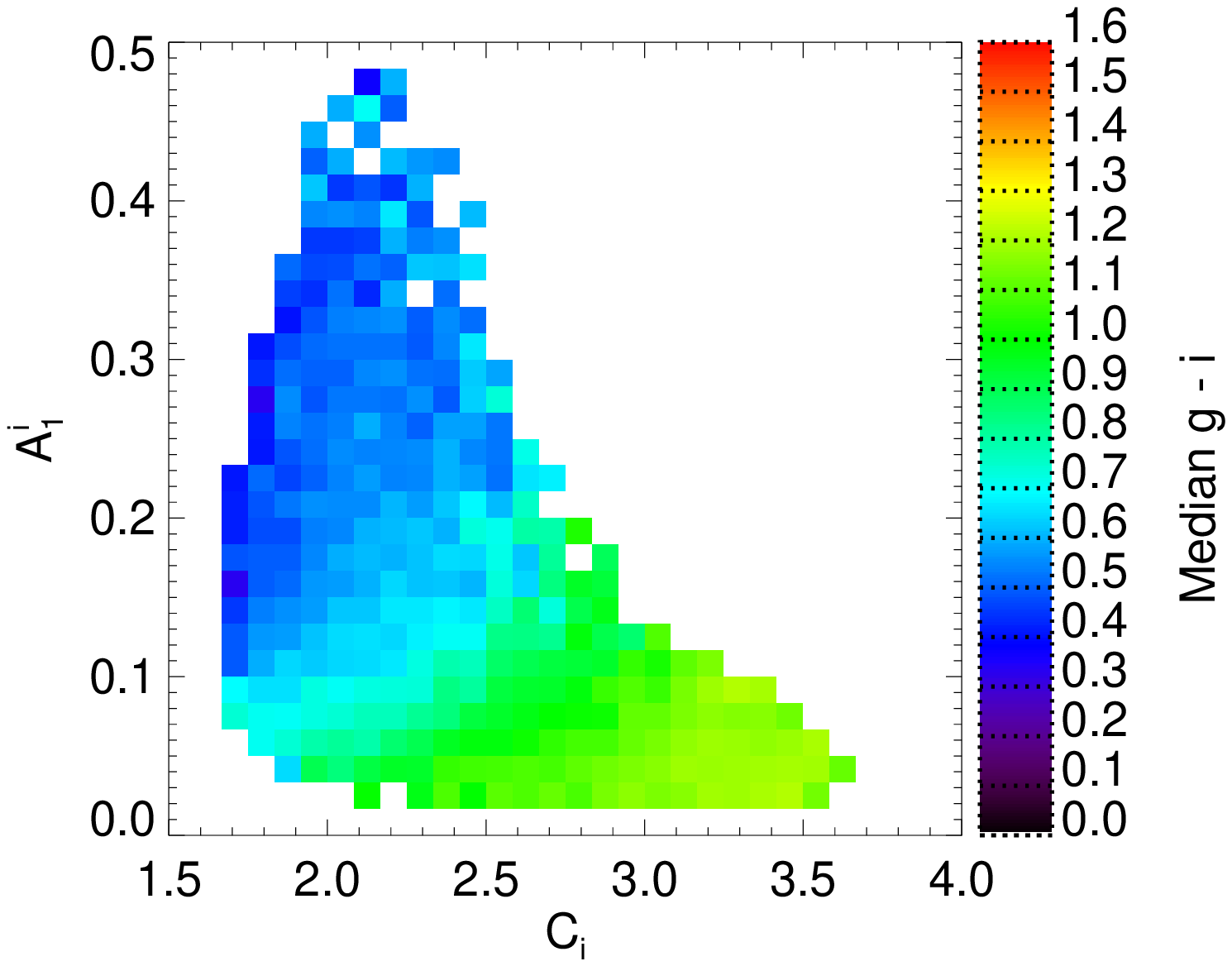}{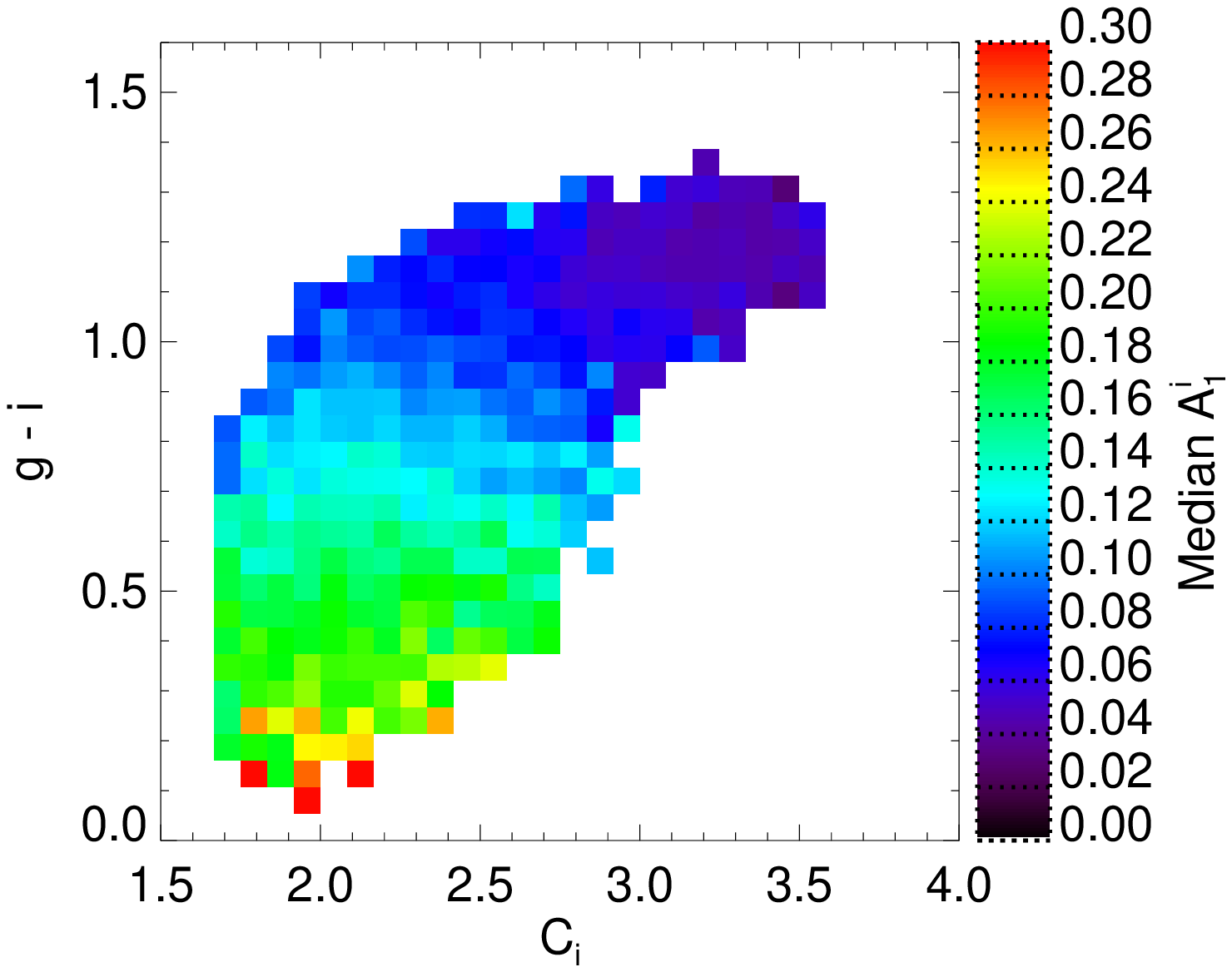}
\caption{Relationships between $g-i$ color and structure shown in several slices of this parameter space.  The color-coding of these two-dimensional histograms indicates the median $g-i$ or $A_1^i$. Galaxies with higher lopsidedness and smaller values of the other structural properties tend to have bluer colors.}
\label{fig:a1-gi}
\end{figure}

We begin our assessment of the interdependence between lopsidedness,
star formation history, and galaxy structure in
Fig.~\ref{fig:s-d4000}. In the three left panels we show the joint
dependence of the median value of the age parameter $\Dbreak$ on
lopsidedness and each of the three structural parameters.  To get a
complementary view, the three panels in the right column show the
joint dependence of the median $A_1^i$ as a function of $\Dbreak$ and
the three structural parameters. The analogous relations using the
post-burst diagnostic $PC_2-PC_1$ and SSFR are shown in
 Fig.~\ref{fig:s-pcs} and Fig.~\ref{fig:a1-ssfr}. These figures refer
to the star formation history in the central region of the galaxy
(inside the fiber). In Figure~\ref{fig:a1-gi} we show the same set of plots, but now using the Petrosian $g-i$ color as an indicator of
the global star formation history of the galaxy. Our result is similar to Woods \& Geller (2007) who find that the star formation becomes more centrally concentrated in interacting galaxies as the pair separation decreases and the specific star formation rate increases. 

In all the plots the overall trends are broadly similar: the left-hand
panels show that galaxies with larger values of lopsidedness and
smaller values of mass, density, and concentration have younger
stellar populations.\footnote{In the left hand panels of
Fig.~\ref{fig:a1-ssfr} we are unable to calculate a meaningful value
for the median SSFR for galaxies lying in the region of low
lopsidedness and high mass, density, or concentration. This is because
the SSFR in most galaxies is too small to be reliably measured. We
leave this region blank, but the low median SSFRs there ($\leq
10^{-12}$ year$^{-1}$) are consistent with what is seen in the figures
using the other age indicators} In each case, for a fixed value of a
structural property (mass, density, or concentration), the stellar
population becomes younger as the lopsidedness increases. Pictorially,
we see that the bands color-coding the stellar age have diagonal or
horizontal orientations. Similarly, the right hand panels in all three
figures show that galaxies with younger stellar populations and lower
masses, densities, and concentrations are more lopsided. In each case,
for a fixed value of a structural property (mass, density, or
concentration), the galaxies become more lopsided as the stellar
population becomes younger: again, we see that the bands color-coding
the lopsidedness have a diagonal or horizontal orientation.

We conclude from this that there is a correlation between lopsidedness
and the star formation history that is independent of any mutual
correlation on the other galaxy structural parameters. To quantify
this result, we turn to a partial correlation analysis to reveal which
correlations are the strongest and perhaps the most fundamental.
Table~\ref{tab:parcor-sfr} shows the correlation coefficients of various
combinations of structural and star formation properties. Values of
$\log_{10} SFR/M_*$ below $-12$ have been set to $-12$ (see
above). The results in this table show that there are separate and
significant correlations between the structural properties themselves
(independent of lopsidedness or star formation history), between
lopsidedness and the other structural parameters (independent of star
formation history), and between star formation history and
lopsidedness (independent of the other structural parameters). These
results are fully consistent with the visual representations in the
figures.

The lower left panels in both Figures~\ref{fig:s-pcs} and
\ref{fig:a1-ssfr} show that the relatively symmetric galaxies with
$A_1^i < 0.15$ have higher values of $PC_2 - PC_1$ and SSFR for less
concentrated galaxies.  However, the trend is reversed for the highly
lopsided galaxies with $A_1^i > 0.2$.  These lopsided galaxies
typically show higher SSFR and larger $PC_2 - PC_1$ at higher
concentration. The most natural interpretation is that these are
objects that are undergoing or have recently undergone central
starbursts (leading to an increased concentration of light). This is
consistent with the fact that an analogous behavior is not clearly
seen in Figure~\ref{fig:a1-gi} where we are investigating the {\it
global} star formation history using the Petrosian $g-i$ color. Here,
the youngest (bluest) galaxies are lopsided but have low
concentration.

\clearpage

\clearpage

\clearpage

\clearpage

\section{Metallicity and Lopsidedness \label{sec:met}}

The results above suggest that the processes that produce lopsidedness
lead to an inflow of gas into the central region where it can fuel
star formation. Any plausible source for this gas (capture of a small galaxy, gas transported from the outer disk to the inner region, the accretion of intergalactic gas) will have a lower metallicity than the pre-existing gas in the central region. Thus - provided that the inflow is rapid enough that the gas is not fully enriched by associated star formation before reaching the center -  this inflow should have a signature in the chemical abundances in the interstellar medium. This has been seen by Kewley et al. (2006) and by Ellison et al. (2008a) for interacting galaxy pairs. Such a signature would also be consistent with Ellison et al. (2008b) who found that residuals in the mass-metallicity relation correlate inversely with the specific star formation rate.

\begin{figure}[ht]
\epsscale{0.6}
\plotone{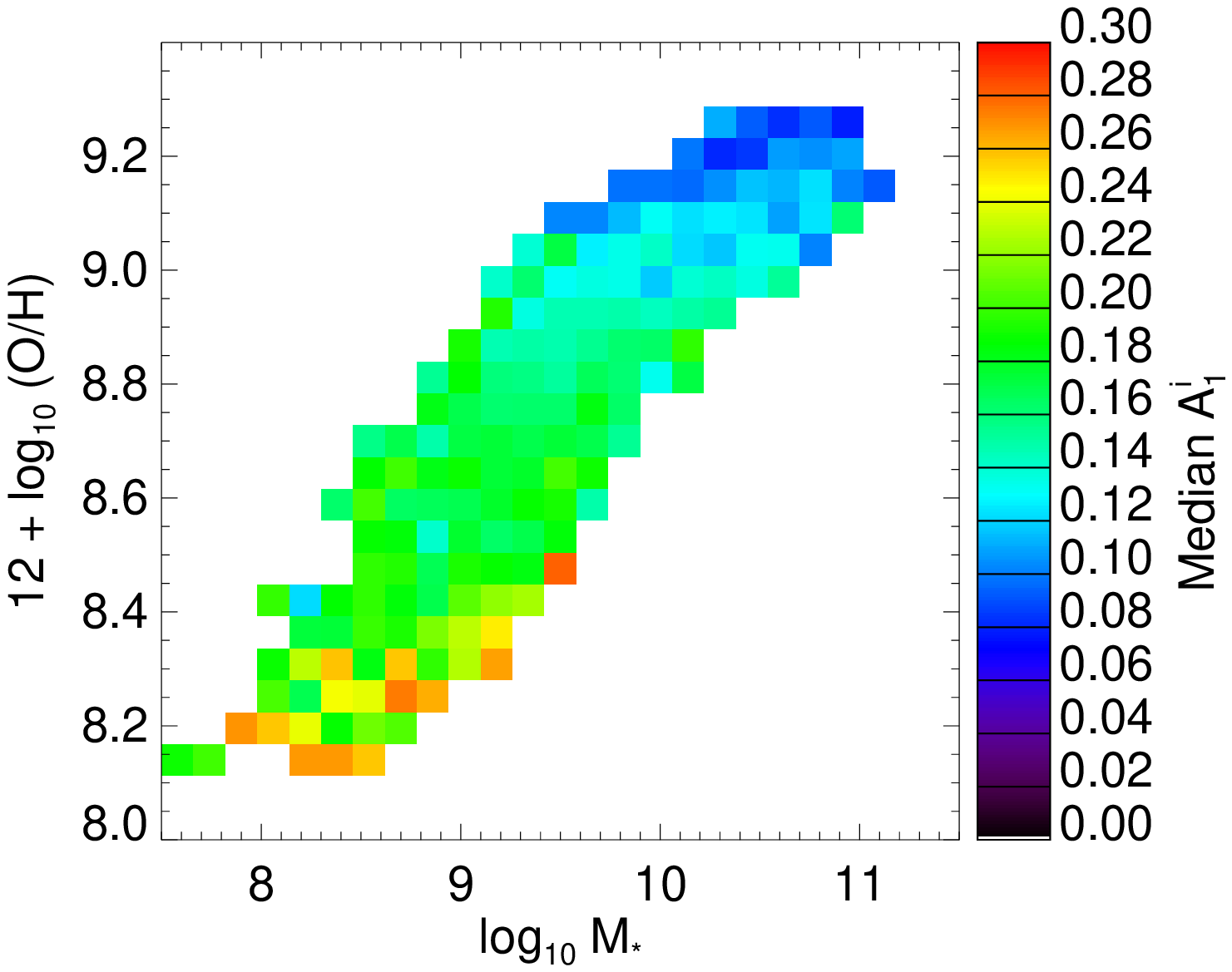}
\plotone{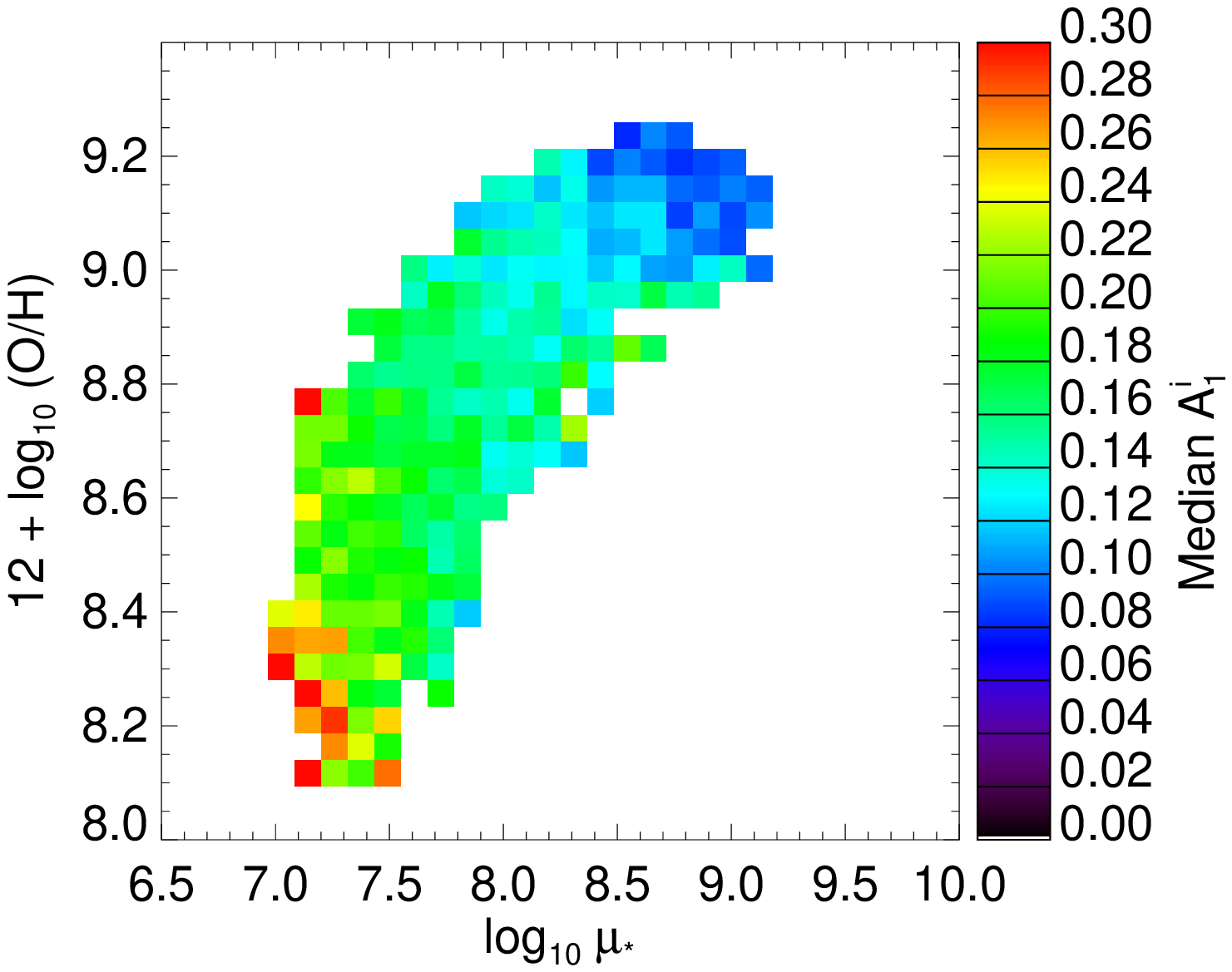}
\label{fig:s-oh}
\caption{Relationships between metallicity and structure shown in two slices of this parameter space.  The color-coding of these two-dimensional histograms indicates the median $A_1^i$. While the main correlation between the three parameters is between lopsidedness and either other structural parameter, there is a residual correlation between lopsidedness and metallicity at fixed mass.}
\end{figure}

To test this idea, in Figure~\ref{fig:s-oh}, we show the
mass-metallicity relation for our sample, color-coded by the median
value for the lopsidedness in each cell. The strongest trend in this
figure is simply the increase in lopsidedness with decreasing galaxy
mass (Paper I). However, there is a weaker residual trend for the
lopsidedness at a given value of stellar mass to increase with
decreasing metallicity. This can also be seen in Table~\ref{tab:parcor-oh} where we list the relevant partial correlation coefficients.

\citet{tre+04} showed that residuals in the mass-metallicity relation 
correlate with the surface mass density (with lower metallicity at
fixed mass for less dense galaxies). We saw in Paper I that there is a
strong inverse correlation between surface mass density and
lopsidedness. Could the apparent correlation between lopsidedness and
metallicity in Figure~\ref{fig:s-oh} be induced through mutual
correlations with surface density? The lower panel in
Figure~\ref{fig:s-oh} shows that at fixed surface mass density
galaxies that have lower metallicity are more lopsided. This is
confirmed by the partial correlation coefficients in Table~\ref{tab:parcor-oh}.  We therefore conclude that there is real correlation between
lopsidedness and metallicity in the sense expected for an inflow of
lower metallicity gas.

\clearpage

\begin{figure}[ht]
\epsscale{0.6}
\plotone{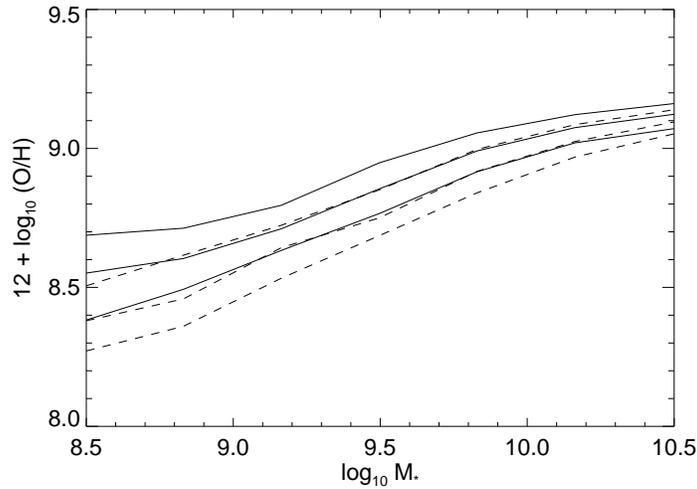}
\label{fig:m-z-q}
\caption{Mass-metallicity relation for the most ($A_1^i > 0.2$, {\em dashed lines}) and least lopsided ($A_1^i < 0.08$, {\em solid lines}) galaxies. The 25th, 50th, and 75th percentile lines are shown for each distribution.  The most lopsided galaxies have a metallicity deficit of 0.05-0.15 dex compared to the least lopsided galaxies at the same mass.  The deficit is greater at low mass than at high mass. }
\end{figure}

We finish this section by quantifying the metallicity deficit between
the most and least lopsided galaxies of our sample.  We use two
subsamples taken from our main sample: low-lopsidedness galaxies
($A_1^i < 0.08$) and high-lopsidedness galaxies ($A_1^i > 0.20$). We
compare the mass-metallicity relations of the two subsamples in
Fig.~\ref{fig:m-z-q}.  The median metallicity of the less lopsided
galaxies is 0.05$-$0.15 dex greater than in the more lopsided galaxies
at the same mass. The metallicity difference is greater at low mass
than at high mass.  This deficit is consistent with the offset found
in \citet{ell+08a} for galaxies with close companions. The relatively small amplitude of the effect would rule out an extreme picture in which very metal poor gas falls into the center and dominates the interstellar medium there.

\clearpage

\section{Nuclear Activity and Lopsidedness \label{sec:agn}}

We start with the simplest measures of the relationship between
AGN and lopsidedness. However, since both lopsidedness (Paper I) and
AGN properties \citep{kau+03c,hec+04,kew+06} correlate separately with
other galaxy structural properties, we follow this with a
multivariate approach to isolate the specific role of lopsidedness. We
also know that AGN with higher luminosity are preferentially located
in galaxies with younger central stellar populations
\citep{kau+03c,kau+07}, and we have seen above that galaxies show a
strong link between lopsidedness and a young central stellar
population. We will attempt to sort out this complex web of
correlations using an approach based on comparing AGN hosts to ``twin''
galaxies without AGN that have been matched to the AGN hosts in their
structure and central stellar population.

\begin{figure}[ht]
\epsscale{0.5}
\plotone{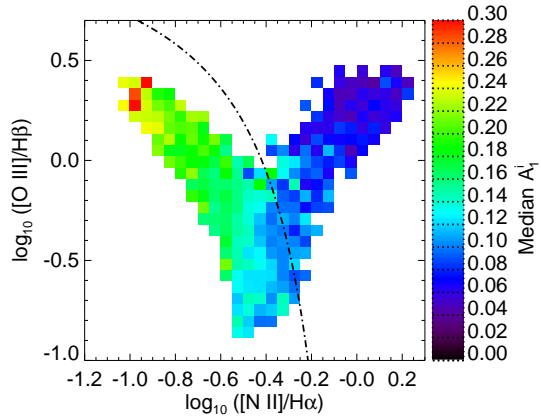}
\caption{BPT diagram color-coded by median $A_1^i$. The strong correlation with lopsidedness is that with  star formation rather than that with AGN activity. The black line separates the AGN from the star-forming galaxies and is adopted from \citet{kau+03c}.}
\label{fig:bpt}
\end{figure}

We begin with Figure~\ref{fig:bpt} in which we plot the distribution of
lopsidedness in the diagnostic line ratio diagram used by
\citet{kau+03c} to select Type 2 AGN (the ``BPT Diagram'', after
\citealt{bal+81}). As discussed in \citet{kau+03c} and \citet{kew+06},
the distribution running from the upper left to the lower right is the
locus of star forming galaxies (with low metallicity at the top left
and high metallicity at the lower right). The spur running to the
upper right off the location of the metal-rich star forming galaxies
is the locus of AGN. It represents a ``mixing line'' with AGN dominated
objects at the upper right and star-forming/AGN composite objects at
the lower left close to where the AGN spur joins the locus of star
forming galaxies.

It is clear from this plot that star forming galaxies are
systematically more lopsided than AGN host galaxies. This is not
surprising, since \citet{kau+03c} show that AGN hosts are generically
massive, dense galaxies (which we have shown in Paper I to typically
have small values for $A_1^i$). In the region occupied by the AGN the
median value of $A_1^i$ is less than 0.1, with a weak trend for
lopsidedness to be larger in the composite objects than in the pure
AGN. This trend is not surprising, given the connection between
lopsidedness and star formation discussed above.

\begin{figure}[ht]
\epsscale{0.5}
\plotone{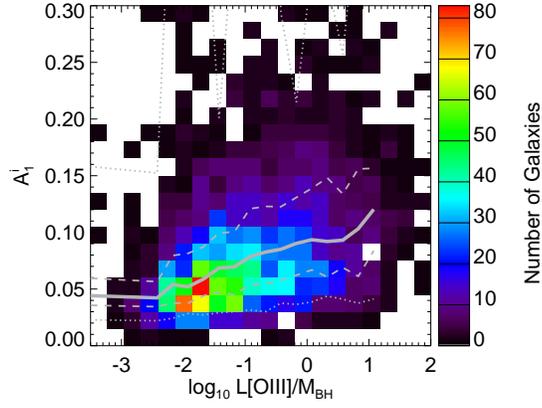}
\epsscale{1.1}
\caption{The distribution of $A_1^i$ and $L$[\ion{O}{3}]$/M_{BH}$ for all galaxies in the sample. The 5th, 25th, 50th, 75th, and 95th percentiles of $A_1^i$ are overplotted in gray. Typical lopsidedness increases with  $L$[\ion{O}{3}]$/M_{BH}$.  Median $A_1^i$ reaches a moderate value ($A_1^i = 0.1$) when $L$[\ion{O}{3}]$/M_{BH}$ is high at 10~$L_{\odot}/M_{\odot}$.}
\label{fig:a1-lo3mbh}
\end{figure}

The Type 2 AGN in our sample span a large range in luminosity,
and it is quite possible that the mechanism responsible for fueling
the growth of the black hole might change systematically as a function
of AGN luminosity. To examine this, in Figure~\ref{fig:a1-lo3mbh} we plot the
distribution of $A_1^i$ vs. the ratio of the extinction-corrected
[\ion{O}{3}]5007 emission line luminosity to the black hole mass. This ratio
will be roughly proportional to the black hole accretion rate relative
to the Eddington limit (see \citealt{hec+04}).

\begin{figure}
\centering
\leavevmode
\columnwidth=.2\columnwidth
\begin{minipage}[b]{1.25in}
\includegraphics[width=1.25in]{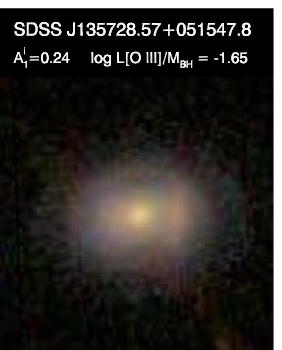}
\end{minipage}
\hfill
\begin{minipage}[b]{1.25in}
\includegraphics[width=1.25in]{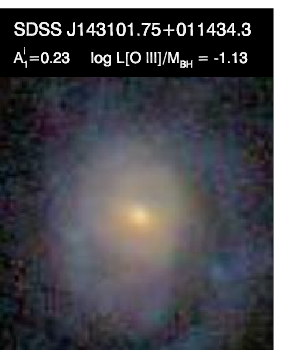}
\end{minipage}
\hfill
\begin{minipage}[b]{1.25in}
\includegraphics[width=1.25in]{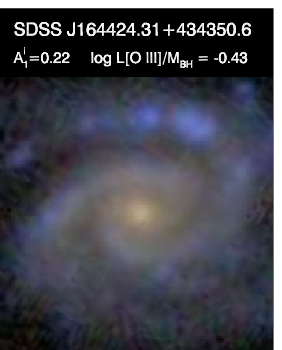}
\end{minipage}
\hfill
\begin{minipage}[b]{1.25in}
\includegraphics[width=1.25in]{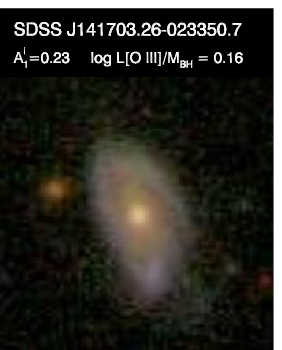}
\end{minipage}
\hfill
\begin{minipage}[b]{1.25in}
\includegraphics[width=1.25in]{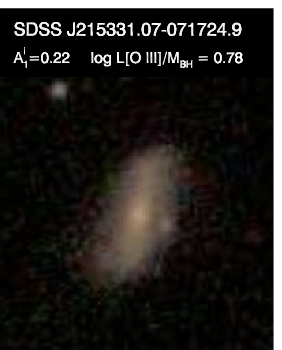}
\end{minipage}
\hfill
\begin{minipage}[b]{1.25in}
\includegraphics[width=1.25in]{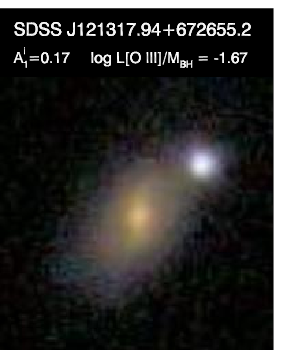}
\end{minipage}
\hfill
\begin{minipage}[b]{1.25in}
\includegraphics[width=1.25in]{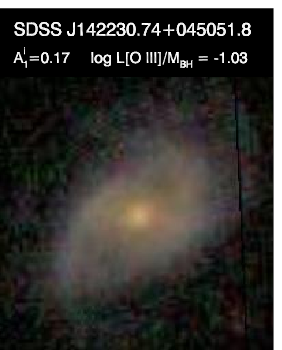}
\end{minipage}
\hfill
\begin{minipage}[b]{1.25in}
\includegraphics[width=1.25in]{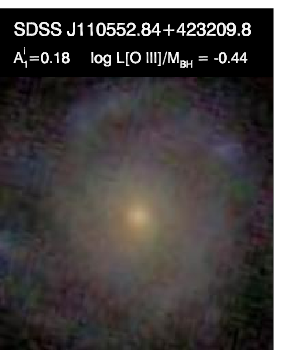}
\end{minipage}
\hfill
\begin{minipage}[b]{1.25in}
\includegraphics[width=1.25in]{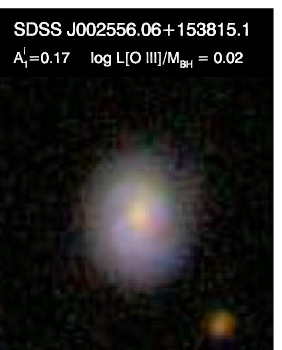}
\end{minipage}
\hfill
\begin{minipage}[b]{1.25in}
\includegraphics[width=1.25in]{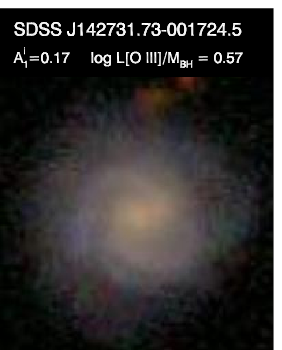}
\end{minipage}
\hfill
\begin{minipage}[b]{1.25in}
\includegraphics[width=1.25in]{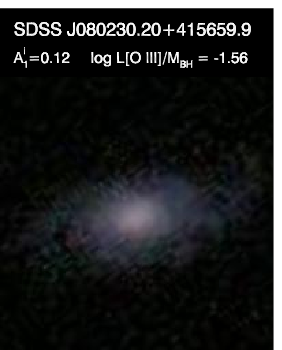}
\end{minipage}
\hfill
\begin{minipage}[b]{1.25in}
\includegraphics[width=1.25in]{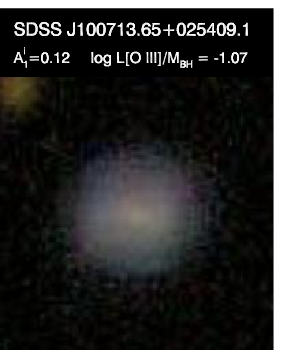}
\end{minipage}
\hfill
\begin{minipage}[b]{1.25in}
\includegraphics[width=1.25in]{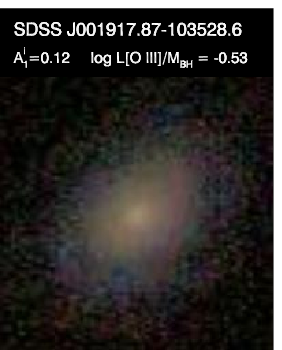}
\end{minipage}
\hfill
\begin{minipage}[b]{1.25in}
\includegraphics[width=1.25in]{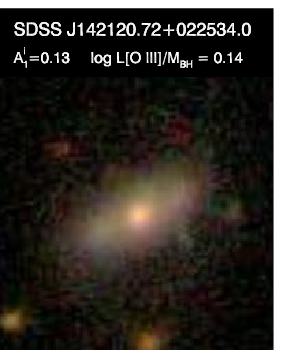}
\end{minipage}
\hfill
\begin{minipage}[b]{1.25in}
\includegraphics[width=1.25in]{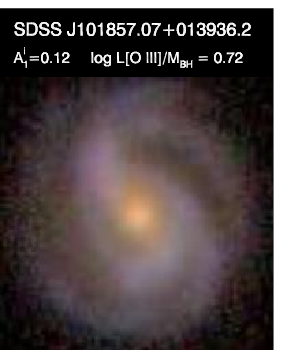}
\end{minipage}
\hfill
\begin{minipage}[b]{1.25in}
\includegraphics[width=1.25in]{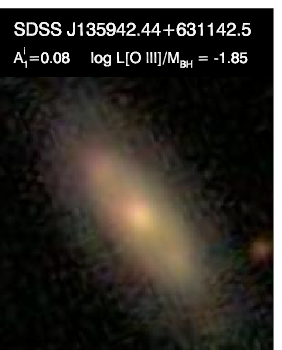}
\end{minipage}
\hfill
\begin{minipage}[b]{1.25in}
\includegraphics[width=1.25in]{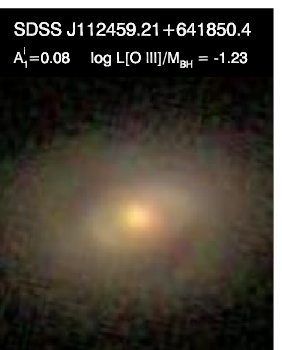}
\end{minipage}
\hfill
\begin{minipage}[b]{1.25in}
\includegraphics[width=1.25in]{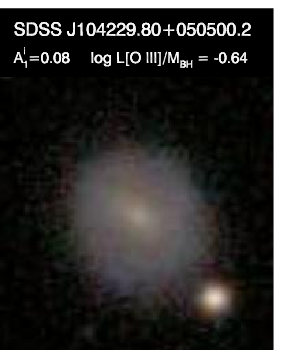}
\end{minipage}
\hfill
\begin{minipage}[b]{1.25in}
\includegraphics[width=1.25in]{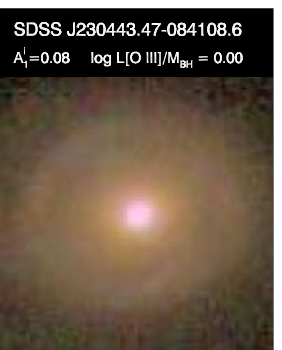}
\end{minipage}
\hfill
\begin{minipage}[b]{1.25in}
\includegraphics[width=1.25in]{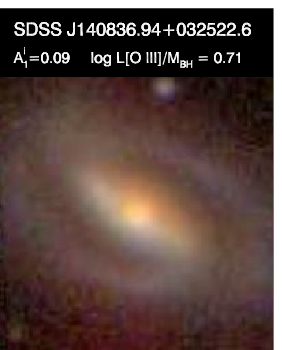}
\end{minipage}
\hfill
\begin{minipage}[b]{1.25in}
\includegraphics[width=1.25in]{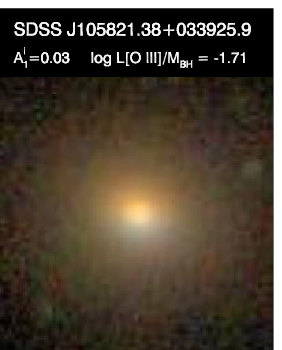}
\end{minipage}
\hfill
\begin{minipage}[b]{1.25in}
\includegraphics[width=1.25in]{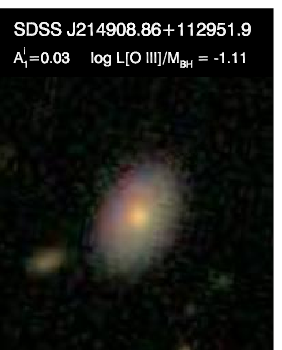}
\end{minipage}
\hfill
\begin{minipage}[b]{1.25in}
\includegraphics[width=1.25in]{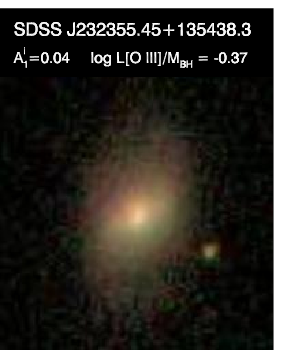}
\end{minipage}
\hfill
\begin{minipage}[b]{1.25in}
\includegraphics[width=1.25in]{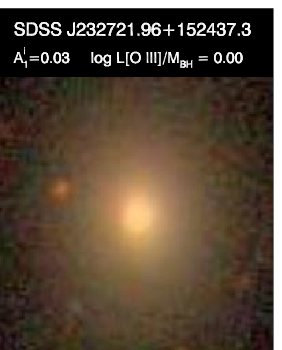}
\end{minipage}
\hfill
\begin{minipage}[b]{1.25in}
\includegraphics[width=1.25in]{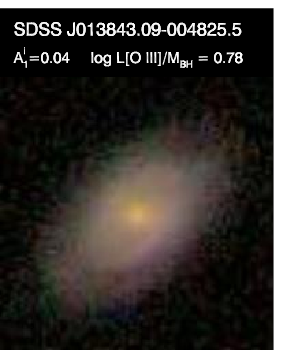}
\end{minipage}
\hfill
\caption{Twenty-five SDSS galaxies with increasing lopsidedness from bottom to top and increasing $L$[\ion{O}{3}]$/M_{BH}$ from left to right. Each image is 30\arcsec $\times$ 30\arcsec, or about 23 kpc $\times$ 23 kpc at $z = 0.04$, a typical redshift of the sample. }
\label{fig:agnmosaic}
\end{figure}

We see in this plot that there is a systematic increase in
lopsidedness with increasing AGN luminosity (increasing black hole
growth rate). Below values of $L$[\ion{O}{3}]$/M_{BH} \sim 10^{-2}
L_{\odot}/M_{\odot}$ (corresponding roughly to accretion rates less
than $10^{-4}$ of the Eddington limit), the median value of $A_1^i$ is
only $\sim$ 0.05 (lopsidedness is essentially undetectably small). The
median lopsidedness then increases with increasing AGN luminosity,
reaching a value of $A_1^i \sim$0.12 at $L$[\ion{O}{3}]$/M_{BH} = 10
L_{\odot}/M_{\odot}$ ($\sim$ 10\% of the Eddington limit). This median
value corresponds to only mild lopsidedness and (even at such high AGN
luminosities) only about 10\% of the AGN hosts are strongly lopsided
($A_1^i > 0.2$). Thus, few of the high luminosity AGN
hosts appear to be undergoing major mergers, but many are mildly
disturbed. To give the reader a visual impression of the
structure of AGN hosts we show a montage of SDSS color images of low-
and high-luminosity AGN in Figure~\ref{fig:agnmosaic}. 

\begin{figure}[ht]
\epsscale{1.1}
\plottwo{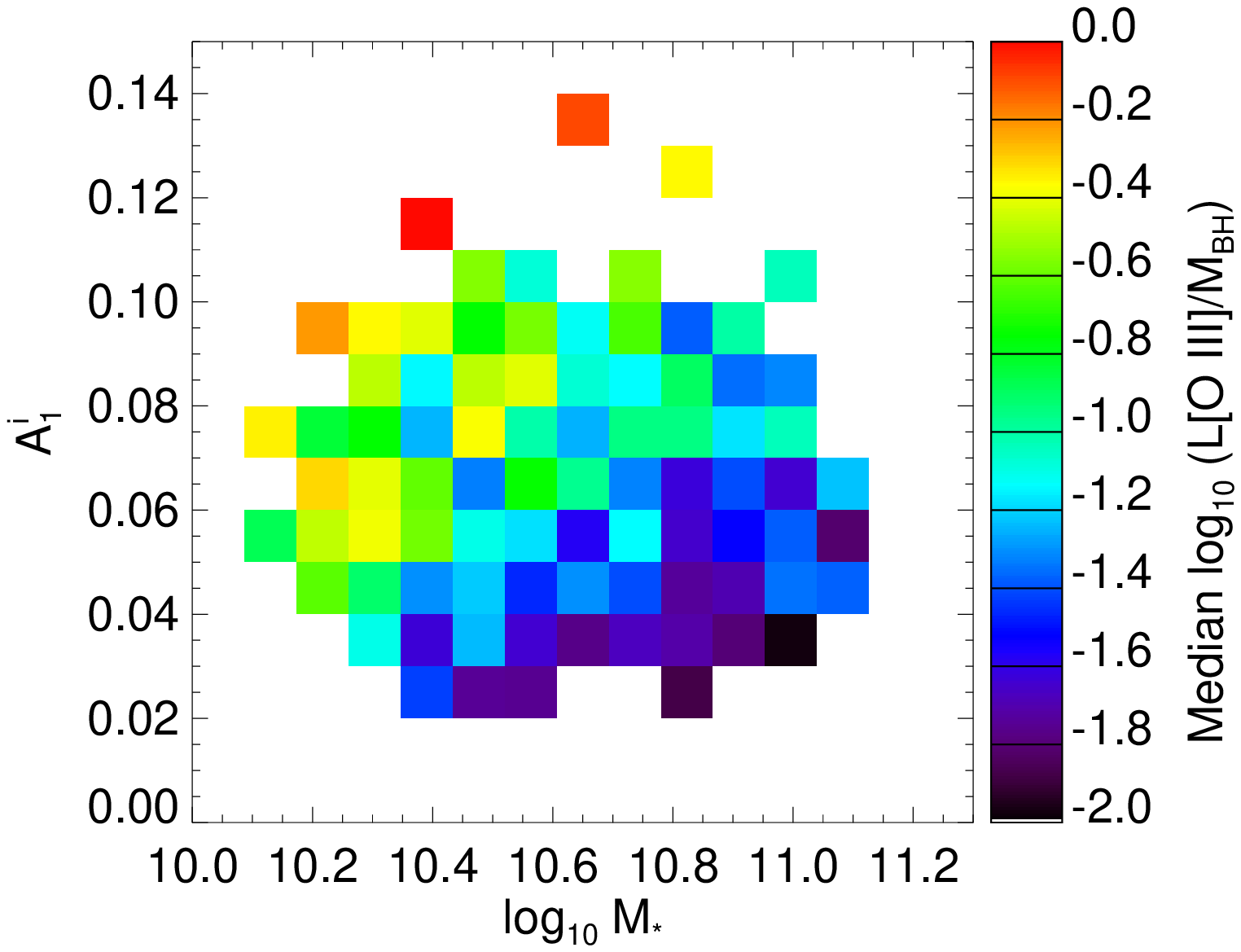}{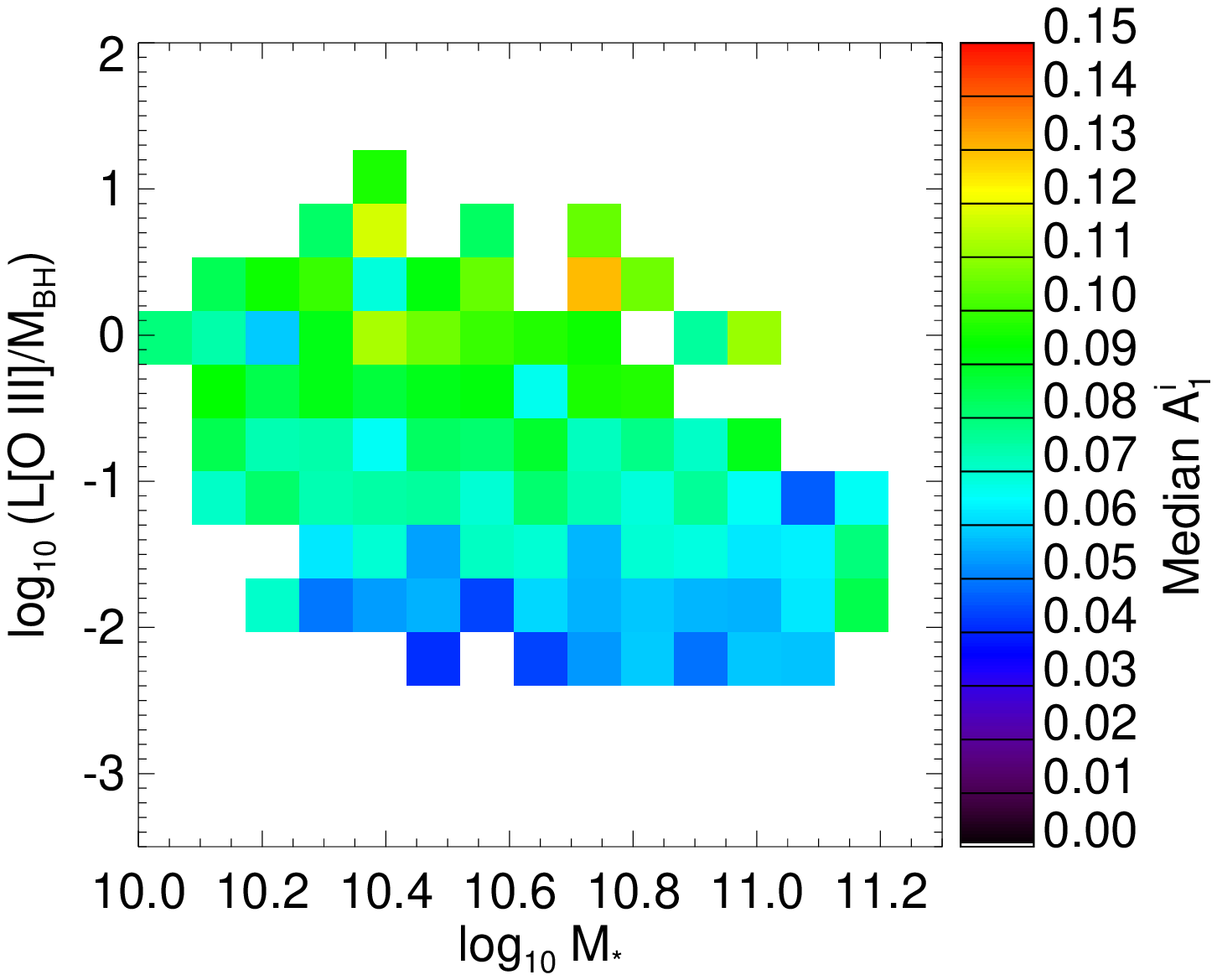}
\plottwo{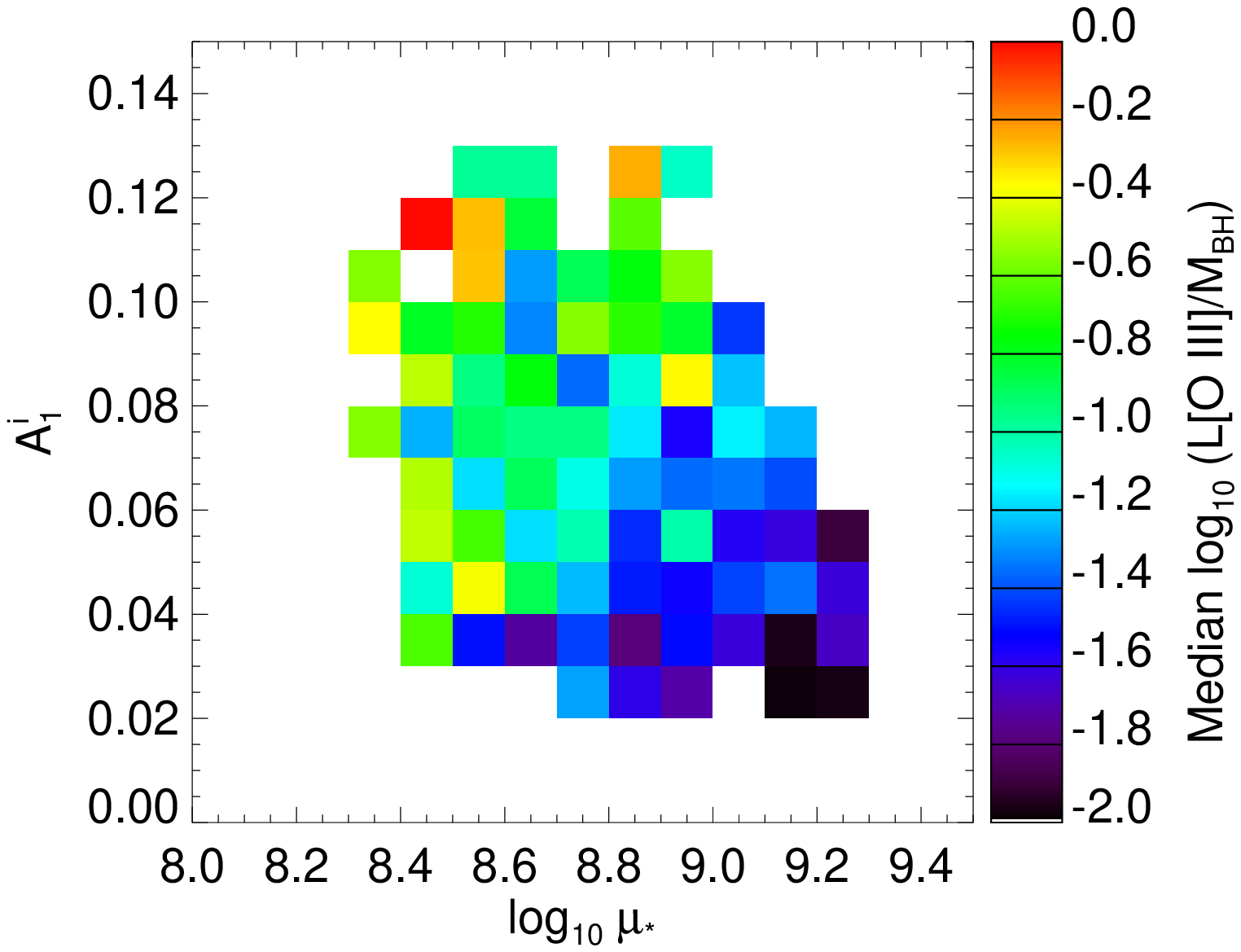}{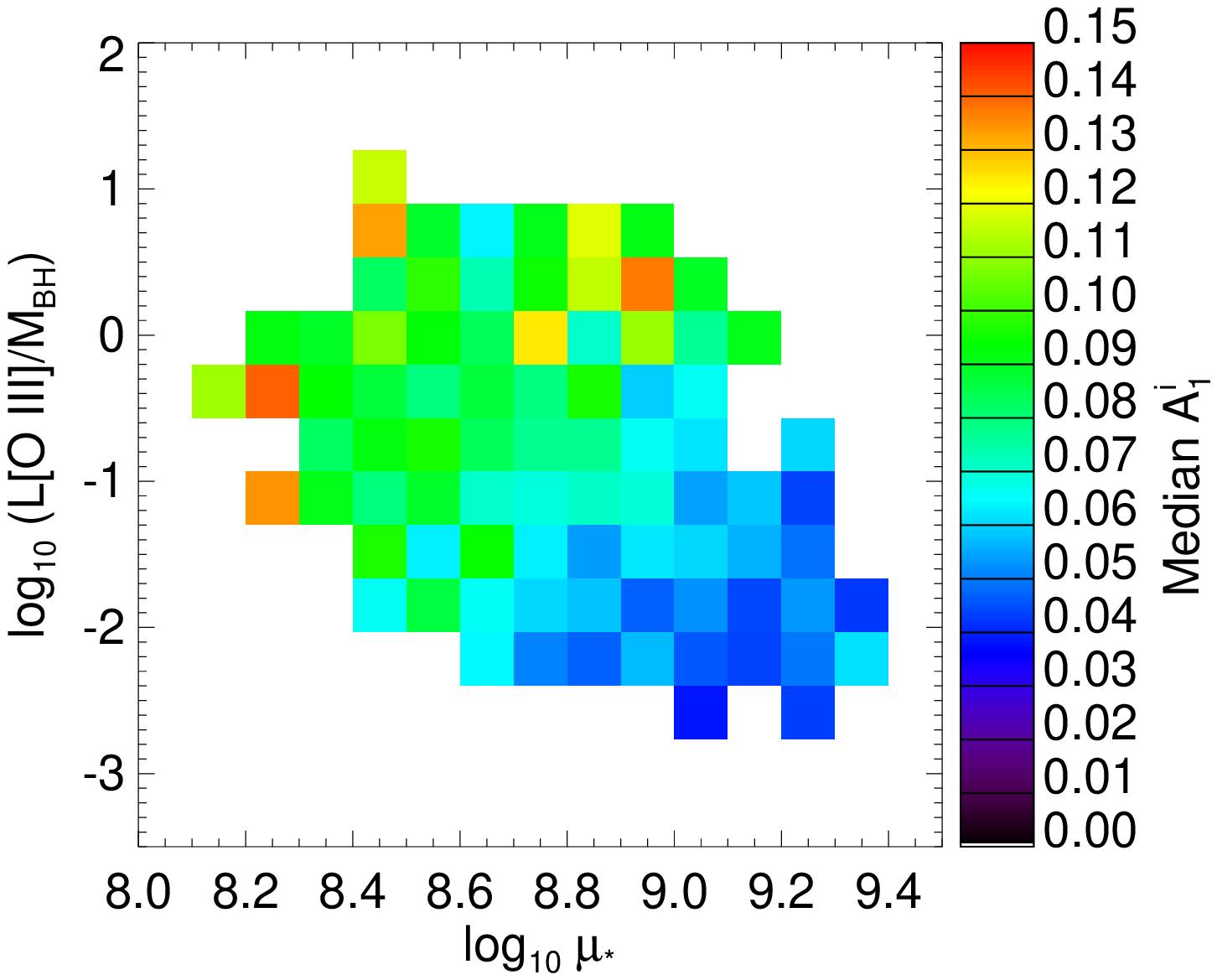}
\plottwo{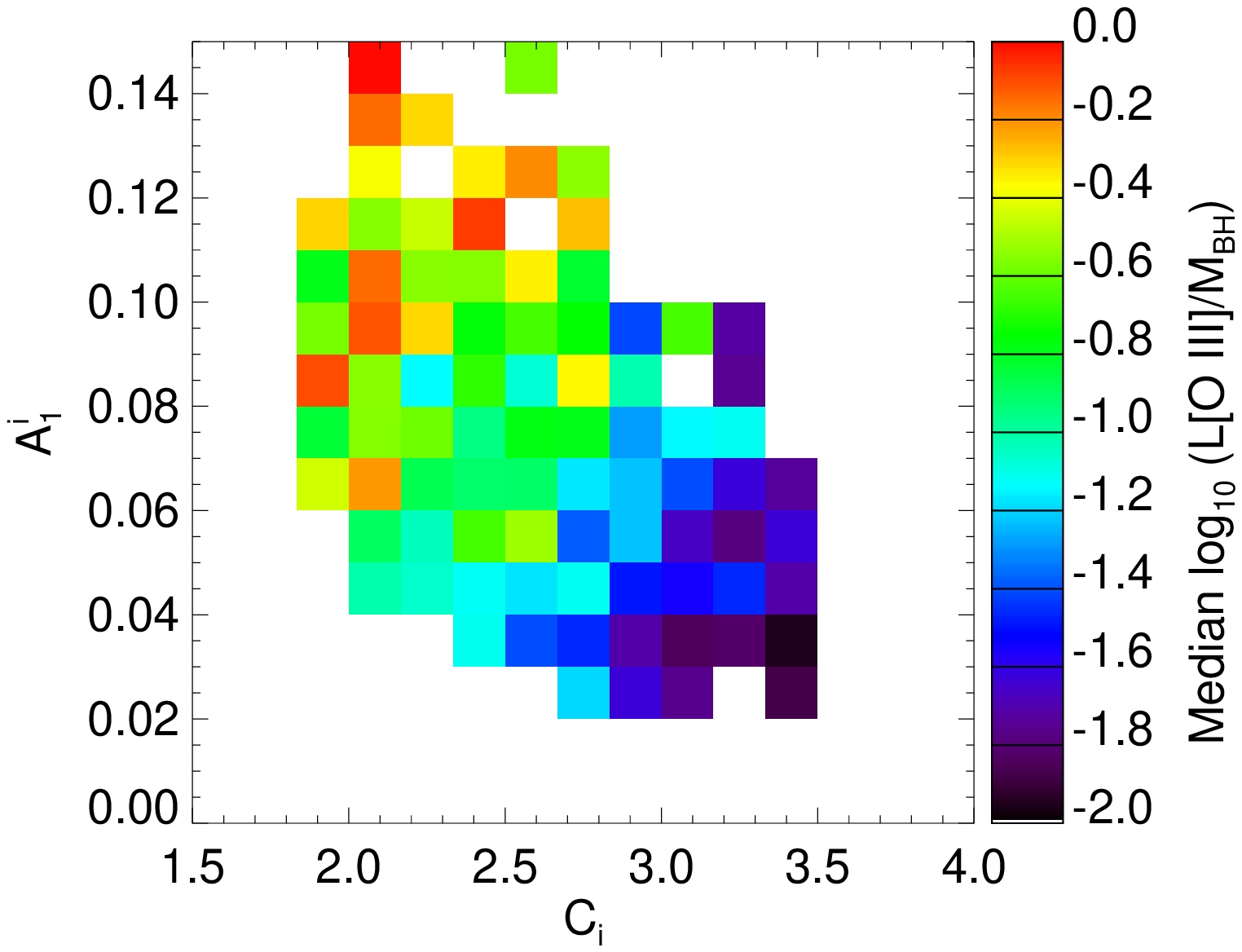}{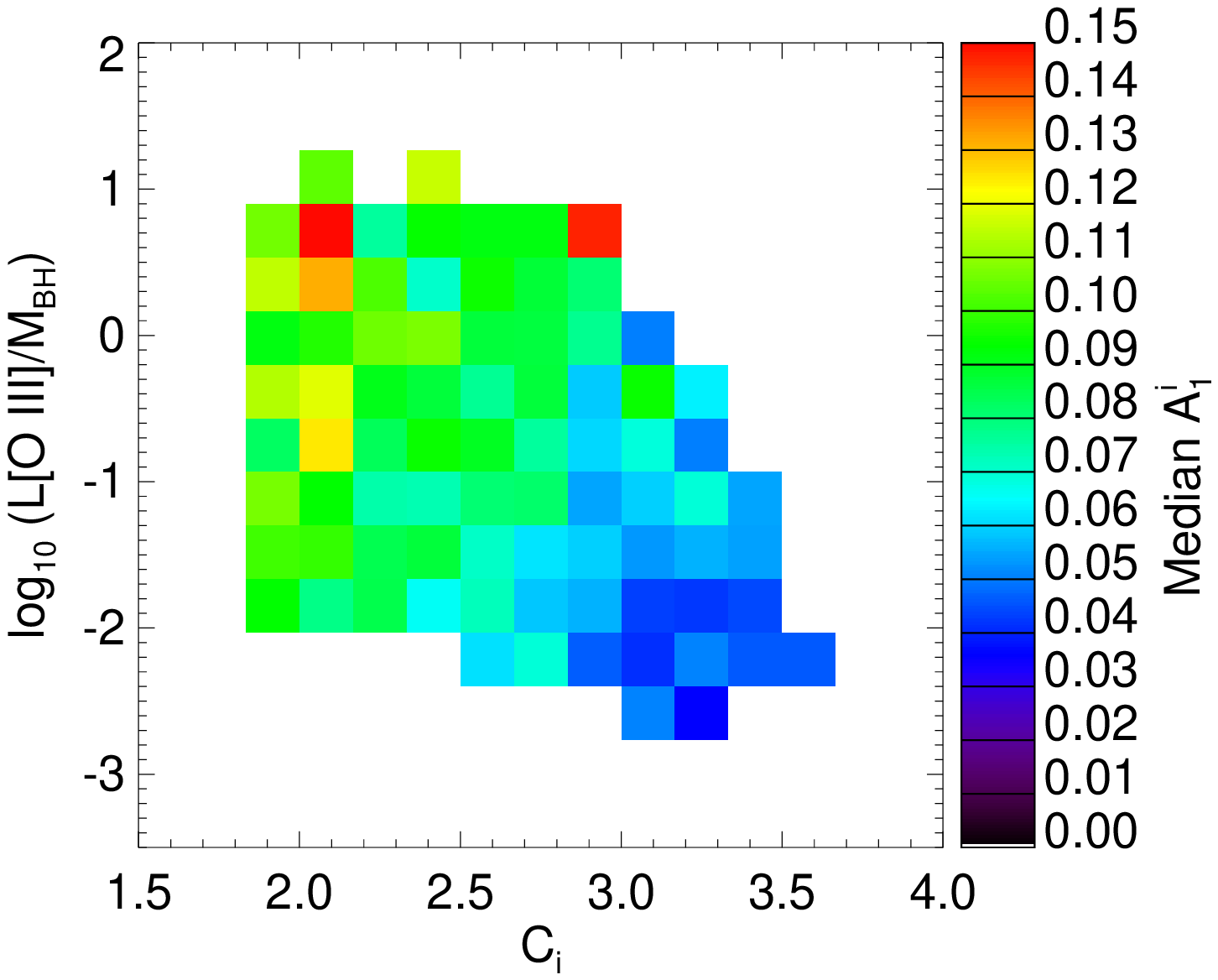}
\caption{Relationships between L[\ion{O}{3}]/$M_{BH}$ and structure shown in several slices of this parameter space.  The color-coding of these two-dimensional histograms indicates the median L[\ion{O}{3}]/$M_{BH}$ or $A_1^i$.  In each case, there is a correlation between lopsidedness and AGN luminosity independent of these structural parameters. }
\label{fig:s-lo3mbh}
\end{figure}

Figure~\ref{fig:a1-lo3mbh} by itself does not establish a direct connection between AGN
luminosity and lopsidedness. As discussed in \citet{kau+03c} and
\citet{kew+06}, the properties of AGN hosts vary systematically as a
function of AGN luminosity, and we have seen that lopsidedness is in
turn strongly linked to galaxy structural properties (Paper I). Thus,
in Figure~\ref{fig:s-lo3mbh} we examine the interdependence between
lopsidedness, galaxy structure, and AGN luminosity. In the left three
panels we see that, at fixed galaxy mass, density, and concentration,
the median AGN luminosity increases with increasing
lopsidedness. Similarly, the right-hand panels show that at fixed
values of galaxy mass, density, and concentration the median value of
lopsidedness increases with increasing AGN luminosity. The trends are
most clear in the panels involving mass, and least clear in those
involving concentration. These results are confirmed by the partial
correlation coefficients listed in Table~\ref{tab:parcor-agn}.

We conclude that there is a link between AGN luminosity and
lopsidedness that is independent of the other galaxy structural
parameters. This is analogous to the results for star formation, but
the trends for the AGN are significantly weaker (the range spanned by
the variation in the median lopsidedness in the AGN hosts is much
smaller than in the star forming galaxies).

\begin{figure}[ht]
\epsscale{0.5}
\plotone{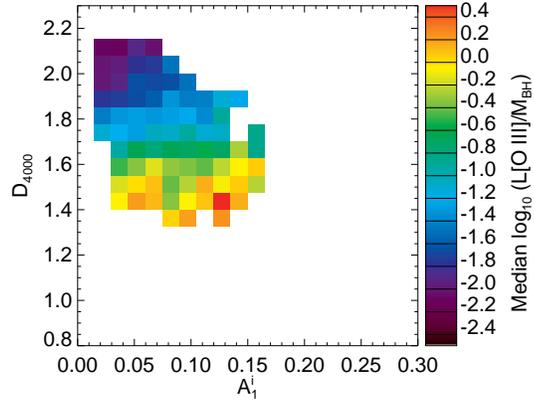}
\caption{Relationship between $L$[\ion{O}{3}]/$M_{BH}$, $D_{4000}$, and structure.  The color-coding of these two-dimensional histograms indicates the median L[\ion{O}{3}]/$M_{BH}$. The primary correlation is between stellar age and $L$[\ion{O}{3}]/$M_{BH}$.  There is no noticeable correlation between $L$[\ion{O}{3}]/$M_{BH}$ and lopsidedness in this data.}
\label{fig:s-lo3mbh-d}
\end{figure}

The age of the stellar population in AGN host galaxies decreases
systematically as a function of increasing AGN luminosity
\citet{kau+03c,kau+07}. We have also seen above that there is a very
strong inverse correlation between lopsidedness and stellar age. Is
the trend for more powerful AGN to be hosted by galaxies that are more
lopsided simply induced by these other correlations?  One way to
visualize this is to plot the joint dependence of AGN luminosity on
both stellar age ($D_{4000}$) and lopsidedness. This is shown in
Figure~\ref{fig:s-lo3mbh-d}. We see clearly that the primary correlation is that between the age of the stellar population and AGN luminosity with little
dependence on lopsidedness (the color-coded bands of AGN luminosity
are nearly horizontal). This is further quantified by the correlation
coefficients in Table~\ref{tab:parcor-agn}.

A more direct way to ask the question is as follows: for fixed values
of both galaxy structural and stellar age parameters, is there a
remaining trend for higher luminosity AGN to be hosted by more
lopsided galaxies? The large number of parametric dimensions involved
in such a test leads us to adopt a ``twinning'' strategy. That is, for
each AGN with a given value of $L$[\ion{O}{3}]$/M_{BH}$ we will find within
our sample another non-AGN galaxy with the same redshift, stellar
mass, surface mass density, and stellar age (same value of
$D_{4000}$). The matching tolerances in redshift, mass, surface mass
density, and $D_{4000}$ were 0.005, 0.05 dex, 0.05 dex, and 0.10, respectively.

\begin{figure}[ht]
\epsscale{1.1}
\plottwo{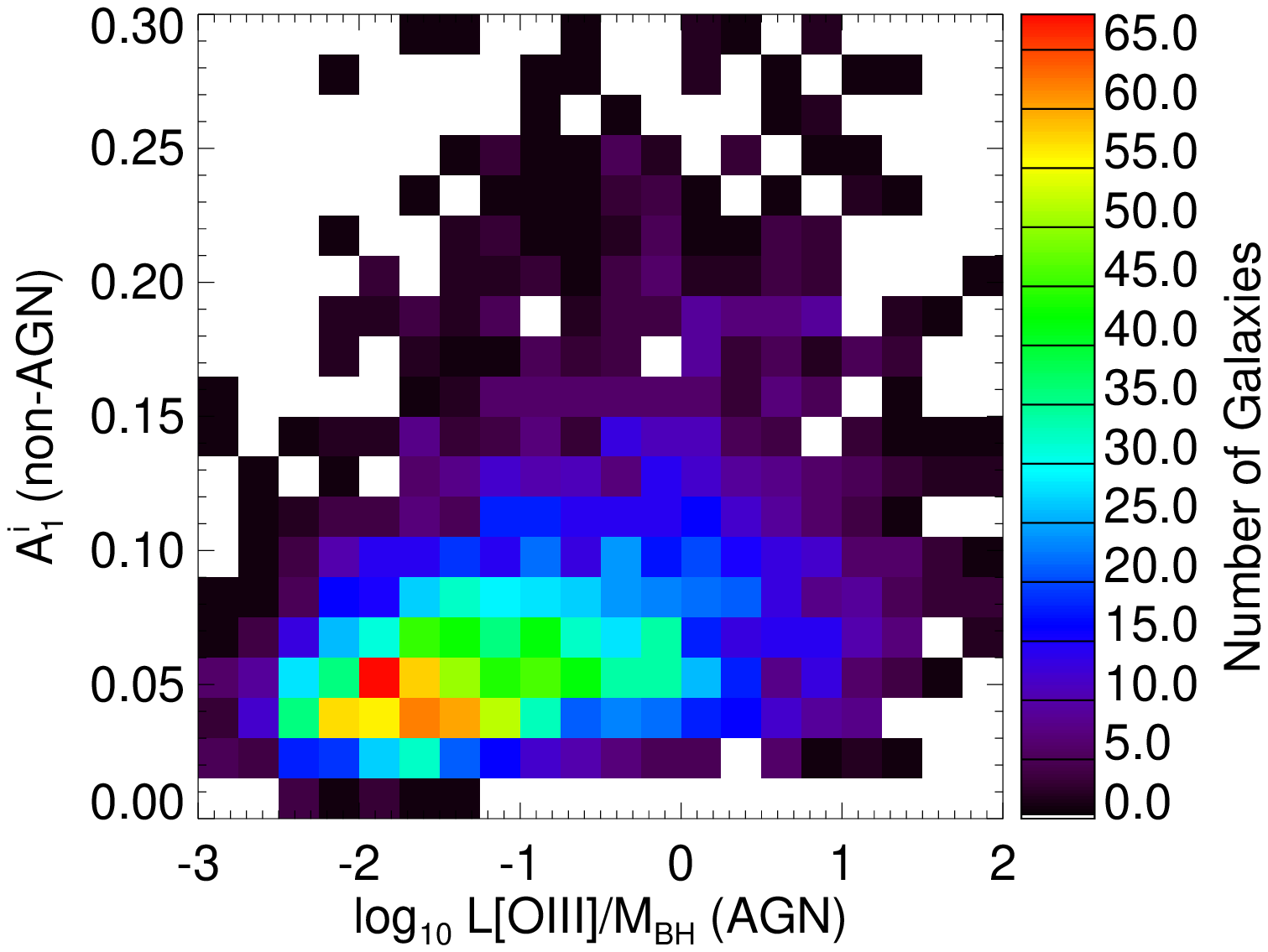}{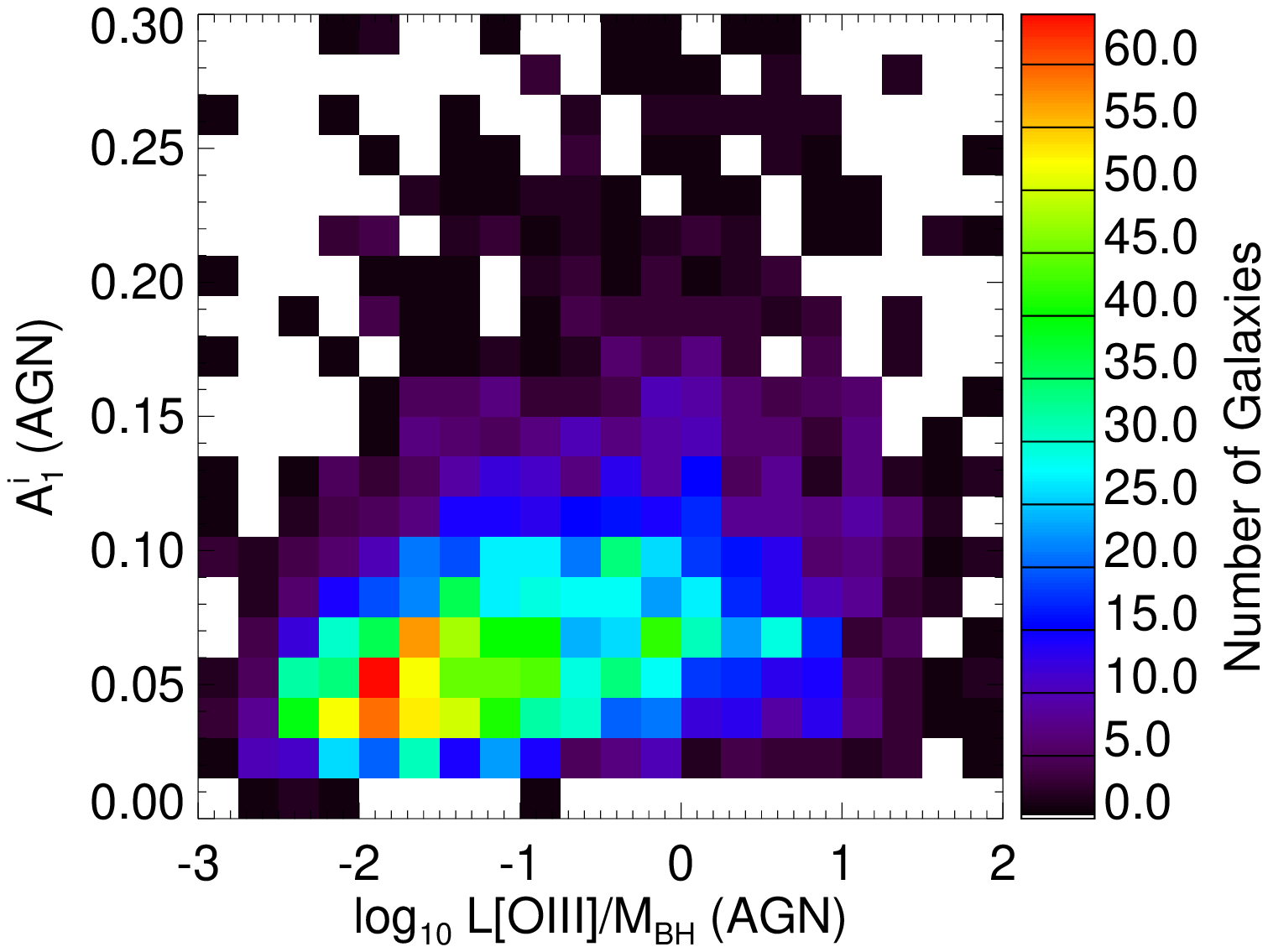}
\epsscale{0.5}
\plotone{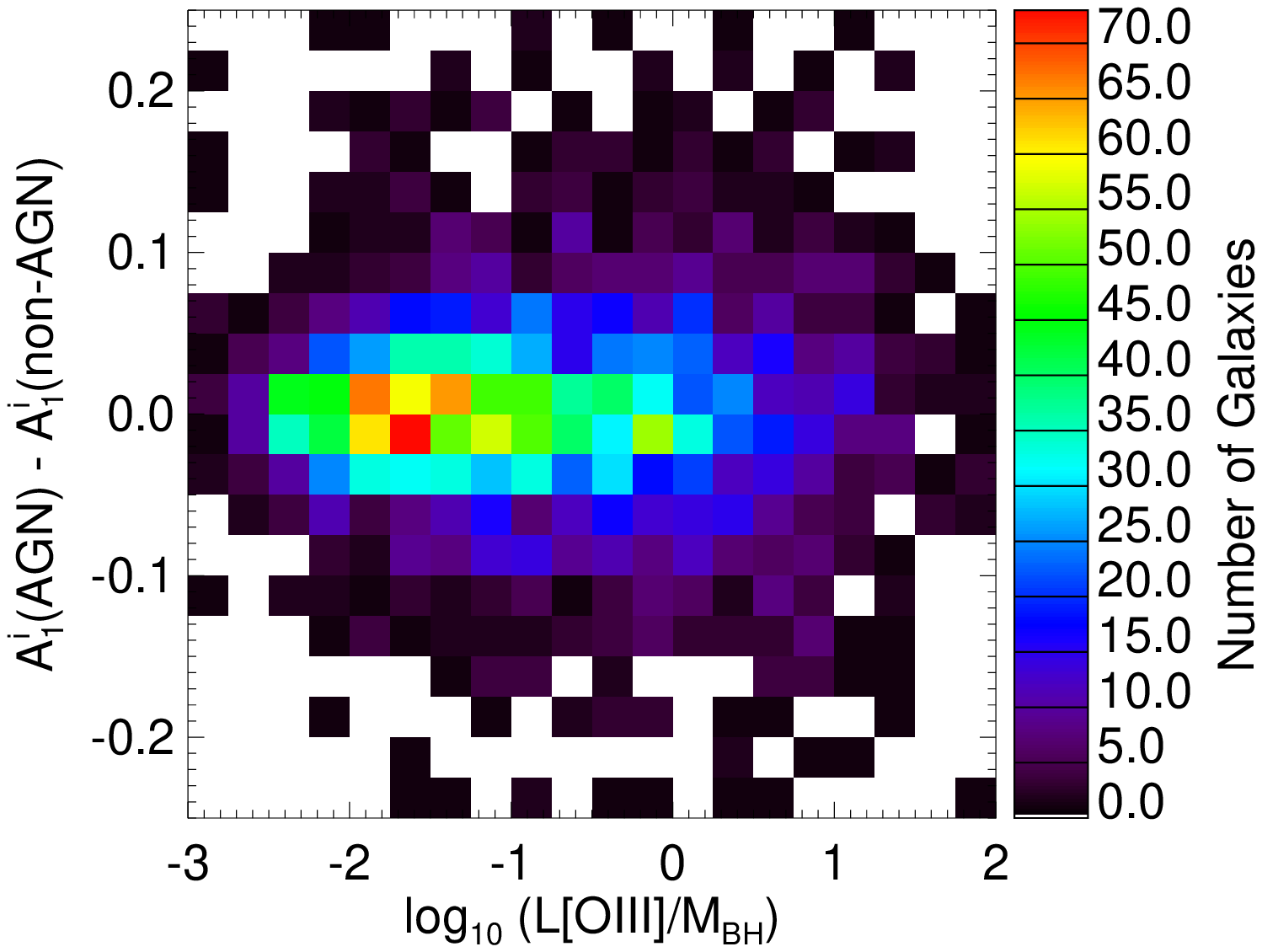}
\caption{{\em Top left and top right:} Two-dimensional distributions of $L$[\ion{O}{3}]/$M_{BH}$ and lopsidedness of non-AGN and AGN galaxy pairs matched in redshift, mass, mass density, and stellar age ($\Dbreak$).  The distributions show no significant difference between the lopsidedness-AGN luminosity relationship. {\em Bottom:} Difference in lopsidedness of the AGN and non-AGN in matched galaxies pairs vs. the AGN luminosity.  No preference is shown for the AGN or non-AGN to be more often more lopsided than the other for any value in our range of $L$[\ion{O}{3}]/$M_{BH}$.}
\label{fig:twins1}
\end{figure}

\begin{figure}[ht]
\epsscale{0.5}
\plotone{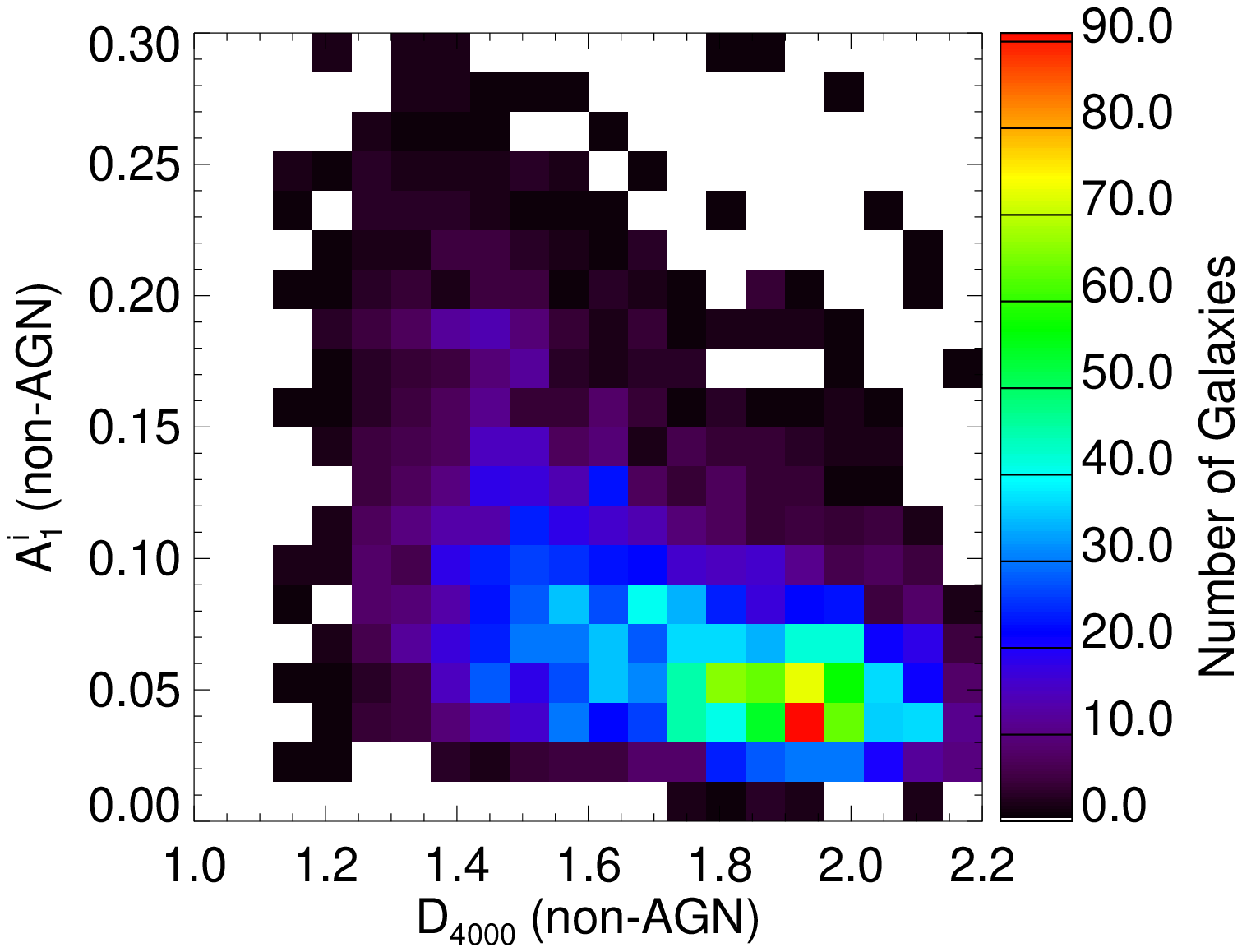}
\plotone{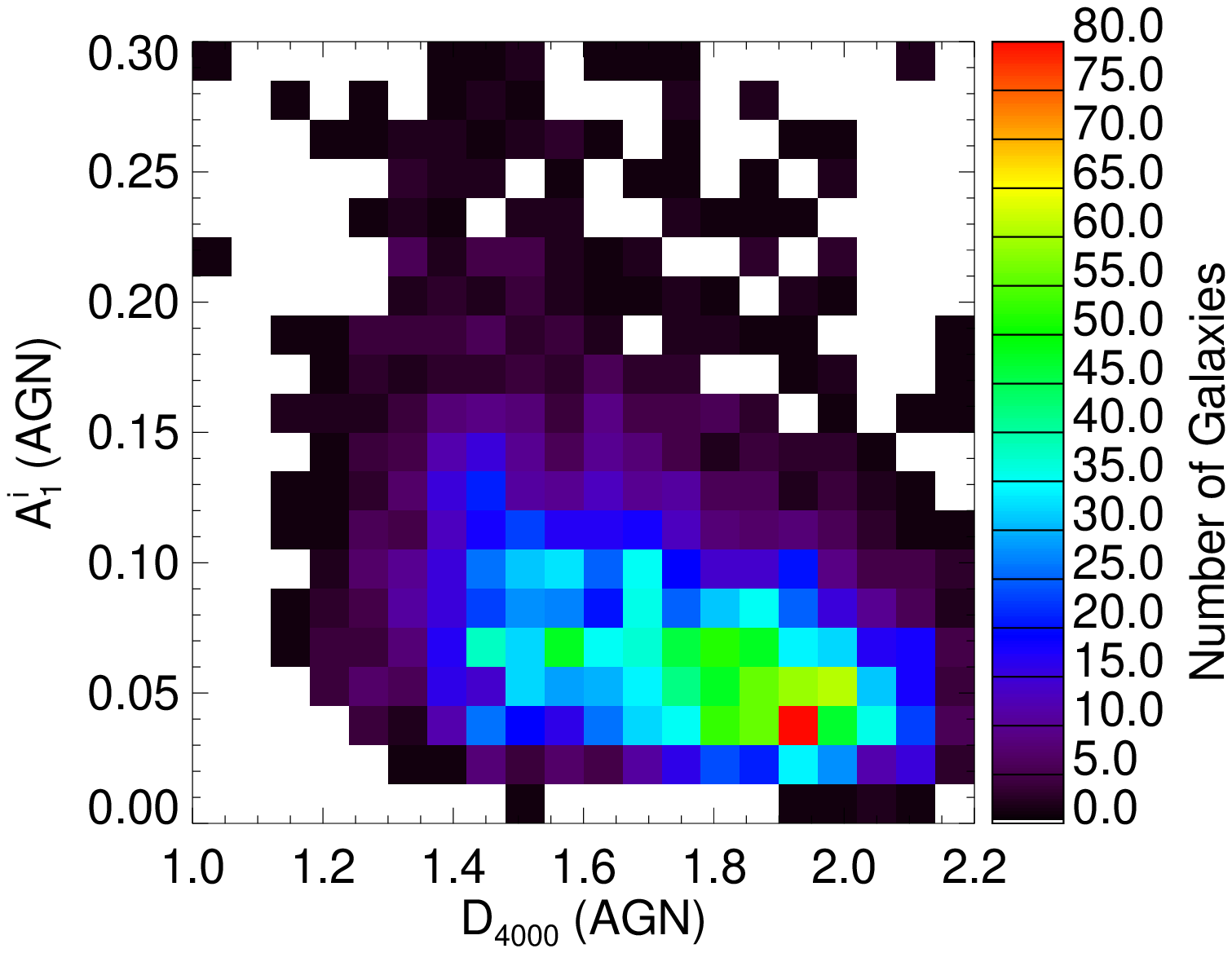}
\caption{Relationship between lopsidedness and mean stellar age ($\Dbreak$) for the AGN and non-AGN in pairs of galaxies matched in redshift, mass, mass density, and mean stellar age.  The star formation-lopsidedness connection is similar for AGN and non-AGN.}
\label{fig:twins2}
\end{figure}

In Figure~\ref{fig:twins1}, we compare the distributions of lopsidedness as a
function of the AGN luminosity for the AGN hosts and their non-AGN
twin galaxies. In this plot, we have assigned each non-AGN twin the
value for $L$[\ion{O}{3}]$/M_{BH}$ of its AGN twin. We have also plotted the
distribution of the difference in lopsidedness (AGN host minus twin)
as a function of AGN luminosity. We see from this figure that there is
no systematic difference between the AGN hosts and their twins. We
conclude that the trend for more powerful AGN to be hosted by galaxies
that are more lopsided is due to the strong link of the age of the
stellar population in the galaxy bulge to both the AGN luminosity and
to the lopsidedness of the host galaxy. This can also be seen in
Figure~\ref{fig:twins2} where we show the very similar dependence between stellar
age and lopsidedness for both the AGN hosts and their twins. These
results are quantified by the partial correlation coefficients in
Table~\ref{tab:parcor-agn}.

This leads us to the following thoughts. It is clear that the
processes that produce lopsidedness facilitate the delivery of gas
into the central regions of galaxies. It is also clear that the
presence of cold gas in the central region (however it arrives there) 
is required for the formation of stars there. The fact that the more rapidly growing black holes are strongly associated with a younger stellar population implies that the presence of cold gas in the central region is also required for rapid fueling of the black hole. The results above then suggest that the once the gas has arrived in the central kpc-scale region, some other process (not directly related to the cause of the lopsidedness) regulates the subsequent transfer of a small fraction of the gas inward by orders-of-magnitude in radius to the black hole accretion disk. 

It is interesting to compare our results to those of \citet{li+08b}, who have examined a very similar SDSS galaxy sample for a
link between the presence of close companion galaxies and the fueling
of black holes. They found that close companions are associated with a
significant boost in the star formation rate, but are not associated
with an enhancement of nuclear activity (compared to galaxies of
similar mass). Ellison et al. (2008a) reached similar conclusions with a different sample drawn from SDSS. These results contrast with our result showing that (at fixed galaxy mass, density, or concentration) higher AGN luminosities are preferentially found in galaxies that are more lopsided.

\citet{li+08b} concluded from their results that the star
formation induced by a close companion and the star formation
associated with black hole accretion are distinct events. They argued
that these events may be part of the same physical process, for
example a merger, provided they are separated in time. They therefore
speculated that accretion onto the black hole and its associated star
formation occurs after the two interacting galaxies have merged. This
scenario is consistent with our results, since lopsidedness will
persist for several galaxy dynamical times after the merger is
complete. In a future paper we will combine our results on close
companions and lopsidedness to try to test this idea.

\clearpage

\clearpage

\clearpage 

\clearpage

\clearpage

\clearpage

\clearpage

\section{Summary \label{sec:conclusions}}

We have studied a sample of approximately 25000 low-redshift ($z <
0.06$) galaxies from the Sloan Digital Sky Survey (SDSS) Data Release
4 to investigate the links between large-scale asymmetries in the
stellar mass distribution in galaxies (lopsidedness), star formation,
the metallicity of the interstellar medium, and the presence of active
galactic nuclei (AGN). Lopsidedness has been defined as the radially
averaged $m=1$ azimuthal Fourier amplitude ($A_1$) measured between
the radii enclosing 50\% and 90\% of the galaxy light in the SDSS
images (i.e. in the outer part of the galaxy). We have previously shown that lopsidedness traces the
distribution of the underlying stellar mass (Paper I). Lopsidedness is
a signpost of a non-equilibrium global dynamical state, and can be
induced by mergers, asymmetric accretion of cold gas, tidal
interactions, or underlying asymmetries related to the dark matter
halo. We have used spectra obtained through the SDSS 3\arcsec diameter
fibers (typical projected diameter of $\sim$ 3 kpc) to characterize
the stars, gas, and AGN in these central regions.

We have used the amplitude of the 4000 \AAA break, the strength of the
high-order Balmer absorption-lines, and the specific star formation
rate derived from the nebular emission-line to characterize the
stellar population in the central few-kpc-scale region. We found
strong links between lopsidedness and recent/on-going central
star-formation: galaxies with younger stellar populations are more
lopsided and more lopsided galaxies have younger stellar
populations. Starburst and post-starburst galaxies are the most
lopsided on average.

We have previously shown that galaxies with lower surface mass density
and mass are more lopsided (Paper I), and that galaxies with lower
mass and lower density have younger stellar populations
\citep{kau+03b,bri+04}. Here, we have shown that there is still a
strong correlation between lopsidedness and the age of the stellar
population even after their mutual dependences on these other
structural parameters have been removed. These results are consistent
with other evidence that mergers and tidal interactions trigger
central star-formation, but place these results on a firm statistical
base. They are also consistent with the idea that star formation in
galaxies today is regulated by the accretion of cold gas
\citep{k+05} which can excite lopsidedness \citep{bou+05}. They imply that the timescale for the transport of gas into the central region can not be significantly longer than the timescale over which lopsidedness persists in the outer disk. This is consistent with recent numerical simulations of star formation in minor mergers and galaxy interactions \citep{dim07,cox08}.

We have also shown that at fixed stellar mass, more lopsided galaxies
have systematically lower gas-phase metallicities by about 0.1 dex on average. This suggests that the processes causing lopsidedness deliver lower metallicity gas into the central region, but the small amplitude of the effect rules out extreme events as being typical.

Finally, we found that there is a trend for the more rapidly growing
black holes to be hosted by galaxies with higher average
lopsidedness. This is true when comparing galaxies at fixed galaxy
mass, surface mass density, and concentration. However, when the AGN
hosts were compared to non-active galaxies that were also matched in
the age of the stellar population in their central region, we found no
excess lopsidedness in the AGN hosts. The strongest link is that
between the youth of the stellar population and the growth rate of the
black hole. The correlations of these two properties with lopsidedness
are weaker. This suggests that the presence of cold gas in the central
few kpc-scale region (which is facilitated through the processes that
produce lopsidedness, and which leads to significant star formation)
is a necessary but not sufficient condition for subsequent fueling of
the growth of the black hole. Other processes are subsequently
required to deliver the gas from scales of a few kpc all the way to
the black hole accretion disk and these would not be directly related to the process(es) that produced lopsidedness.

Combining our results with the analysis by \citet{li+08b} and Ellison et al. (2008a) of the role of close companion galaxies in driving star formation and fueling black holes, suggests that the period of black hole growth may be
preferentially associated with the end stages of a minor merger. This
idea will be tested in a future paper.

We would like to thank Christy Tremonti for reading a draft of this
manuscript.  Funding for the SDSS and SDSS-II has been provided by the
Alfred P. Sloan Foundation, the Participating Institutions, the
National Science Foundation, the U.S. Department of Energy, the
National Aeronautics and Space Administration, the Japanese
Monbukagakusho, the Max Planck Society, and the Higher Education
Funding Council for England. The SDSS Web Site is
http://www.sdss.org/.

The SDSS is managed by the Astrophysical Research Consortium for the
Participating Institutions. The Participating Institutions are the
American Museum of Natural History, Astrophysical Institute Potsdam,
University of Basel, University of Cambridge, Case Western Reserve
University, University of Chicago, Drexel University, Fermilab, the
Institute for Advanced Study, the Japan Participation Group, Johns
Hopkins University, the Joint Institute for Nuclear Astrophysics, the
Kavli Institute for Particle Astrophysics and Cosmology, the Korean
Scientist Group, the Chinese Academy of Sciences (LAMOST), Los Alamos
National Laboratory, the Max-Planck-Institute for Astronomy (MPIA),
the Max-Planck-Institute for Astrophysics (MPA), New Mexico State
University, Ohio State University, University of Pittsburgh,
University of Portsmouth, Princeton University, the United States
Naval Observatory, and the University of Washington.

\begin{deluxetable}{cclr}
\tabletypesize{\small}
\tablecolumns{4}
\tablewidth{0pc}
\tablecaption{Partial Correlation Coefficients: Lopsidedness and Star Formation}
\tablehead{
	\colhead{} & \colhead{} & \colhead{Dependence} & \colhead{Partial}
	\\ \colhead{Par. 1} & \colhead{Par. 2} & \colhead{Removed} &
	\colhead{Corr. Coeff.} }
\startdata
$\log A_1^i$ & $D_{4000}$ & \nodata & $-0.58$ \\
$\log A_1^i$ & $H\delta_A$ & \nodata & $ 0.52$ \\
$\log A_1^i$ & $PC_1$ & \nodata & $-0.60$ \\
$\log A_1^i$ & $PC_2 - PC_1$ & \nodata & $ 0.46$ \\
$\log A_1^i$ & $\log SFR/M_*$ & \nodata & $ 0.33$ \\
$\log A_1^i$ & $g-i$ & \nodata & $-0.57$ \\
\hline
$\log A_1^i$  & $D_{4000}$ & $\log M_*$, $\log \mu_*$, $C_i$ & $-0.30$ \\
$\log A_1^i$  & $H\delta_A$ & $\log M_*$, $\log \mu_*$, $C_i$ & $ 0.27$ \\
$\log A_1^i$  & $PC_1$ & $\log M_*$, $\log \mu_*$, $C_i$ & $-0.33$ \\
$\log A_1^i$  & $PC_2 - PC_1$ & $\log M_*$, $\log \mu_*$, $C_i$ & $ 0.22$ \\
$\log A_1^i$  & $\log SFR/M_*$ & $\log M_*$, $\log \mu_*$, $C_i$ & $ 0.15$ \\
$\log A_1^i$  & $g-i$ & $\log M_*$, $\log \mu_*$, $C_i$ & $-0.21$ \\
\hline
$\log A_1^i$ & $\log M_*$ & $D_{4000}$ & $-0.14$ \\
$\log A_1^i$ & $\log \mu_*$ & $D_{4000}$ & $-0.26$ \\
$\log A_1^i$ & $C_i$ & $D_{4000}$ & $-0.17$ \\
\hline
$\log A_1^i$ & $H\delta_A$ & $D_{4000}$ & $ 0.11$ \\
$\log A_1^i$ & $PC_2$ & \nodata & $-0.20$ \\

\enddata
\label{tab:parcor-sfr}
\end{deluxetable}

\begin{deluxetable}{cclr}
\tabletypesize{\small}
\tablecolumns{4}
\tablewidth{0pc}
\tablecaption{Partial Correlation Coefficients: Lopsidedness and Metallicity}
\tablehead{
	\colhead{} &
	\colhead{} &
	\colhead{Dependence} &
	\colhead{Partial} \\
   \colhead{Par. 1} & 
	\colhead{Par. 2} & 
	\colhead{Removed} & 
	\colhead{Corr. Coeff.} }
\startdata
$\log A_1^i$ & $12 + \log (O/H)$ & $\log M_*$ & $-0.15$ \\
$\log A_1^i$ &  $12 + \log (O/H)$  & $\log M_*$, $\log \mu_*$ & $-0.10$ \\
$\log \mu_*$ & $12 + \log (O/H)$ & $\log M_*$ & $ 0.20$ \\
$\log \mu_*$ &  $12 + \log (O/H)$  & $\log A_1^i$, $\log M_*$ & $ 0.19$ \\
$\log SFR/M_*$ & $12 + \log (O/H)$ & $\log M_*$ & $ 0.02$ \\
\enddata
\label{tab:parcor-oh}
\end{deluxetable}

\begin{deluxetable}{cclr}
\tabletypesize{\small}
\tablecolumns{4}
\tablewidth{0pc}
\tablecaption{Partial Correlation Coefficients: Lopsidedness and AGN Activity}
\tablehead{
	\colhead{} &
	\colhead{} &
	\colhead{Dependence} &
	\colhead{Partial} \\
   \colhead{Par. 1} & 
	\colhead{Par. 2} & 
	\colhead{Removed} & 
	\colhead{Corr. Coeff.} }
\startdata
$\log A_1^i$ & $\log (L$[\ion{O}{3}]$/M_{BH})$ & \nodata & $ 0.32$ \\
\hline
$\log A_1^i$ & $\log (L$[\ion{O}{3}]$/M_{BH})$ & $\log M_*$ & $ 0.30$ \\
$\log A_1^i$ & $\log (L$[\ion{O}{3}]$/M_{BH})$ & $\log \mu_*$ & $ 0.23$ \\
$\log A_1^i$ & $\log (L$[\ion{O}{3}]$/M_{BH})$ & $C_i$ & $ 0.19$ \\
$\log A_1^i$ & $\log (L$[\ion{O}{3}]$/M_{BH})$ & $D_{4000}$ & $ 0.09$ \\
\hline
$\log A_1^i$  & $\log (L$[\ion{O}{3}]$/M_{BH})$ & $\log M_*$, $\log \mu_*$, $C_i$ & $ 0.22$ \\
$\log A_1^i$ &  $\log (L$[\ion{O}{3}]$/M_{BH})$  & $\log M_*$, $D_{4000}$ & $ 0.09$ \\
$\log A_1^i$ &  $\log (L$[\ion{O}{3}]$/M_{BH})$  & $\log \mu_*$, $D_{4000}$ & $ 0.07$ \\
$\log A_1^i$ &  $\log (L$[\ion{O}{3}]$/M_{BH})$  & $C_i$, $D_{4000}$ & $ 0.05$ \\
$\log A_1^i$ & $\log (L$[\ion{O}{3}]$/M_{BH})$ & $\log M_*$, $\log \mu_*$, $C_i$, $D_{4000}$ & $ 0.06$ \\

\enddata
\label{tab:parcor-agn}
\end{deluxetable}

\end{document}